\documentclass[final,5p,times,twocolumn]{elsarticle}
\usepackage{amsmath}
\usepackage{mathtools, cuted}
\usepackage{setspace}
\usepackage{flushend, cuted}
\usepackage{lipsum}
\usepackage{lingmacros}
\usepackage{tree-dvips}
\usepackage{varioref}
\usepackage{colortbl}
\usepackage{url}
\usepackage{hyperref}
\usepackage{balance}

\usepackage{graphicx}
\usepackage{epstopdf}
\usepackage{color,soul}
\usepackage{amsfonts}
\usepackage{subfigure}
\usepackage{multicol}
\usepackage{multirow}
\usepackage{hhline}
\usepackage{amssymb}
\usepackage{newlfont}
\usepackage{mathrsfs}
\usepackage{mathtools}
\usepackage{mathrsfs}
\usepackage{multirow}
\usepackage{nomencl}   
\usepackage{fancyhdr}
\usepackage{caption}
\usepackage{slashbox}
\usepackage{float}
\usepackage{placeins}
\usepackage{tabularx}
\usepackage[table]{xcolor}
\usepackage{physics}
\usepackage[linesnumbered,ruled]{algorithm2e}
\SetKwRepeat{Do}{do}{while}%

\DeclareRobustCommand{\frac}[3][0pt]{%
	{\begingroup\hspace{#1}#2\hspace{#1}\endgroup\over\hspace{#1}#3\hspace{#1}}}

\let\norm\undefined 
\DeclarePairedDelimiter\norm{\lVert}{\rVert}

\begin{document}

\pagestyle{myheadings}

\begin{frontmatter}
	
\title{Ultrasound Strain Imaging using ADMM}


\author[add1]{Md~Ashikuzzaman}
\ead{m\_ashiku@encs.concordia.ca}
\author[add1]{Hassan~Rivaz}
\ead{hrivaz@ece.concordia.ca}

\address[add1]{Department of Electrical and Computer Engineering, Concordia University, Montreal, QC, H3G 1M8, Canada.}
\begin{abstract}
Ultrasound strain imaging, which delineates mechanical properties to detect tissue abnormalities, involves estimating the time-delay between two radio-frequency (RF) frames collected before and after tissue deformation. The existing regularized optimization-based time-delay estimation (TDE) techniques suffer from at least one of the following drawbacks: 1) The regularizer is not aligned with tissue deformation physics due to taking only the first-order displacement derivative into account. 2) The $L2$-norm of the displacement derivatives, which oversmooths the estimated time-delay, is utilized as the regularizer. 3) The absolute value function should be approximated by a smooth function to facilitate the optimization of $L1$-norm. Herein, to resolve these shortcomings, we propose employing the alternating direction method of multipliers (ADMM) for optimizing a novel cost function consisting of $L2$-norm data fidelity term and $L1$-norm first- and second-order spatial continuity terms. ADMM empowers the proposed algorithm to use different techniques for optimizing different parts of the cost function and obtain high-contrast strain images with smooth background and sharp boundaries. We name our technique \textbf{A}DMM for tota\textbf{L} varia\textbf{T}ion \textbf{R}eg\textbf{U}lar\textbf{I}zation in ultrasound \textbf{ST}rain imaging (ALTRUIST). In extensive simulation, phantom, and \textit{in vivo} experiments, ALTRUIST substantially outperforms GLUE, OVERWIND, and $L1$-SOUL, three recently-published TDE algorithms, both qualitatively and quantitatively. ALTRUIST yields $118\%$, $104\%$, and $72\%$ improvements of contrast-to-noise ratio over $L1$-SOUL for simulated, phantom, and \textit{in vivo} liver cancer datasets, respectively. We will publish the ALTRUIST code after the acceptance of this paper at \url{ http://code.sonography.ai}.                                        
\end{abstract}


\end{frontmatter}

                  
\section{Introduction}
Ultrasound is a widely-used medical imaging modality due to its cost-effectiveness, ease of use, non-invasiveness, and portability. One of the eminent diagnostic applications of ultrasound imaging is elastography~\cite{ophir1}, which comprises mapping the mechanical properties of tissue. Elastography distinguishes between healthy and diseased tissues by displaying their elastic contrast, presuming that particular pathologies like a tumor, cancer, benign lesion, cyst, \textit{etc.}, modify tissue elasticity. Thus far, ultrasound elastography has been employed in breast tissue classification~\cite{hall2003vivo,hybrid,jiang_2015,rglue_ius}, liver health assessment~\cite{DPAM,guest,tang2015ultrasound,soul}, ablation monitoring~\cite{varghese2003elastographic,rivaz2008ablation,mariani2014real}, overseeing vascular~\cite{de2000characterization, li2019two, maurice2004noninvasive} and cardiac~\cite{chen2009ultrasound, strachinaru2017cardiac, konofagou2002myocardial} health, and numerous other clinical applications. This work focuses on free-hand palpation quasi-static strain imaging which entails acquiring time-series radio-frequency (RF) frames from a tissue being deformed by a quasi-static force created with a hand-held probe. The deformation field between the pre- and post-deformed frames is obtained by a certain time-delay estimation (TDE) technique. The spatial derivative of the displacement field provides the strain image where changes in strain values can potentially distinguish the pathologic tissue from the healthy one.

The non-trivial task of TDE is accomplished by tracking the displacement between the frames under consideration. Three mainstream classes of speckle-tracking techniques have been proposed: window-based~\cite{luo2010fast, hall2003vivo, Zahiri_2006, Xunchang_2004, sharmin_2013}, deep learning-based~\cite{kibria2018gluenet,gao2019learning,wu2018direct,peng2020neural,tehrani2022bi,ali_2020}, and energy function optimization-based~\cite{rpca_glue,intro14,mglue,pan2014two,rglue_tuffc, islam2018new,guest}. This work concerns energy-based speckle tracking techniques, which obtain the displacement estimates by optimizing a non-linear cost function enforcing data fidelity and spatial continuity constraints. Although these algorithms yield attractive noise-suppression ability, they are computationally expensive. However, the issue of computational load has been resolved by a series of works~\cite{DP,intro14,DPAM,GLUE}. Along this line, Second-order Ultrasound Elastography (SOUL)~\cite{soul} has been proposed, which penalizes both first- and second-order displacement derivatives to formulate a physics-based regularization constraint and render a high-contrast strain image. However, like its closely-related previous work GLobal Ultrasound Elastography (GLUE)~\cite{GLUE}, SOUL uses the $L2$-norm to formulate the regularizer, which blurs the boundary between different tissue types. Total Variation Regularization and Window-based TDE (OVERWIND)~\cite{overwind} has addressed the blurring issue by penalizing the $L1$-norm of displacement derivatives instead of the $L2$-norm. Although OVERWIND obtains sharper strain images than GLUE, it incorporates a non-physics-based first-order regularizer. Our recently published technique $L1$-SOUL~\cite{soulmate} combines the advantages of SOUL and OVERWIND by penalizing the $L1$-norms of both first- and second-order displacement derivatives. However, $L1$-SOUL defines the $L1$-norm in terms of a smooth approximation of the absolute value function to facilitate the simultaneous optimizations of $L2$ data and $L1$ continuity norms. Such a coarse approximation hinders the optimality of $L1$-SOUL's displacement tracking performance.

In this paper, we propose using the alternating direction method of multipliers (ADMM)~\cite{boyd2011distributed,parikh2014proximal,eckstein2012augmented,wahlberg2012admm,giselsson2016linear} to optimize a cost function consisting of $L2$-norm data fidelity term and $L1$-norm first- and second-order continuity terms. ADMM splits the target variable into two variables and couples them using a valid constraint to decompose the original optimization problem into two simpler subproblems. These subproblems are alternatingly solved for the associated variables, providing two major advantages. First, it allows using different optimization methods for solving different parts of the TDE problem, and therefore, an approximation of the $L1$-norm is not required. Second, it establishes a sample-wise regularization strategy, which exploits the spatial distribution of elasticity in each iteration to control the level of smoothness. Combining the aforementioned features, the proposed technique aims at attaining the maximum potential of a physics-based regularizer in ultrasonic strain elastography. Unlike previous works~\cite{mohammed2021multifrequency,otesteanu2018fem,mohammed20192d} employing ADMM in the reconstruction of tissue elasticity modulus, this work applies ADMM in ultrasound strain imaging for optimizing a $L1$-norm (alternatively known as total variation) regularizer containing both first- and second-order displacement derivatives. We name our technique ALTRUIST: \textbf{A}DMM for tota\textbf{L} varia\textbf{T}ion \textbf{R}eg\textbf{U}lar\textbf{I}zation in ultrasound \textbf{ST}rain imaging. To the best of our knowledge, ALTRUIST is the first technique to employ ADMM in ultrasonic strain elastography. We have validated ALTRUIST with simulated, phantom, and \textit{in vivo} liver cancer datasets. Similar to our previous work~\cite{guest,RAPID_TMI,soul}, the ALTRUIST code will be published at \url{http://code.sonography.ai} after the acceptance of this paper.          

\begin{algorithm}[tb]
	\SetKwInOut{Input}{Input}
	\SetKwInOut{Output}{Output}
	\KwIn{Pre- and post-deformed RF frames $I_{1}$ and $I_{2}$, DP initial estimate $d$}
	\KwOut{Total displacement field $d + \Delta d$, axial strain image}
	
	Formulate the cost function $C$ containing $L2$-norm data and $L1$-norm first- and second-order continuity terms (Eq.~\ref{eq:admm_c2})\;
	Split $\Delta d$ into two variables $\Delta d$ and $\nu$ and devise a constrained optimization problem (Eq.~\ref{eq:constrained})\;
	Formulate the augmented Lagrangian corresponding to the constrained problem (Eq.~\ref{eq:augmented})\;
	Alternatively solve for $\Delta d$ and $\nu$ following an iterative scheme (Eqs.~\ref{eq:solvex} - \ref{eq:updateu})\;
	Add optimal $\Delta d$ to $d$ for obtaining the final displacement field\;
	Differentiate the axial displacement field spatially to find the axial strain image\; 
	
	\caption{Workflow of the ALTRUIST algorithm}
	\label{algo1}
\end{algorithm}

\section{Methods}
Let $I_{1} \in \mathbb{R}^{m \times n}$ and $I_{2} \in \mathbb{R}^{m \times n}$ be two time-series RF frames collected from a tissue being deformed by an external or internal force. Our goal is to obtain the axial strain map between $I_{1}$ and $I_{2}$. To that end, we calculate the initial axial and lateral displacement fields $a_{i,j}$ and $l_{i,j}$, $1 \leq i \leq m$ and $1 \leq j \leq n$, using DP~\cite{DP}. The most vital step is to refine this initial displacement estimate using a TDE algorithm. This section presents ALTRUIST, the proposed TDE technique in detail.

\subsection{ADMM for Total Variation Regularization in Ultrasound Strain Imaging (ALTRUIST)}
Our recently published $L1$-norm optimization-based TDE algorithm $L1$-SOUL~\cite{soulmate} approximates the absolute value function with a differentiable function to facilitate the simultaneous optimizations of $L2$ data norm and $L1$ continuity norm. This inexact approximation of $L1$-norm limits the TDE technique's ability to render sharp transitions at strain boundaries, while maintaining a smooth background. In addition, the regularization scheme of $L1$-SOUL treats every sample equally, which hinders a sharp transition at inclusion edges. To resolve these shortcomings of $L1$-SOUL, ALTRUIST employs the powerful technique ADMM for optimizing the penalty function containing data fidelity and total variation regularization terms. ADMM facilitates the use of different techniques for optimizing $L2$- and $L1$-norms and therefore, an inexact approximation of the $L1$-norm is no longer necessary. In addition, ADMM can handle every RF sample individually and adaptively impose different levels of smoothing based on the spatial distribution of elasticities, allowing sharp strain transitions at target boundaries.

ALTRUIST organizes the cost function as Eq.~\ref{eq:admm_c} so that it suits the incorporation of ADMM.

\begin{equation}
\begin{aligned}
&C (\Delta a_{1,1},...,\Delta a_{m,n},\Delta l_{1,1},...,\Delta l_{m,n}) =\\
&\frac{1}{2}\norm{D_{I}(i,j,a_{i,j},l_{i,j},\Delta a_{i,j},\Delta l_{i,j})}_{2}^{2} + \norm{D_{R} \Delta d + D_{R} d + \mathcal{E}}_{1}
\end{aligned}
\label{eq:admm_c}
\end{equation}

\noindent
\noindent
where $d \in \mathbb{R}^{2mn \times 1}$ and $\Delta d \in \mathbb{R}^{2mn \times 1}$ stack the DP initial estimates and the fine-tuning displacement estimates, respectively. $\mathcal{E} \in \mathbb{R}^{(8mn + 2n) \times 1}$ contains the adaptive regularization terms~\cite{guest,soul}. The data constancy term $D_{I}(i,j,a_{i,j},l_{i,j},\Delta a_{i,j}, \Delta l_{i,j})$ is defined as follows.

\begin{equation}
\begin{aligned}
&D_{I}(i,j,a_{i,j},l_{i,j},\Delta a_{i,j},\Delta l_{i,j})=\\
&I_{1}(i,j)-I_{2}(i+a_{i,j}+\Delta a_{i,j},j+l_{i,j}+\Delta l_{i,j})
\end{aligned}
\label{eq:data}
\end{equation}

We approximate $I_{2}$ by its first-order Taylor series expansion to remove the non-linearity present in the data function:

\begin{equation}
\begin{aligned}
&I_{2}(i+a_{i,j}+\Delta a_{i,j},j+l_{i,j}+\Delta l_{i,j}) \approx \\
&I_{2}(i+a_{i,j},j+l_{i,j})+\Delta a_{i,j}I_{2,a}^{'}+\Delta l_{i,j}I_{2,l}^{'} 
\end{aligned}
\label{eq:i2_taylor}
\end{equation}

\noindent
where $I_{2,a}^{'}$ and $I_{2,l}^{'}$, respectively, denote the axial and lateral derivatives of $I_{2}$ at $(i+a_{i,j},j+l_{i,j})$. Now the cost function can be formulated as follows.

\begin{equation}
\begin{aligned}
&C (\Delta a_{1,1},...,\Delta a_{m,n},\Delta l_{1,1},...,\Delta l_{m,n}) =\\
&\frac{1}{2}\norm{\Xi - D^{'} \Delta d}_{2}^{2} + \norm{D_{R} \Delta d + D_{R} d + \mathcal{E}}_{1}
\end{aligned}
\label{eq:admm_c2}
\end{equation}

\noindent
Note that $C(\cdot)$ is a summation of $L2$-data and $L1$-continuity norms, both convex functions, and therefore, ADMM is guaranteed to converge~\cite{boyd2011distributed,eckstein2012augmented}. $D^{'}$ and $\Xi$ are defined by Eqs.~\ref{eq:dprime} and \ref{eq:datadiff}, respectively.

\begin{equation}
\Xi =\begin{bmatrix}
g_{1,1} & g_{1,2} & g_{1,3} & \dots & g_{m,n}
\end{bmatrix}^T
\label{eq:datadiff}
\end{equation}

\noindent
Here, $g_{i,j}$ is defined as follows.

\begin{equation}
g_{i,j}=
I_{1}(i,j)-I_{2}(i+a_{i,j},j+l_{i,j}) 
\label{eq:datadiff2}
\end{equation}

\noindent
$D_{R} \in \mathbb{R}^{(8mn + 2n) \times 2mn}$ embeds the weighted first- and second-order derivative operators in axial and lateral directions, and is defined as:

\begin{equation}
D_{R} =\begin{bmatrix}
D_{f,a}\\
D_{1,a}\\
D_{1,l}\\
D_{2,a}\\
D_{2,l}
\end{bmatrix}
\label{eq:dr}
\end{equation}

\noindent
where $D_{f,a} \in \mathbb{R}^{2n \times 2mn}$ is given by:
$D_{f,a} = \begin{bmatrix} D_{f,a,1} & O & O & \dots & O \end{bmatrix}$.

\noindent
Here, $O \in \mathbb{R}^{2n \times 2n}$ is a zero matrix and $D_{f,a,1} \in \mathbb{R}^{2n \times 2n}$ is defined as:  
$D_{f,a,1} = diag(\gamma, 0, \gamma, 0, \dots, \gamma, 0)$

\noindent
where $\gamma$ denotes a continuity weight. $D_{1,a} \in \mathbb{R}^{2mn \times 2mn}$ represents the first-order derivative in the axial direction and is defined as:
\begin{equation}
D_{1,a}=
\begin{bmatrix}
O & O & O & \dots & O \\
-B_{1,a} & B_{1,a} & O & \ddots & O \\
O & -B_{1,a} & B_{1,a} & \ddots & O \\
\vdots & \ddots & \ddots & \ddots & \vdots \\
O & \dots & O & -B_{1,a} & B_{1,a}
\label{eq:d1a}
\end{bmatrix}
\end{equation}

\noindent
where $B_{1,a} \in \mathbb{R}^{2n \times 2n}$ is defined as:
$B_{1,a} = diag(\alpha_{1},\beta_{1},\alpha_{1},\beta_{1},\dots,\alpha_{1},\beta_{1})$.

Here, $\alpha_{1}$ and $\beta_{1}$ refer to the axial and lateral regularization parameters, respectively. $D_{1,l} \in \mathbb{R}^{2mn \times 2mn}$ denotes the first-order derivative in the lateral direction and is given by:

$D_{1,l} = diag(B_{1,l},B_{1,l},B_{1,l},\dots,B_{1,l})$

\noindent
where $B_{1,l} \in \mathbb{R}^{2n \times 2n}$ is defined as:

\begin{equation}
B_{1,l}=
\begin{bmatrix}
0 & 0 & 0 & 0 & \dots & \dots & 0 \\
0 & 0 & 0 & 0 & \dots & \dots & 0 \\
-\alpha_{2} & 0 & \alpha_{2} & 0 & \dots & \dots & 0\\
0 & -\beta_{2} & 0 & \beta_{2} & \dots & \dots & 0\\
\vdots & \ddots & \ddots & \ddots & \ddots & \dots & \vdots\\
0 & \dots & \dots & -\alpha_{2} & 0 & \alpha_{2} & 0\\
0 & \dots & \dots & 0 & -\beta_{2} & 0 & \beta_{2}
\label{eq:b1l}
\end{bmatrix}
\end{equation}

Here, $\alpha_{2}$ and $\beta_{2}$, respectively, denote the axial and lateral continuity weights. $D_{2,a} \in \mathbb{R}^{2mn \times 2mn}$ formulates the second-order derivative in the axial direction and is defined as Eq.~\ref{eq:d2a}.

\begin{small}
\begin{equation}
D_{2,a}=
\begin{bmatrix}
O & O & O & O & \dots & O \\
B_{2,a} & -2B_{2,a} & B_{2,a} & O & \dots & O \\
O & B_{2,a} & -2B_{2,a} & B_{2,a} & \dots & O \\
\vdots & \ddots & \ddots & \ddots & \ddots & \vdots\\
O & \dots & \dots & B_{2,a} & -2B_{2,a} & B_{2,a}\\
O & \dots & \dots & O & O & O
\label{eq:d2a}
\end{bmatrix}
\end{equation}
\end{small}

\noindent
where $B_{2,a} = diag(\theta_{1},\lambda_{1},\theta_{1},\lambda_{1},\dots,\theta_{1},\lambda_{1})$. Here, $\theta_{1}$ and $\lambda_{1}$ stand for the axial and lateral regularization weights, respectively. $D_{2,l} \in \mathbb{R}^{2mn \times 2mn}$ presents the second-order derivative in the lateral direction which is defined as follows:
$D_{2,l} = diag(B_{2,l},B_{2,l},B_{2,l},\dots,B_{2,l})$


\noindent
where
\begin{small}
\begin{equation}
B_{2,l}=
\begin{bmatrix}
0 & 0 & 0 & 0 & 0 & 0 & \dots & \dots & 0 \\
0 & 0 & 0 & 0 & 0 & 0 & \dots & \dots & 0 \\
\theta_{2} & 0 & -2\theta_{2} & 0 & \theta_{2} & 0 & \dots & \dots & 0 \\
0 & \lambda_{2} & 0 & -2\lambda_{2} & 0 & \lambda_{2} & \dots & \dots & 0 \\
\vdots & \ddots & \ddots & \ddots & \ddots & \ddots & \ddots & \ddots & \vdots\\
0 & \dots & \dots & \theta_{2} & 0 & -2\theta_{2} & 0 & \theta_{2} & 0\\
0 & \dots & \dots & 0 & \lambda_{2} & 0 & -2\lambda_{2} & 0 & \lambda_{2}\\
0 & \dots & \dots & 0 & 0 & 0 & 0 & 0 & 0\\
0 & \dots & \dots & 0 & 0 & 0 & 0 & 0 & 0
\label{eq:b2l}
\end{bmatrix}
\end{equation}
\end{small}

\noindent
$\theta_{2}$ and $\lambda_{2}$, respectively, denote the axial and lateral continuity parameters. Now we split $\Delta d$ into two variables $\Delta d$ and $\nu$ and impose $D_{R} \Delta d + D_{R} d + \mathcal{E} = \nu$. Then the optimization problem is transformed to:

\begin{equation}
\begin{aligned}
(\Delta \hat{d}, \hat{\nu}) = &\underset{\Delta d, \nu}{\arg\min} \frac{1}{2}\norm{\Xi - D^{'} \Delta d}_{2}^{2} + \norm{\nu}_{1}\\
&s.t. \quad D_{R} \Delta d + D_{R} d + \mathcal{E} = \nu
\end{aligned}
\label{eq:constrained}
\end{equation}

We convert the constrained optimization problem into an unconstrained one by formulating the augmented Lagrangian:

\begin{equation}
\begin{aligned}
(\Delta \hat{d}, \hat{\nu}) = &\underset{\Delta d, \nu}{\arg\min} \frac{1}{2}\norm{\Xi - D^{'} \Delta d}_{2}^{2} + \norm{\nu}_{1} +\\ &\frac{\zeta}{2}\norm{D_{R} \Delta d + D_{R} d + \mathcal{E} - \nu + u}_{2}^{2}
\end{aligned}
\label{eq:augmented}
\end{equation}

\noindent
where $\zeta$ is a tunable parameter and $u$ is the Lagrange multiplier. Now Eq.~\ref{eq:augmented} is solved alternatively to obtain the optimal solutions for $\Delta d$ and $\nu$:

\pagebreak
\begin{strip}
	\noindent\makebox[\linewidth]{\rule{18cm}{0.4pt}}
	\begin{equation}
	D^{'}=
	\begin{bmatrix}
	I_{2,a}^{'}(1, 1) & I_{2,l}^{'}(1, 1) & 0 & 0 & \dots & 0 & 0\\
	0 & 0 & I_{2,a}^{'}(1, 2) & I_{2,l}^{'}(1, 2) & \dots & 0 & 0\\
	\dots & \dots & \ddots & \ddots & \dots & \dots & \dots\\
	0 & 0 & \dots & \dots & \dots & I_{2,a}^{'}(m, n) & I_{2,l}^{'}(m, n)\\ 
	\end{bmatrix}
	\label{eq:dprime}
	\end{equation}
	\noindent\makebox[\linewidth]{\rule{18cm}{0.4pt}}
\end{strip}

For \textit{K} iterations \{
\begin{equation}
\begin{aligned}
\Delta \hat{d} \leftarrow \underset{\Delta d}{\arg\min} &\{\frac{1}{2}\norm{\Xi - D^{'} \Delta d}_{2}^{2} +\\ &\frac{\zeta}{2}\norm{D_{R} \Delta d + D_{R} d + \mathcal{E} - \hat{\nu} + u}_{2}^{2}\}
\end{aligned}
\label{eq:solvex}
\end{equation}

\begin{equation}
\begin{aligned}
\hat{\nu} \leftarrow &\underset{\nu}{\arg\min} \frac{1}{2}\norm{\nu - (D_{R} \Delta \hat{d} + D_{R} d + \mathcal{E} + u)}_{2}^{2} + \frac{1}{\zeta}\norm{\nu}_{1}
\end{aligned}
\label{eq:solvev}
\end{equation}

\begin{equation}
\begin{aligned}
u \leftarrow  u + D_{R} \Delta \hat{d} + D_{R} d + \mathcal{E} - \hat{\nu}
\end{aligned}
\label{eq:updateu}
\end{equation}

\}


The quadratic cost function in Eq.~\ref{eq:solvex} is optimized in the similar fashion as SOUL~\cite{soul}. The Lagrange multiplier is updated in Eq.~\ref{eq:updateu}. The classic form of the cost function in Eq.~\ref{eq:solvev} is optimized using the shrinkage function:

\begin{equation}
\hat{\nu} \leftarrow S_{\frac{1}{\zeta}}(D_{R} \Delta \hat{d} + D_{R} d + \mathcal{E} + u)
\label{eq:shrinkage}
\end{equation}

\noindent
where the shrinkage function $S_{\frac{1}{\zeta}}(\cdot)$ is defined as follows.

\begin{equation}
S_{\frac{1}{\zeta}}(\cdot) = \text{sign}(\cdot)\max\{\abs{\cdot} - \frac{1}{\zeta}, 0\}
\label{eq:shrinkage_def}
\end{equation}

The optimal refinement field is added to the DP initial estimate to find the final displacement field. A spatial differentiation of the final displacement estimate is performed to obtain the strain map. The workflow of ALTRUIST is summarized in Algorithm~\ref{algo1}.

	\begin{figure}[h]
	\centering
	\subfigure[GLUE, 3]{{\includegraphics[width=.16\textwidth, height=.14\textwidth]{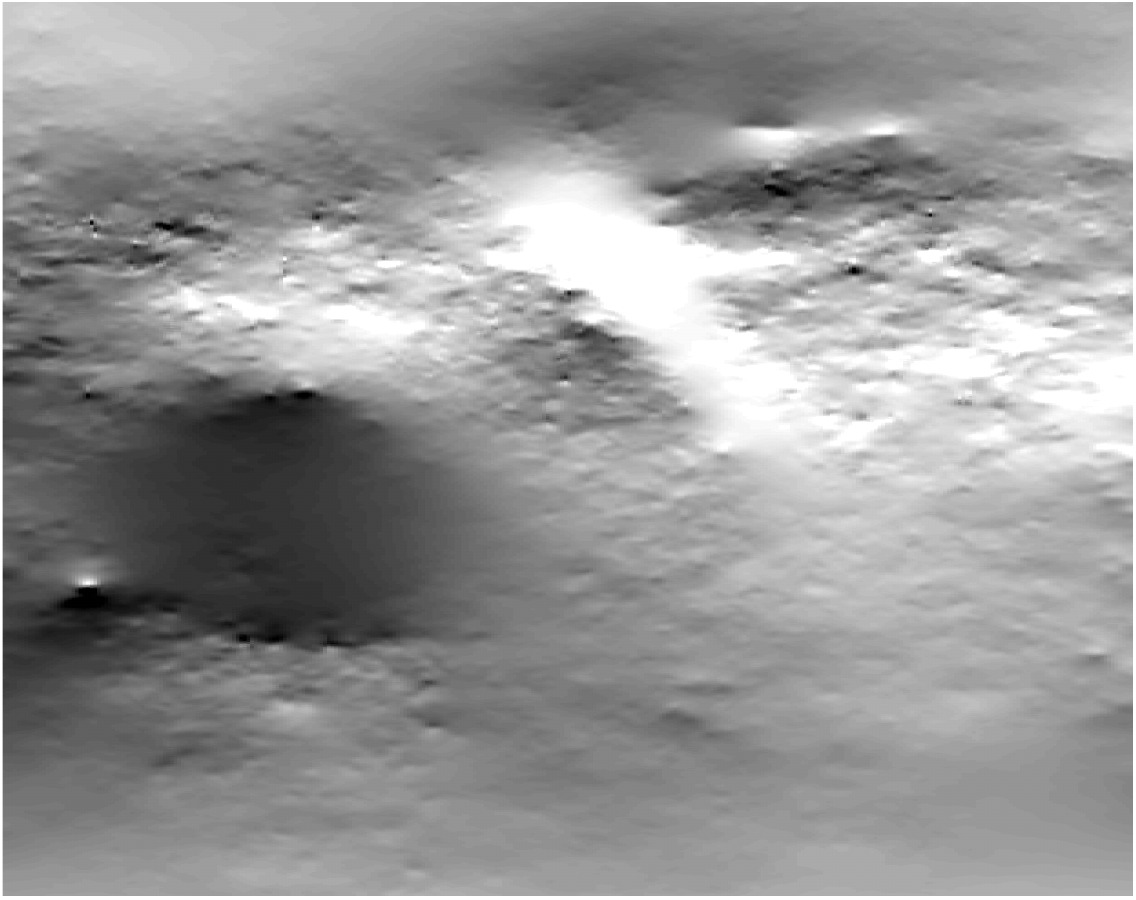}}}%
	\subfigure[GLUE, 43]{{\includegraphics[width=.16\textwidth, height=.14\textwidth]{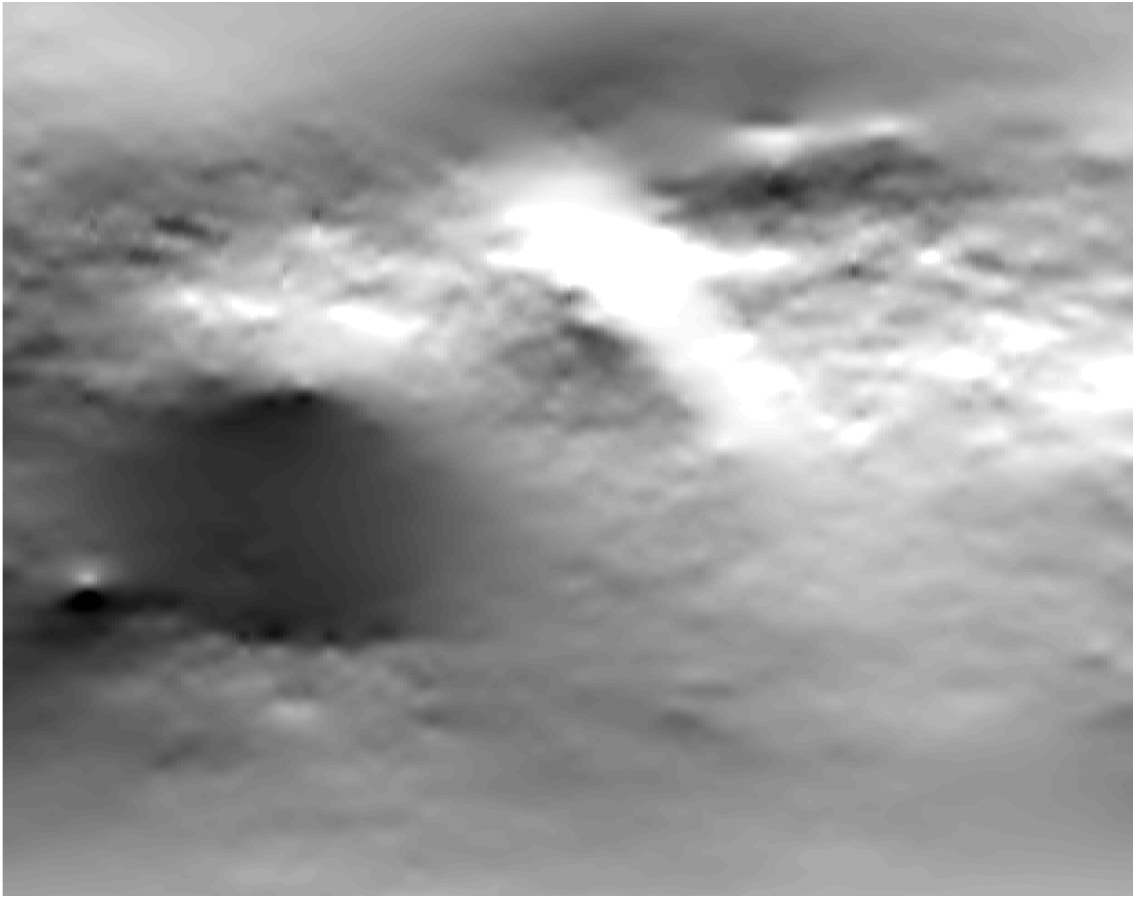}}}%
	\subfigure[GLUE, 63]{{\includegraphics[width=.16\textwidth, height=.14\textwidth]{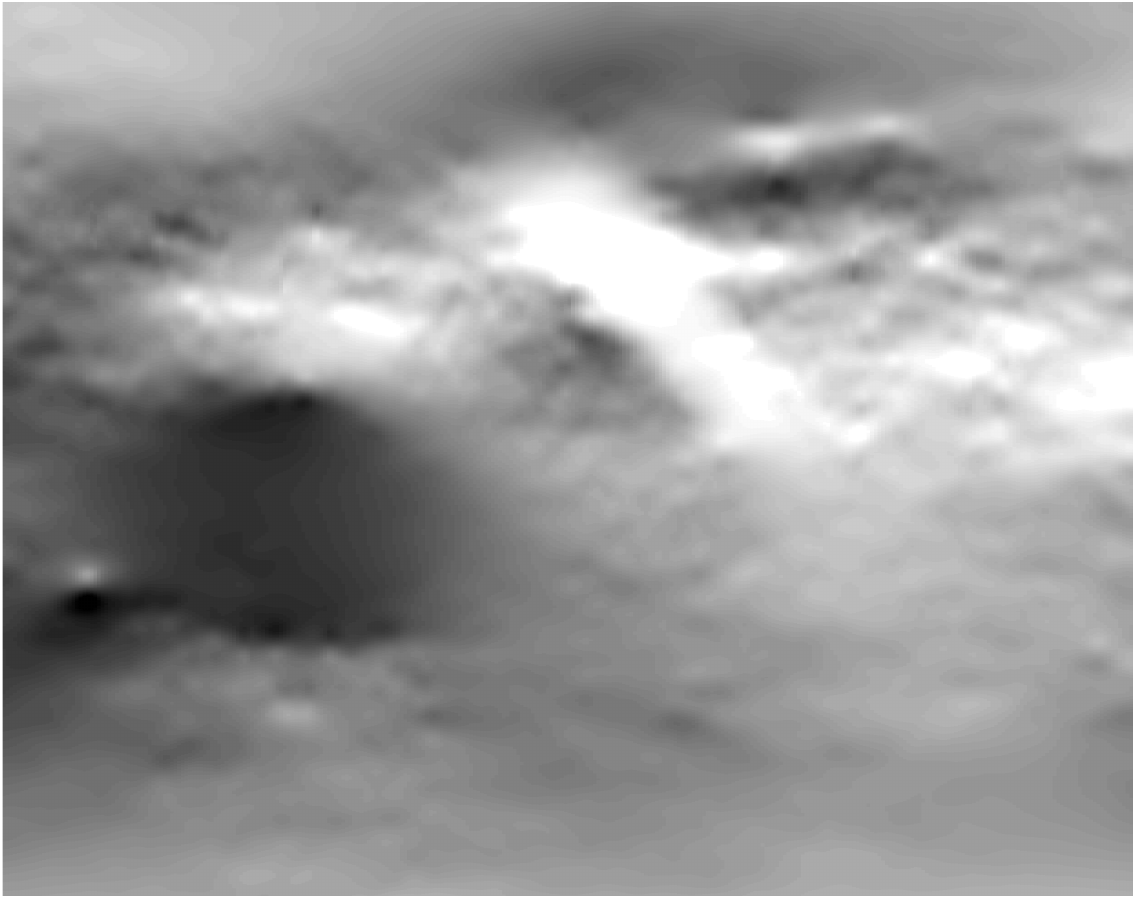}}}
	\subfigure[ALTRUIST, 3]{{\includegraphics[width=.16\textwidth, height=.14\textwidth]{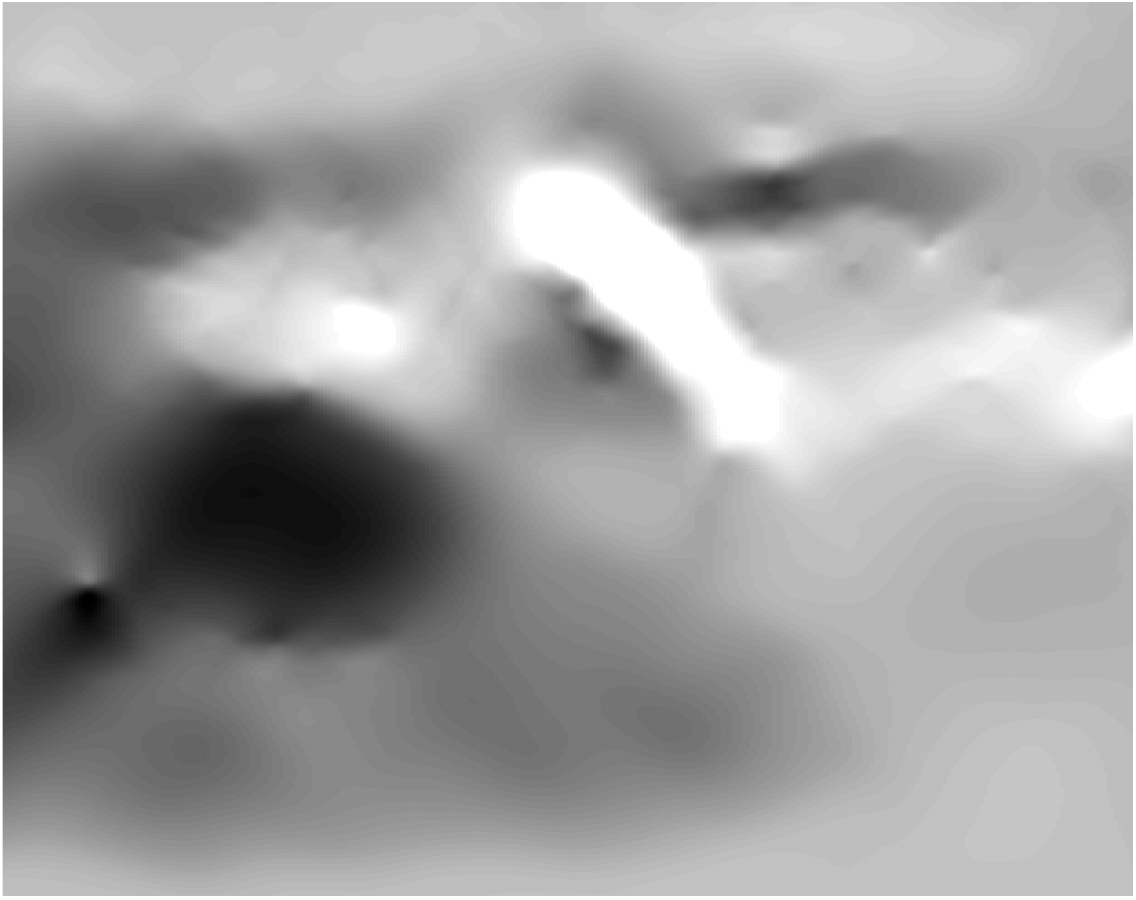}}}%
	\subfigure[ALTRUIST, 43]{{\includegraphics[width=.16\textwidth, height=.14\textwidth]{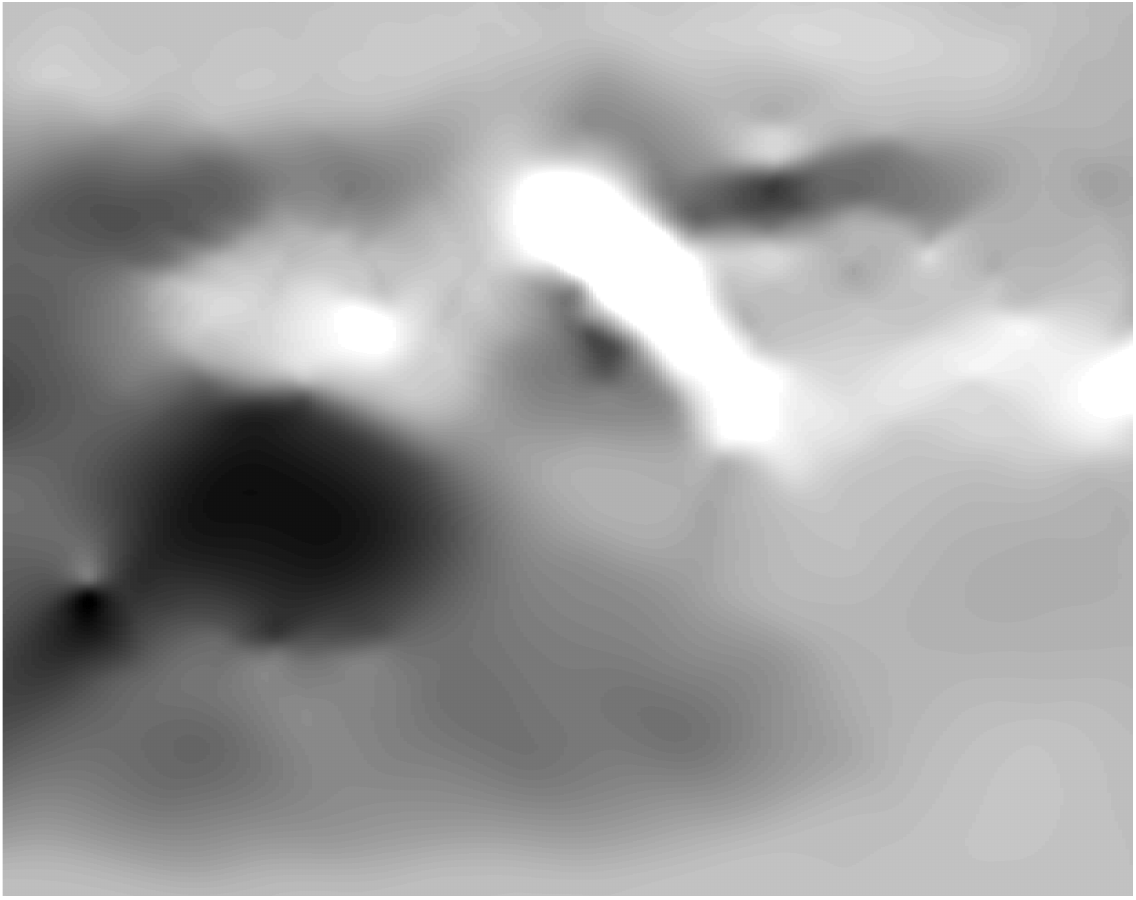}}}%
	\subfigure[ALTRUIST, 63]{{\includegraphics[width=.16\textwidth, height=.14\textwidth]{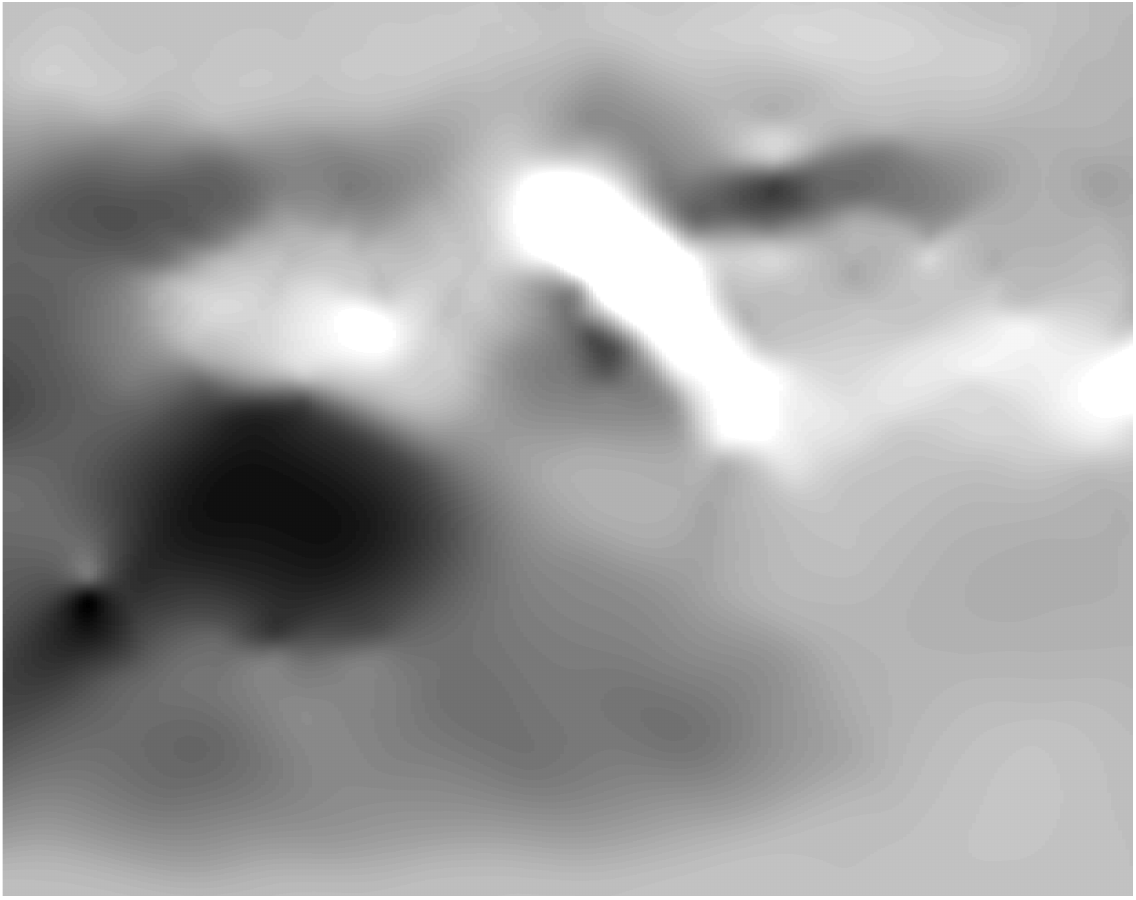}}}
	\subfigure[Axial strain]{{\includegraphics[width=.25\textwidth]{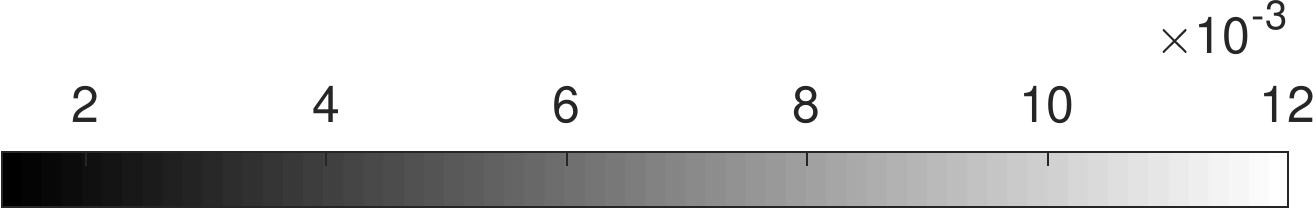}}}%
	\caption{Axial strain results using different sizes of the differentiation kernel. Rows 1 and 2 correspond to GLUE and ALTRUIST, respectively, whereas columns 1-3 correspond to kernel lengths of 3, 43, and 63 RF samples, respectively.}
	\label{diff_ker}
\end{figure}

\begin{figure*}[h]
	
	\centering
	\subfigure[Ground truth]{{\includegraphics[width=0.18\textwidth]{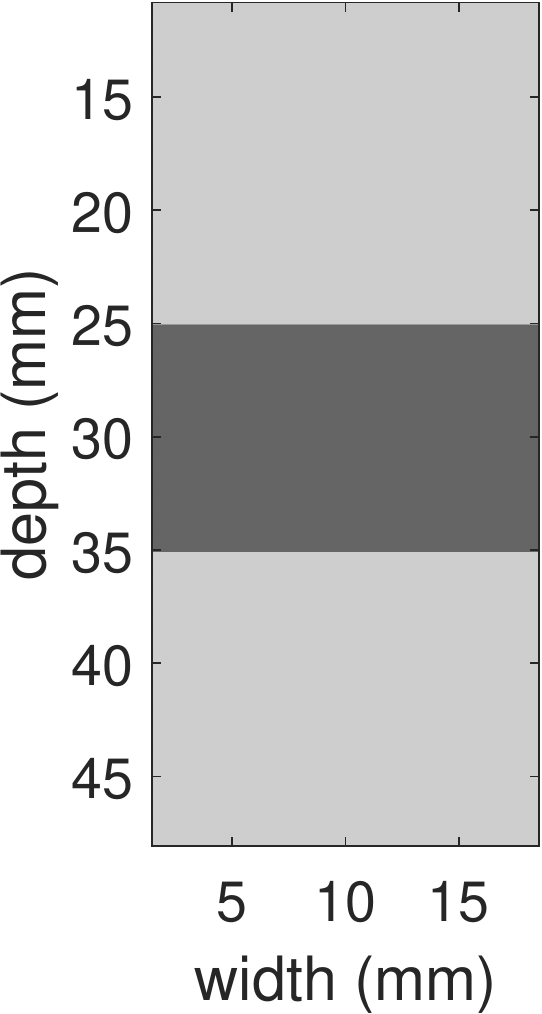}}}%
	\subfigure[GLUE]{{\includegraphics[width=0.18\textwidth]{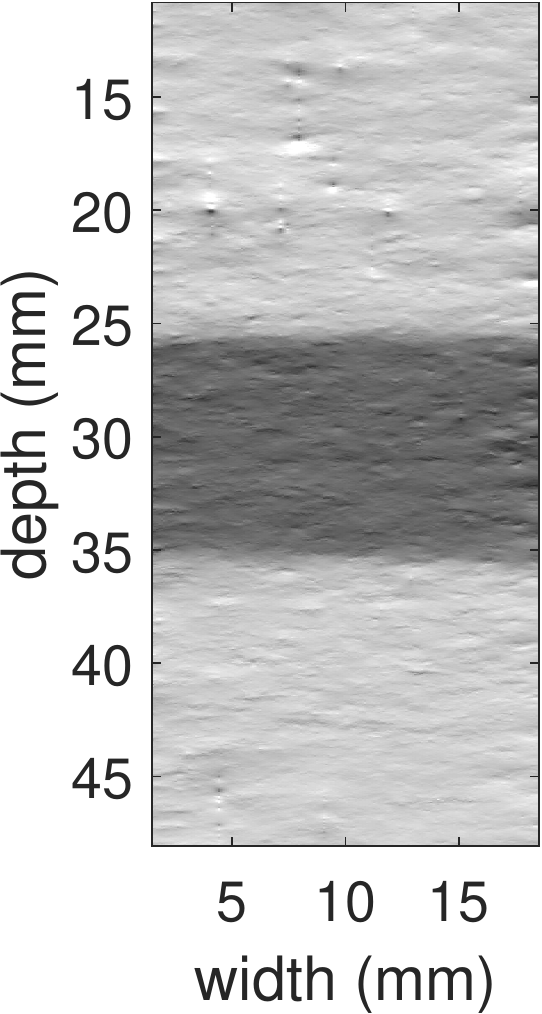}}}%
	\subfigure[OVERWIND]{{\includegraphics[width=0.18\textwidth]{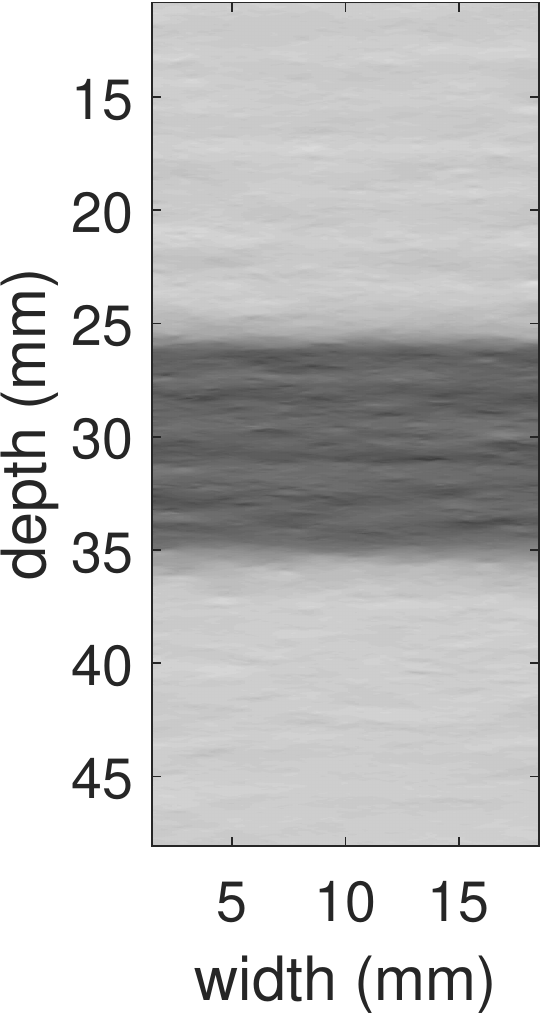} }}%
	\subfigure[$L1$-SOUL]{{\includegraphics[width=0.18\textwidth]{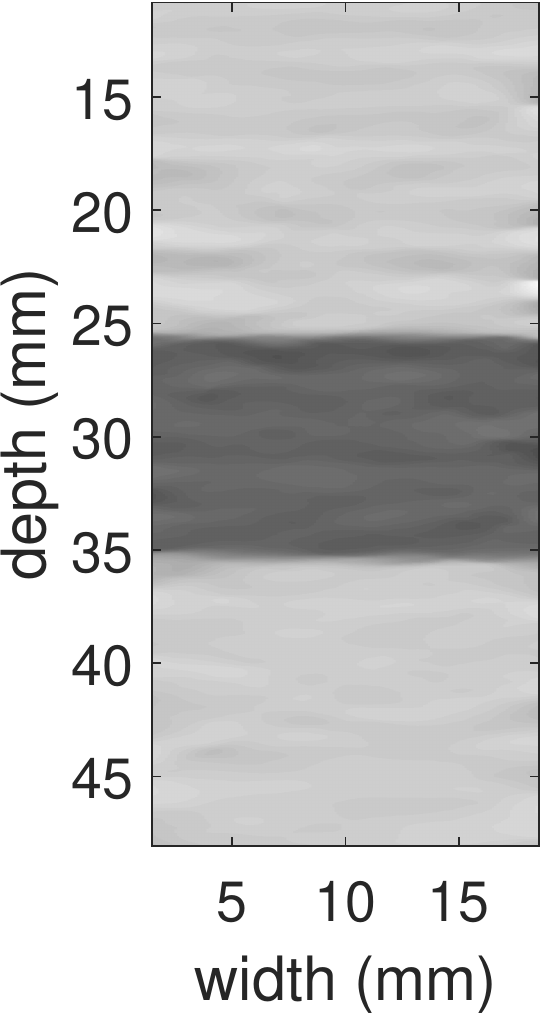} }}%
	\subfigure[ALTRUIST]{{\includegraphics[width=0.18\textwidth]{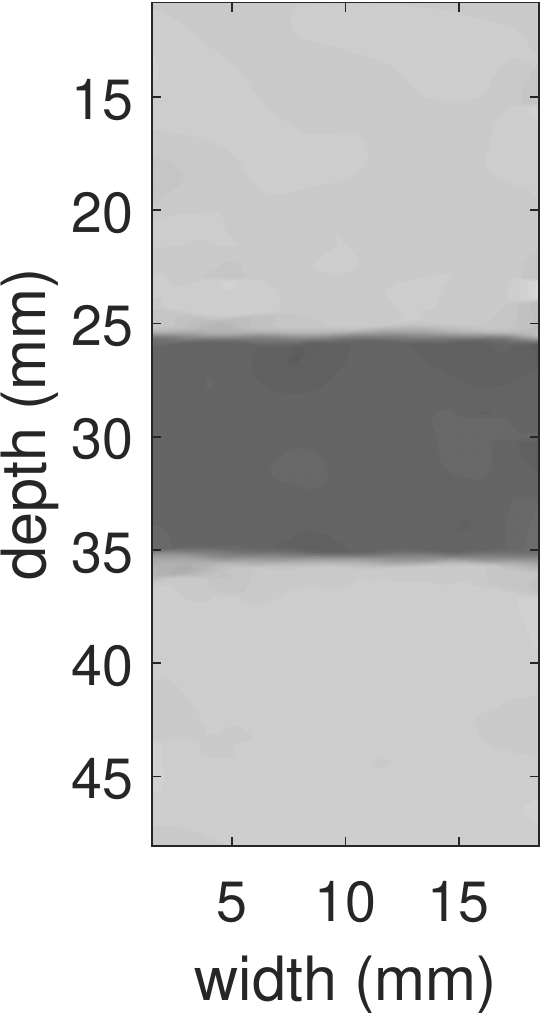} }}
	\subfigure[Ground truth]{{\includegraphics[width=0.18\textwidth]{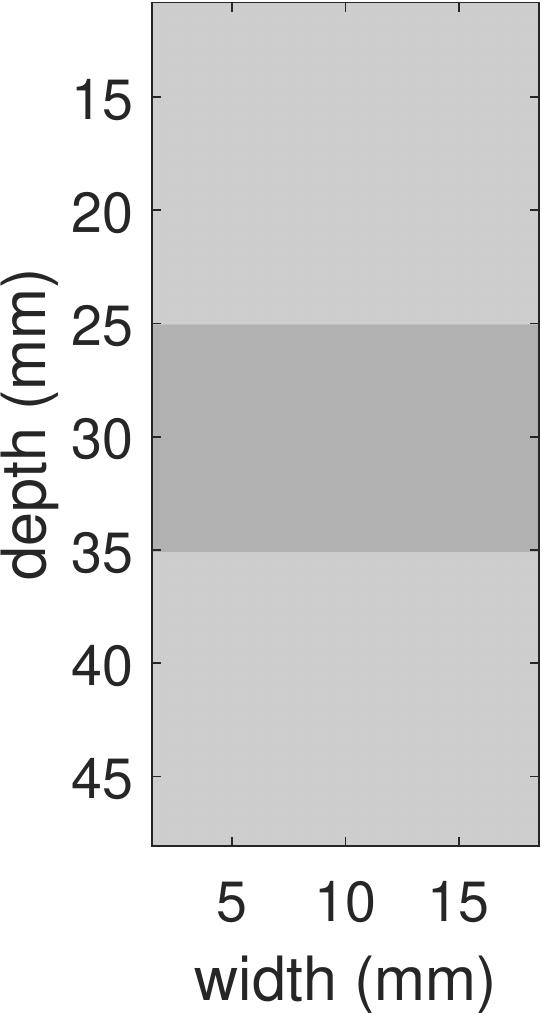}}}%
	\subfigure[GLUE]{{\includegraphics[width=0.18\textwidth]{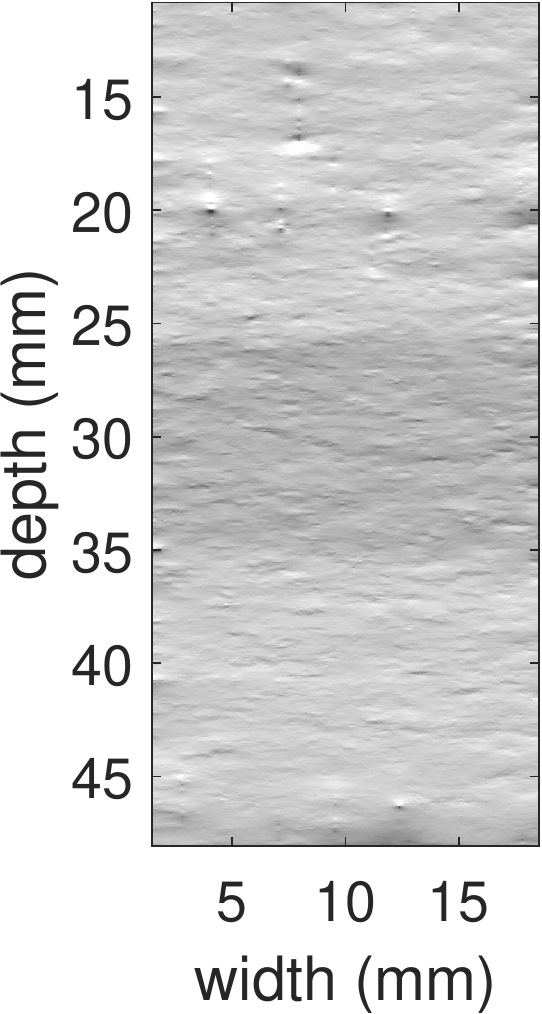}}}%
	\subfigure[OVERWIND]{{\includegraphics[width=0.18\textwidth]{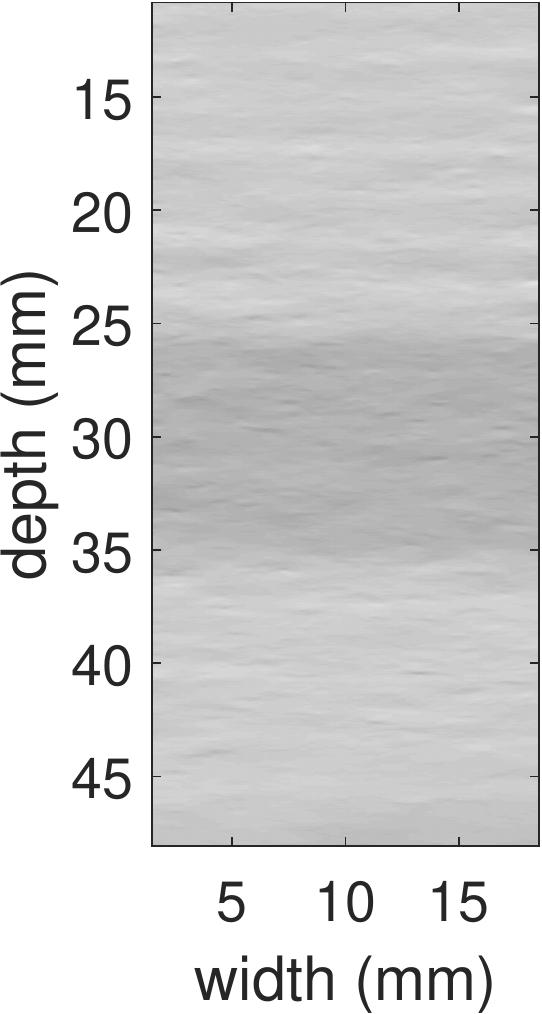} }}%
	\subfigure[$L1$-SOUL]{{\includegraphics[width=0.18\textwidth]{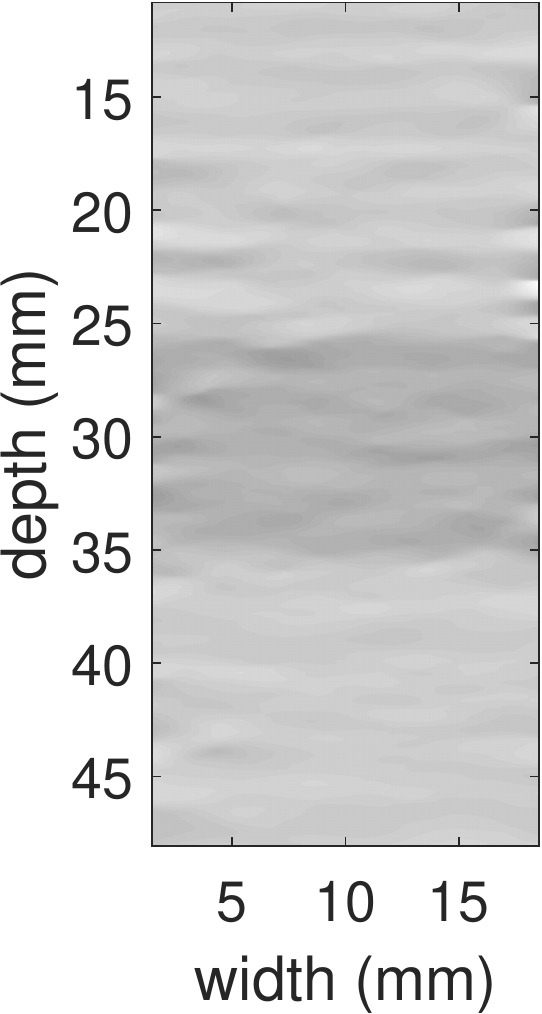} }}%
	\subfigure[ALTRUIST]{{\includegraphics[width=0.18\textwidth]{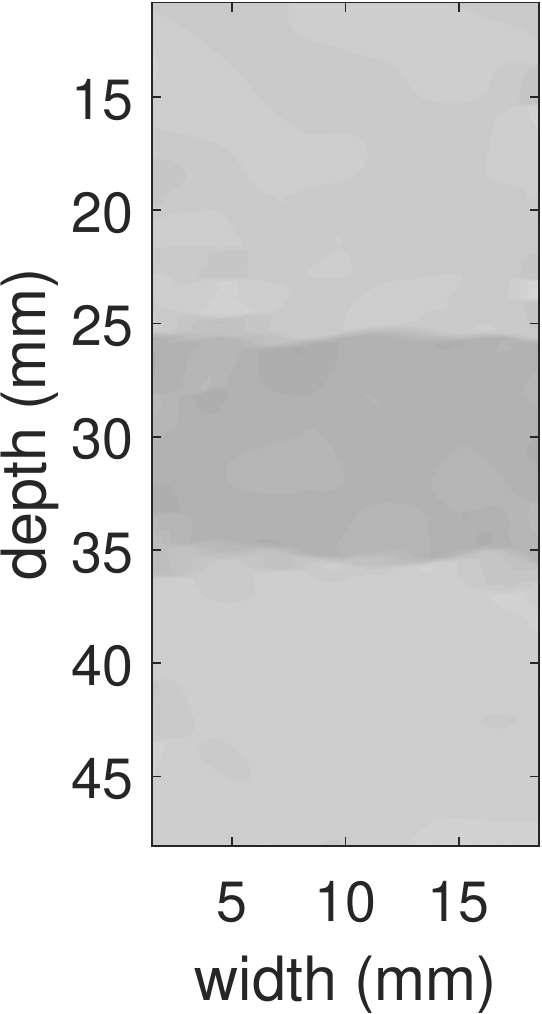} }}
	\subfigure[Axial strain]{{\includegraphics[width=.48\textwidth]{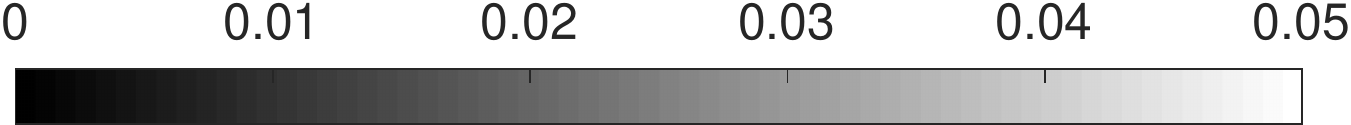}}}
	\caption{Axial strain images obtained from the simulated layer phantoms. Rows 1 and 2 correspond to high and low contrast, respectively, whereas, columns 1 to 5 correspond to the ground truth and axial strain maps estimated by GLUE, OVERWIND, $L1$-SOUL, and ALTRUIST, respectively.}
	\label{layer}
\end{figure*}

\begin{figure*}
	\centering
	\subfigure[Ground truth]{{\includegraphics[width=.18\textwidth]{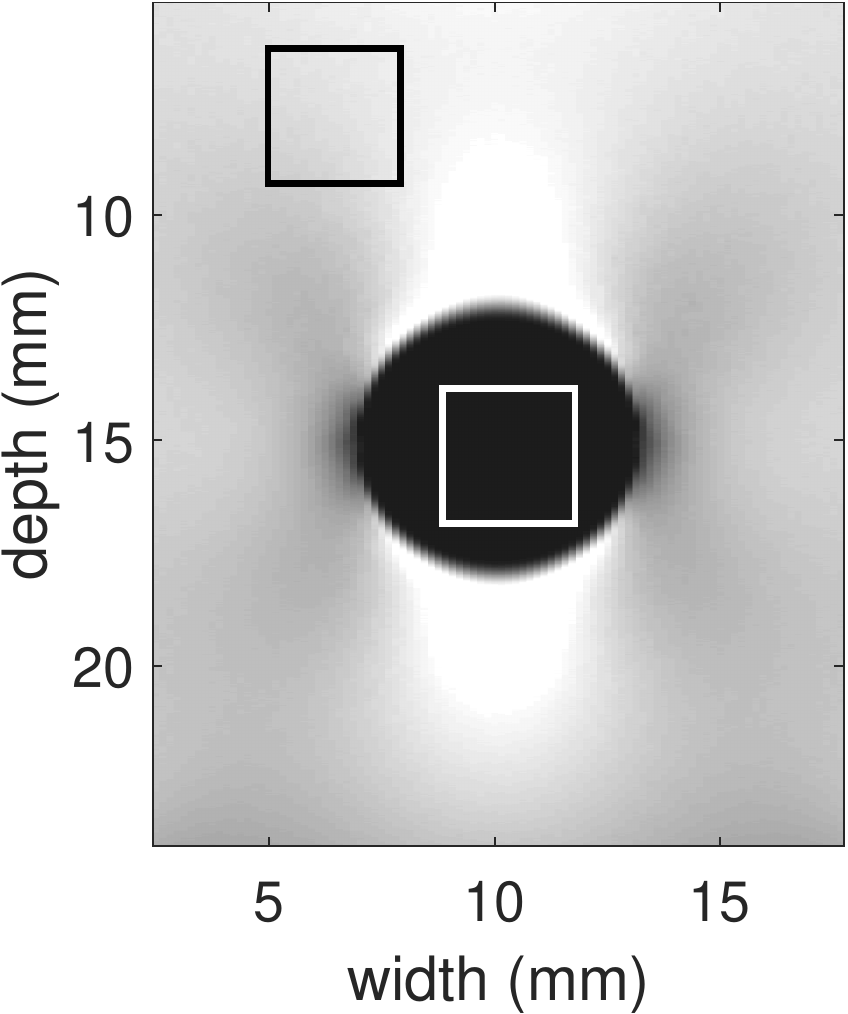}}}%
	\subfigure[GLUE]{{\includegraphics[width=.18\textwidth]{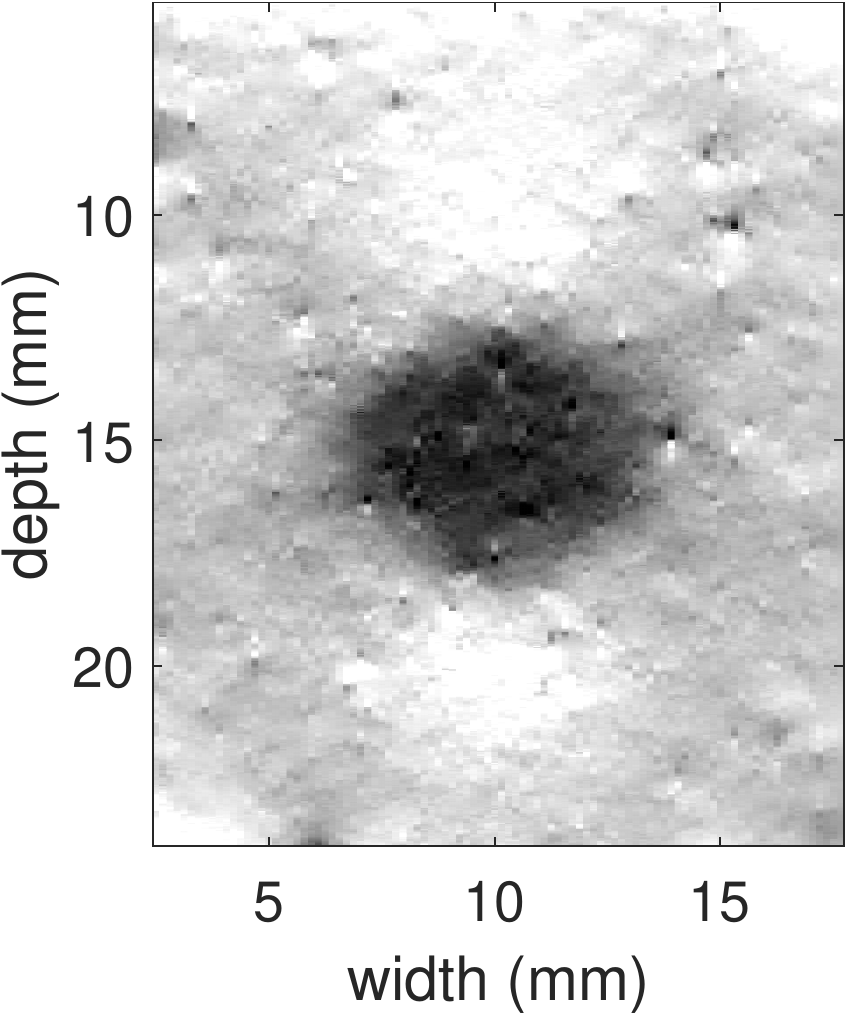}}}%
	\subfigure[OVERWIND]{{\includegraphics[width=.18\textwidth]{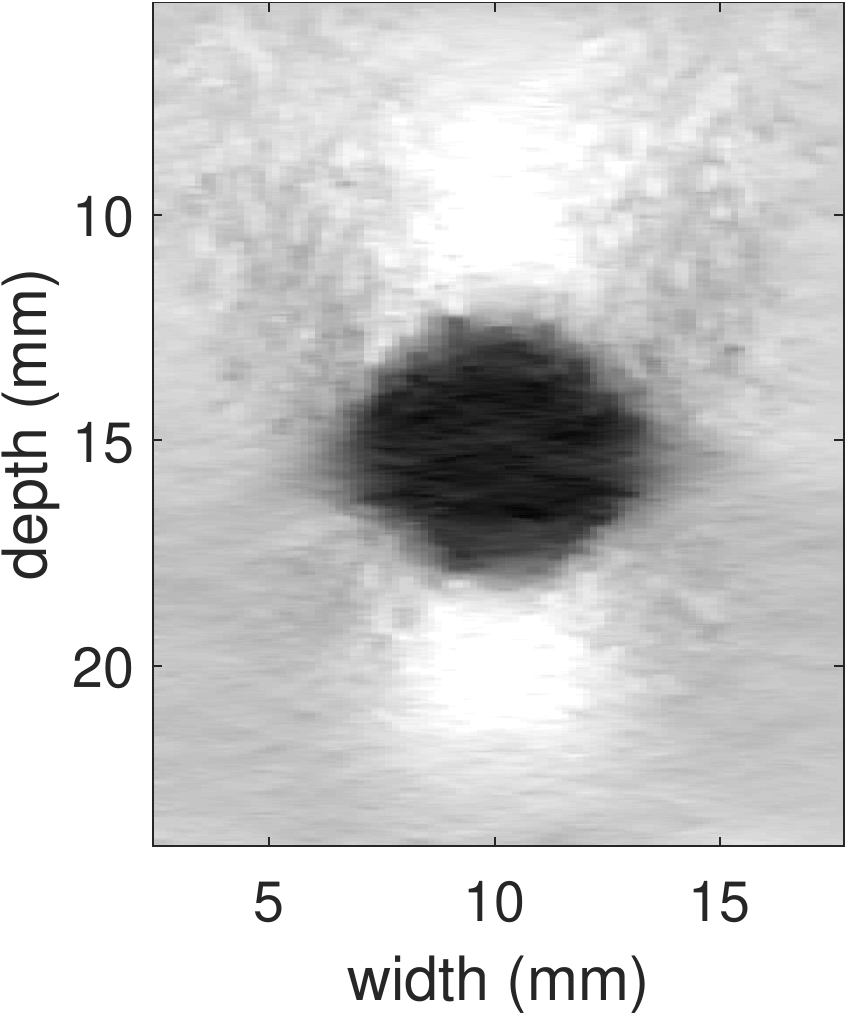} }}%
	\subfigure[$L1$-SOUL]{{\includegraphics[width=.18\textwidth]{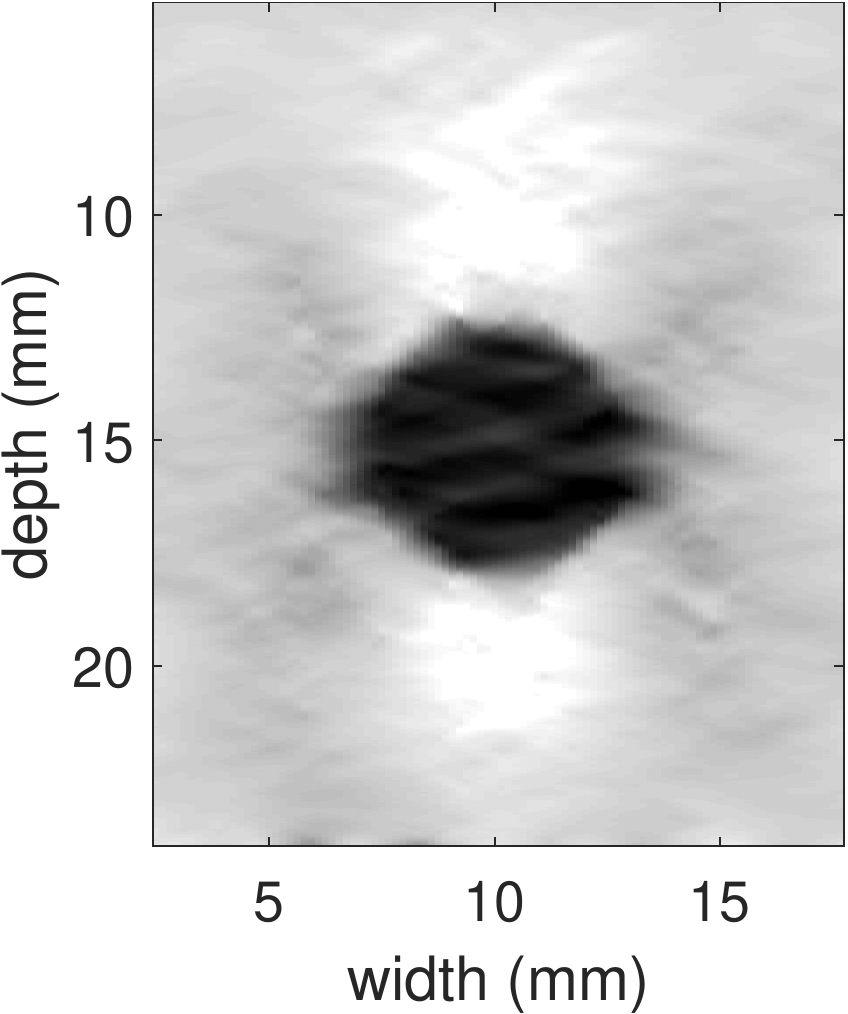} }}%
	\subfigure[ALTRUIST]{{\includegraphics[width=.18\textwidth]{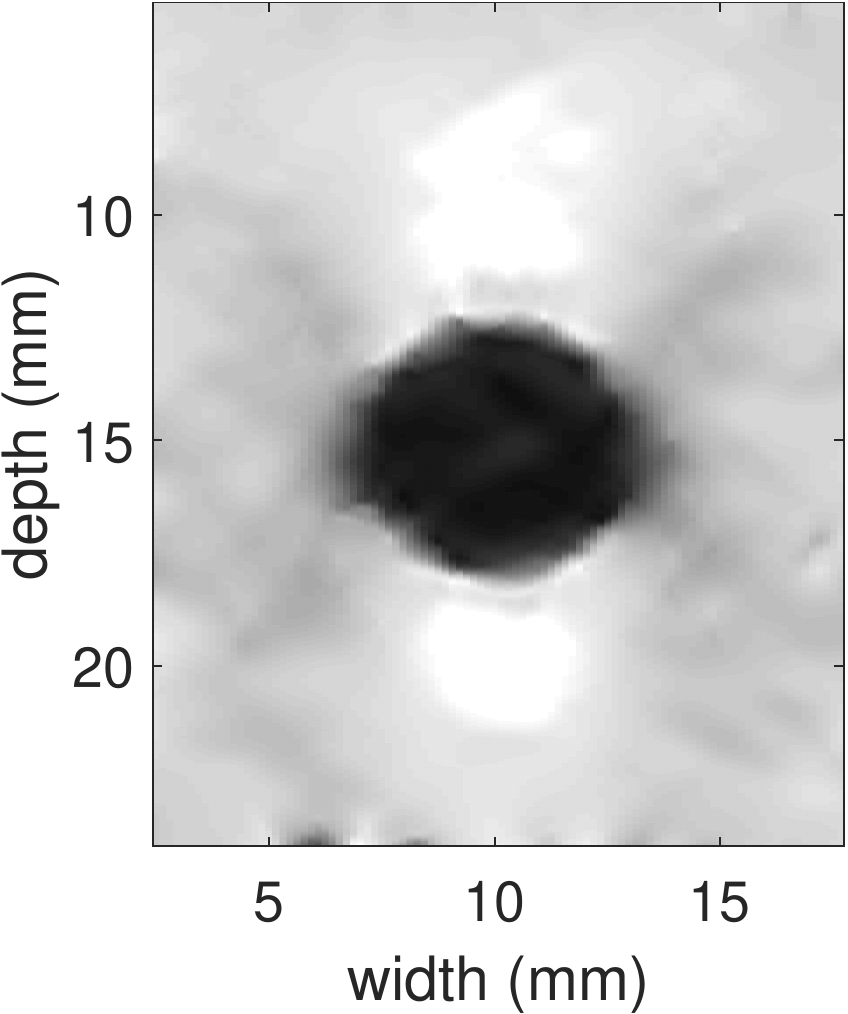} }}
	\subfigure[Axial strain]{{\includegraphics[width=.48\textwidth]{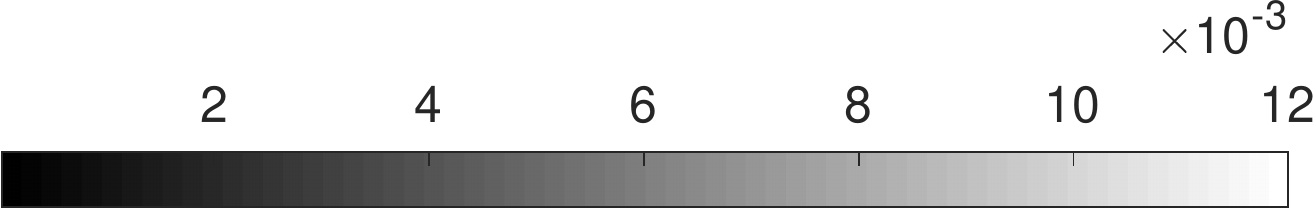}}}
	\caption{Axial strain results obtained from the simulated FEM phantom with added Gaussian noise. (a)-(e) correspond to the FEM strain and the axial strain images produced by GLUE, OVERWIND, $L1$-SOUL, and ALTRUIST, respectively. The foreground and background windows for calculating SNR, CNR, and SR are shown in (a).}
	\label{hard_simu}
\end{figure*}

\subsection{Ultrasound Simulation and Data Acquisition}
We simulated two phantoms containing three layers with varying elasticities. The target-background elastic contrast for the first and second layer phantoms are high and low, respectively. In addition to the layer phantoms, we simulated a homogeneous phantom containing a stiff inclusion. The real validation experiments were performed with one set of breast phantom and two sets of \textit{in vivo} liver cancer data. The specifics of the simulated, phantom, and \textit{in vivo} datasets are outlined below, whereas ultrasound simulation and imaging settings are provided in Table~\ref{table_imaging_params}.

\subsubsection{Simulated Layer Phantoms}
Two homogeneous phantoms, each containing a stiff layer, were compressed by $4\%$ using closed-form equations as described in \cite{guest}. The elastic moduli of the background and target tissue layers, respectively, were set to $20$ kPa and $40$ kPa for the first phantom and $20$ kPa and $22.86$ kPa for the second phantom. The ultrasound simulation software Field II~\cite{field2} was used to generate the RF frames. Random Gaussian noise with $20$ dB peak SNR (PSNR) was added to the RF data to mimic real data acquisition environment.

\subsubsection{Hard-inclusion Simulated Phantom}
A uniform phantom of $4$ kPa Young's modulus containing a hard inclusion of $40$ kPa elastic modulus was simulated. A $1\%$ compression was applied to the simulated phantom using the finite element (FEM) package ABAQUS (Providence, RI). Pre- and post compressed RF frames were simulated with Field II. We corrupted the RF data with added Gaussian noise of $24$ dB PSNR.

\subsubsection{Experimental Breast Phantom}
Pre- and post-deformed RF data were acquired from an experimental breast phantom (Model 059, CIRS: Tissue Simulation \& Phantom Technology, Norfolk, VA). The background Young's modulus was $20 \pm 5$~kPa, whereas the hard inclusion's elasticity modulus was at least twice as large as that of the background. See Table~\ref{table_imaging_params} for the imaging system and settings. 

\subsubsection{\textit{In vivo} Liver Cancer Datasets}
RF datasets were collected from three liver cancer patients at the Johns Hopkins Hospital (Baltimore, MD). All procedures aligned with the ethics approval obtained from the Institutional Review Board. In addition, all patients provided written consent to this \textit{in vivo} study. Further details of this experiment can be found in Table~\ref{table_imaging_params} and \cite{DPAM}.

\subsection{Quantitative Metrics}
The strain imaging quality of ALTRUIST is compared to those of GLUE, OVERWIND, and $L1$-SOUL using Mean Structural SIMilarity (MSSIM)~\cite{Zhou_2004}, root-mean-square error (RMSE), signal-to-noise ratio (SNR)\cite{varghese1997theoretical,ophir1999elastography}, contrast-to-noise ratio (CNR)\cite{varghese1997theoretical,ophir1999elastography}, and strain ratio (SR). RMSE is defined as follows.

\begin{equation}
\textrm{RMSE}=\sqrt{\frac{\sum\limits_{j=1}^n \sum\limits_{i=1}^m (\hat{s}_{i,j}-s_{i,j})^{2}}{mn}} 
\end{equation}

\noindent
where $\hat{s}_{i,j}$ and $s_{i,j}$ stand for the estimated and ground truth strains corresponding to the RF sample at $(i,j)$.


\begin{table*}[tb]  
	\centering
	\caption{Ultrasound simulation and imaging settings.} 
	\label{table_imaging_params}
	\begin{tabular}{c c c c c c c c c} 
		\hline
		$ $  $ $& Ultrasound system & Probe & Center frequency (MHz) & Sampling rate (MHz)\\
		\hline
		Layer phantoms & - & - & 7.27 &  40\\
		Hard-inclusion phantom & - & - & 7.27 &  100\\
		Experimental phantom & Alpinion E-cube R12 & L3-12H Linear array & 10 &  40\\
		\textit{In vivo} Liver & Antares Siemens & VF 10-5 linear array & 6.67  & 40\\
		\hline
	\end{tabular}
\end{table*}

\begin{table}[tb]  
	\centering
	\caption{MSSIM and RMSE values for the simulated high-contrast layer phantom dataset.} 
	\label{table_mssim_rmse_high}
	\begin{tabular}{c c c c c c c} 
		\hline
		$ $  $ $&  MSSIM & RMSE\\
		\hline
		GLUE & 0.05 &  $3.4 \times 10^{-3}$\\
		OVERWIND & 0.39 &  $2.6 \times 10^{-3}$\\
		$L1$-SOUL & 0.71 &  $2.6 \times 10^{-3}$\\
		ALTRUIST & \textbf{0.92}  & $\mathbf{2.4 \times 10^{-3}}$\\
		\hline
	\end{tabular}
\end{table}

\begin{table}[tb]  
	\centering
	\caption{MSSIM and RMSE values for the simulated low-contrast layer phantom dataset.} 
	\label{table_mssim_rmse_low}
	\begin{tabular}{c c c c c c c} 
		\hline
		$ $  $ $&  MSSIM & RMSE\\
		\hline
		GLUE & 0.004 &  $2.9 \times 10^{-3}$\\
		OVERWIND & 0.06 &  $1.1 \times 10^{-3}$\\
		$L1$-SOUL & 0.23 &  $1.3 \times 10^{-3}$\\
		ALTRUIST & \textbf{0.71}  & $\mathbf{7.05 \times 10^{-4}}$\\
		\hline
	\end{tabular}
\end{table}

\section{Results}
ALTRUIST and the three competing techniques obtained the optimal parameter sets using an \textit{ad hoc} technique. All techniques' results for different combinations of parameter values were generated, which were visually compared to each other to select the best one depending on the edge sharpness, background smoothness, and visual contrast. As illustrated in our previous work~\cite{soul,soulmate}, the strain imaging performance is not sensitive to a moderate alteration in the continuity parameter values.

The differentiation kernel length was set to 3 RF samples for all validation experiments. As shown in the representative example in Figure~\ref{diff_ker}, a large differentiation kernel does not improve the image features of interest such as the contrast or the edge-sharpness. Instead, it blurs the strain image while masking the samples with erroneous TDE. Therefore, a minimal-length differentiation kernel was used in this study to prevent an artificial oversmoothing of strain images and establish a fair comparison among different TDE techniques' performance. Detailed analyses of the kernel length and the techniques' performance are presented in the Discussion Section. 

This work calculates the single values of SNR, CNR, and SR using $3$ mm $\times$ $3$ mm windows placed at spatially smooth locations. To investigate the robustness of the techniques' quantitative performance to the selection of region-of-interest (ROI), alongside a single CNR value, we report the histogram of 120 CNR values calculated between 6 target and 20 background strain windows.

\subsection{Simulated Layer Phantoms}
Figure~\ref{layer} shows the axial strain results (see Figure 1 of the Supplementary Material for jet color map) for the high- and low-contrast layer phantoms. For both phantoms, GLUE obtains the noisiest strain images. OVERWIND yields a better noise suppression performance. In addition, the OVERWIND strains present clearer layer boundaries than GLUE. However, the estimated elastic contrast between different tissue layers are not satisfactory. $L1$-SOUL exhibits better target-background contrast than GLUE and OVERWIND. Moreover, it yields a sharper layer edge. Nevertheless, the performance of $L1$-SOUL in the uniform tissue regions is not up to the mark. ALTRUIST produces the best strain images by ensuring smoothness in the uniform tissue region and sharp transition at the layer boundaries. The advantage of ALTRUIST is even more evident in case of the low-contrast phantom since it is a highly challenging scenario. The MSSIM and RMSE values reported in Tables~\ref{table_mssim_rmse_high} and \ref{table_mssim_rmse_low} support our visual inference. It is worth noting that all four techniques slightly distort the layer edges in case of the low-contrast phantom (also see Figure 5 of the Supplementary Material). This might originate from the TDE techniques' limited ability to discern tissues with negligible elastic difference.   


\begin{table}[tb]  
	\centering
	\caption{MSSIM and RMSE values for the hard-inclusion simulated phantom.} 
	\label{table_mssim_rmse_hard}
	\begin{tabular}{c c c c c c c} 
		\hline
		$ $  $ $&  MSSIM & RMSE\\
		\hline
		GLUE & 0.19 &  $1.4 \times 10^{-3}$\\
		OVERWIND & 0.43 &  $9.93 \times 10^{-4}$\\
		$L1$-SOUL & 0.59 &  $9.46 \times 10^{-4}$\\
		ALTRUIST & \textbf{0.66}  & $\mathbf{9.35 \times 10^{-4}}$\\
		\hline
	\end{tabular}
\end{table}

\begin{table}[tb]
	\centering
	\caption{SNR, CNR, and SR for the hard-inclusion simulated phantom dataset. CNR and SR are calculated between the white target and black background windows depicted in Figure~\ref{hard_simu}(a), whereas SNR is obtained from the background window only.}
	\label{table_hard}
	\begin{tabular}{c c c c c c c} 
		\hline
		$ $  $ $&    SNR & CNR & SR\\
		\hline
		GLUE &  13.84 &  10.70 & 0.20\\
		OVERWIND & 24.57 &  22.04 & 0.15\\
		$L1$-SOUL & 33.51 &  17.84 & \textbf{0.12}\\
		ALTRUIST & \textbf{54.65}  & \textbf{38.84} & \textbf{0.12}\\
		\hline
	\end{tabular}
\end{table}

\begin{figure*}
	\centering
	\subfigure[B-mode]{{\includegraphics[width=.2\textwidth]{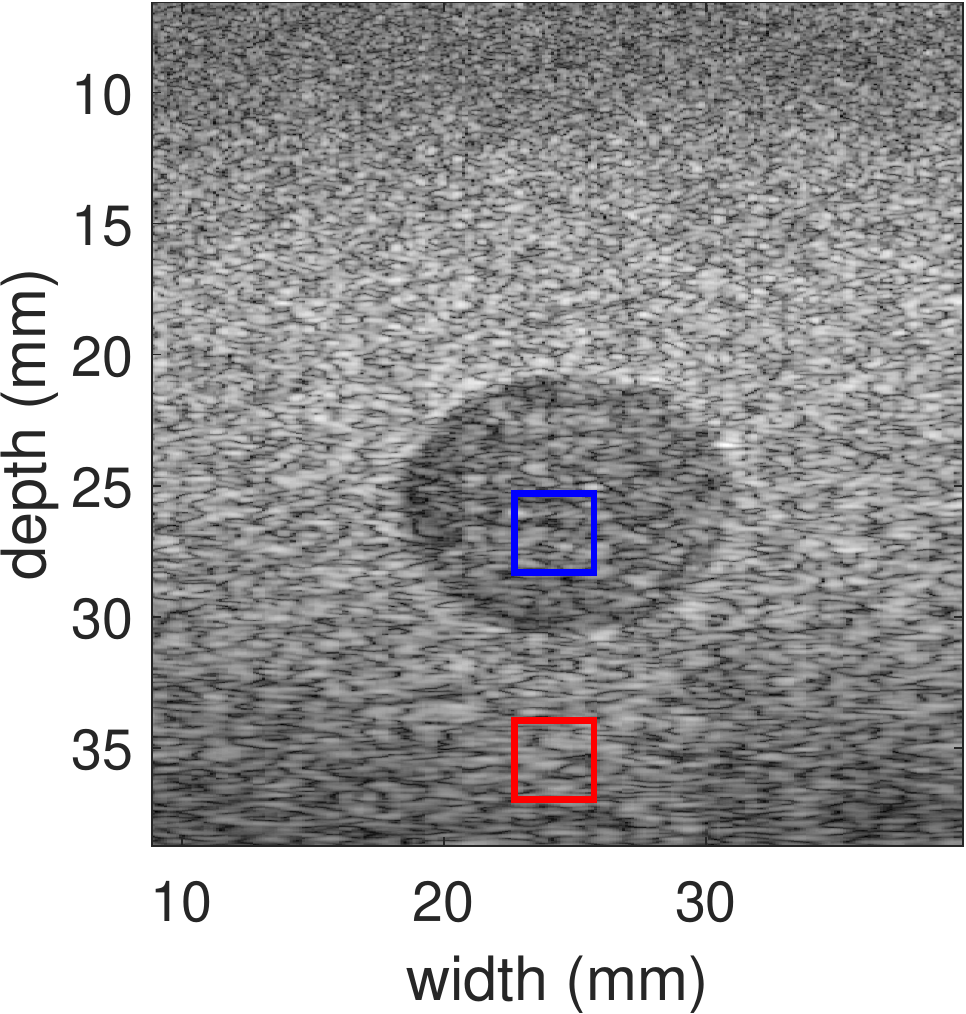}}}%
	\subfigure[GLUE]{{\includegraphics[width=.2\textwidth]{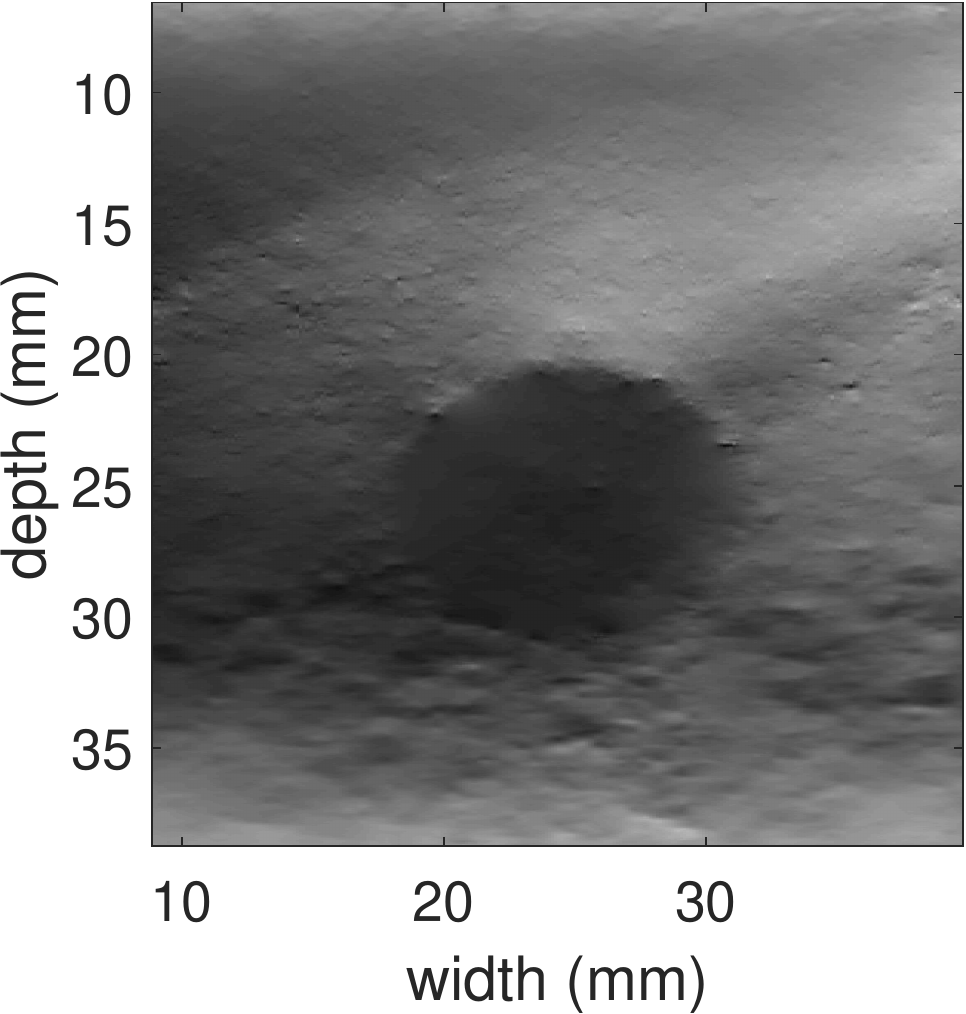}}}%
	\subfigure[OVERWIND]{{\includegraphics[width=.2\textwidth]{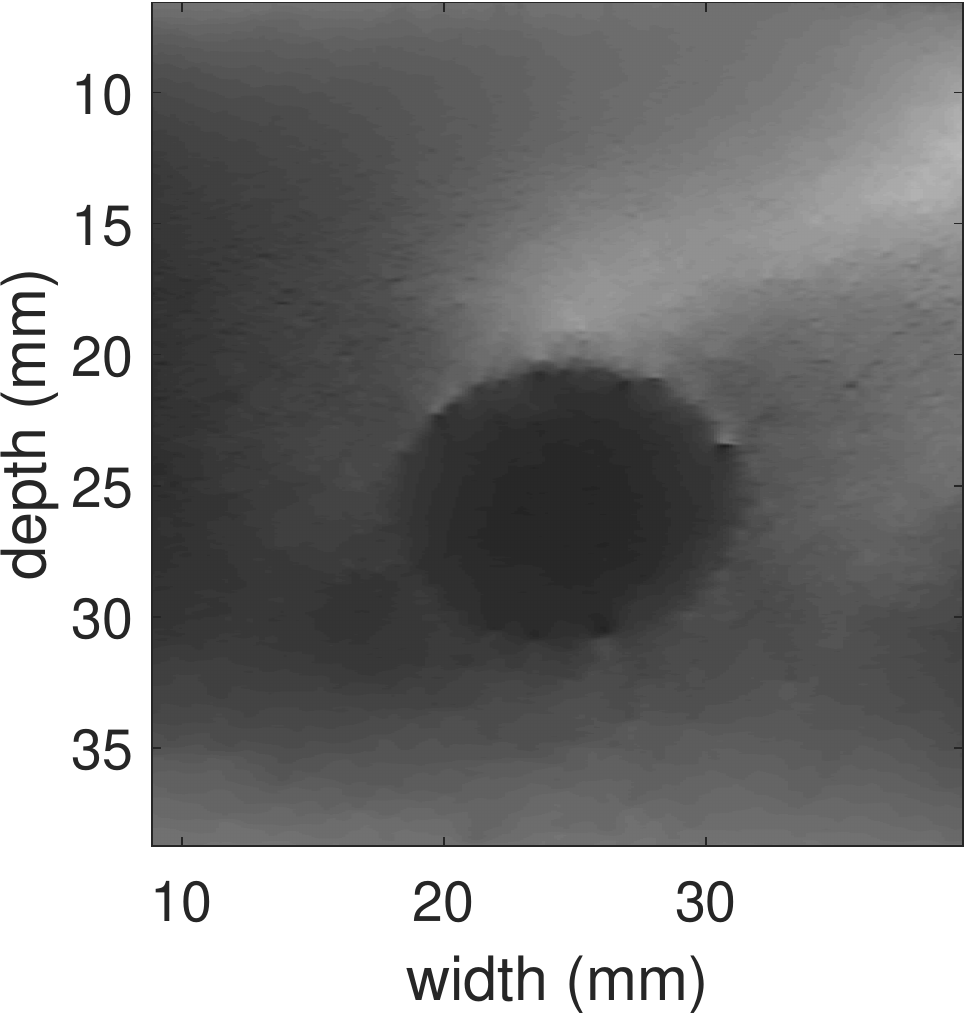} }}%
	\subfigure[$L1$-SOUL]{{\includegraphics[width=.2\textwidth]{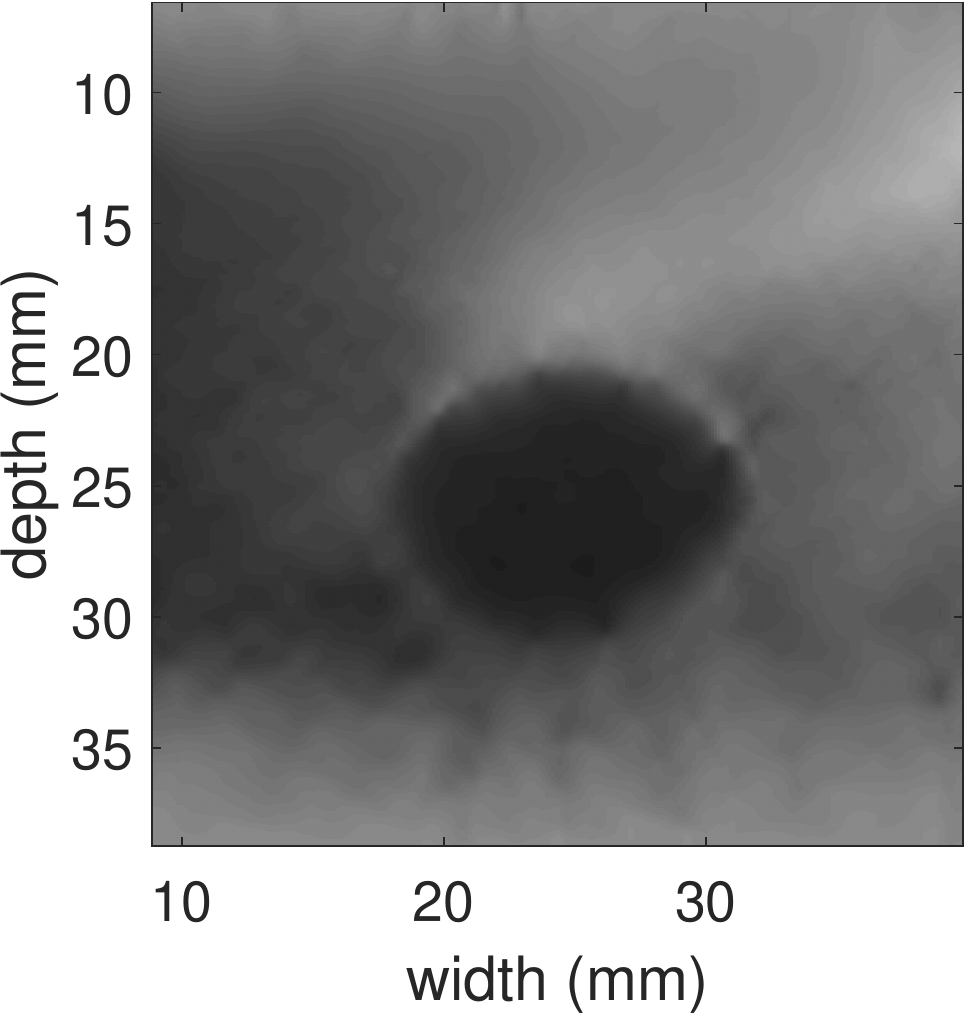} }}%
	\subfigure[ALTRUIST]{{\includegraphics[width=.2\textwidth]{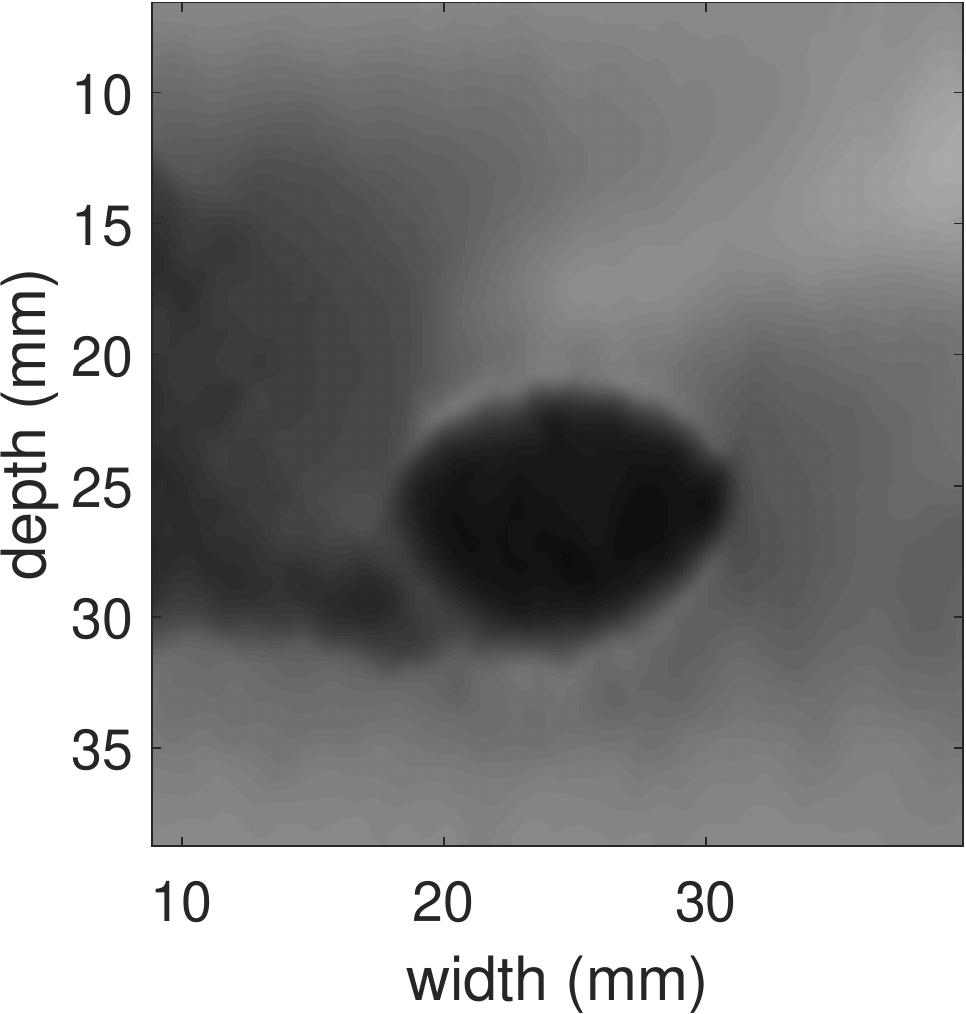} }}
	\subfigure[Axial strain]{{\includegraphics[width=.48\textwidth]{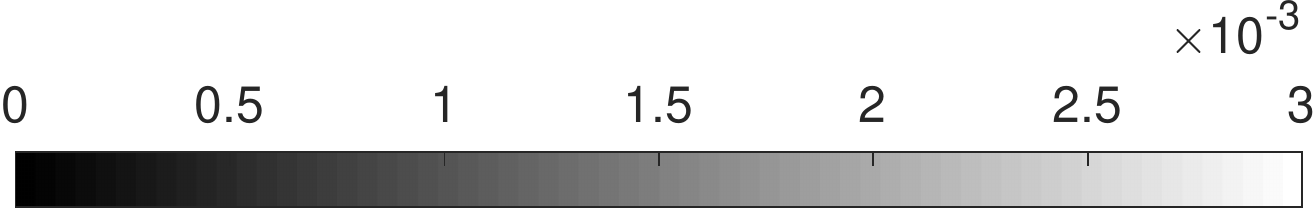}}}
	\caption{Axial strain results obtained from the experimental breast phantom. Columns 1 to 5 correspond to the B-mode image and the axial strain images produced by GLUE, OVERWIND, $L1$-SOUL, and ALTRUIST, respectively. The foreground and background strain windows for calculating SNR, CNR, and SR are shown on the B-mode image.}
	\label{perform_phan}
\end{figure*}

\begin{table}[tb]
	\centering
	\caption{Quantitative results for the breast phantom dataset. CNR and SR are calculated incorporating the blue target and red background windows shown in Figure~\ref{perform_phan}(a). SNR is calculated on the background window.} 
	\label{table_phan}
		\begin{tabular}{c c c c c c c} 
			\hline
			$ $  $ $&    SNR & CNR & SR\\
			\hline
			GLUE &  5.06 &  4.39 & 0.38\\
			OVERWIND & 11.77 &  9.01 & 0.45\\
			$L1$-SOUL & 10.34 &  10.07 & 0.31\\
			ALTRUIST & \textbf{18.79} & \textbf{20.51} & \textbf{0.17}\\
			\hline
		\end{tabular}
\end{table}

\subsection{Hard-inclusion Simulated Phantom}
Figure~\ref{hard_simu} depicts the FEM and the estimated axial strain images (see Figure 2 of the Supplementary Material for jet color map) for the hard-inclusion simulated phantom. GLUE provides a noisy strain estimate with low target-background contrast. OVERWIND yields a more precise inclusion boundary than GLUE, while improving the contrast. However, the background strain imaging performance of OVERWIND is unsatisfactory. By providing a smoother background and preserving the contrast, $L1$-SOUL addresses this issue of OVERWIND. It is worth mentioning that $L1$-SOUL exhibits spurious edges inside the stiff inclusion which stem from the estimation noise. In addition, a small region on the right of the inclusion edge is slightly broken. ALTRUIST substantially outperforms GLUE, OVERWIND, and $L1$-SOUL both in the inclusion border and the uniform areas. In comparison to other three techniques, the ALTRUIST inclusion edge is more circular and closer to the ground truth, which further endorses ALTRUIST's better edge-sharpening and shape-preserving abilities. It is apparent from the MSSIM and RMSE values in Table~\ref{table_mssim_rmse_hard} that ALTRUIST closely resembles the FEM strain. The SNR, CNR, and SR values reported in Table~\ref{table_hard} corroborate that ALTRUIST is superior to the other three techniques.

The histograms of 120 CNR values (see Figure~\ref{cnr_histograms}(a)) demonstrate that ALTRUIST occupies most of the high CNR values and quantitatively outperforms the other techniques throughout the strain image. The paired $t$-test between the calculated CNR distributions indicates that ALTRUIST is significantly better than GLUE, OVERWIND, and $L1$-SOUL with $p$-values of $7.08 \times 10^{-33}$, $7.35 \times 10^{-16}$, and $1.62 \times 10^{-23}$, respectively.      

\subsection{Breast Phantom Dataset}
The axial strain results for the experimental breast phantom dataset have been reported in Figure~\ref{perform_phan} (gray colormap) and Figure 3 of the Supplementary Material (jet color map). GLUE exhibits extensive background noise and blurred inclusion boundary. Although the inclusion edge is clearer in the OVERWIND strain, the background tissue region appears to be dark which might originate from the underestimation of strain. $L1$-SOUL outperforms GLUE and OVERWIND in terms of both edge-clarity and visual contrast. However, the background, especially in the deep tissue regions, is still noisy and the inclusion edge is not sufficiently sharp. The combination of physics-based second-order constraint and $L1$-norm regularization empowers $L1$-SOUL and ALTRUIST to achieve substantially better visual contrast than GLUE and OVERWIND. However, due to making a smooth approximation of the $L1$-norm, $L1$-SOUL's contrast still has room for improvement. ALTRUIST resolves the issues associated with GLUE, OVERWIND, and $L1$-SOUL by providing the highest contrast, smooth background, and sharp target-background boundary. This visual assessment is supported by the quantitative values reported in Table~\ref{table_phan}. It is noticeable that $L1$-SOUL and ALTRUIST exhibit less circular inclusion than the B-mode image, which might be a side-effect of edge-sharpening. Nevertheless, since the ground truth is unknown and the B-mode echogenic contrast is not a true indicator of elastic contrast, a conclusive statement cannot be provided.

The histogram of 120 CNR values (see Figure~\ref{cnr_histograms}(b)) shows that ALTRUIST obtains majority of the high CNR values. The paired $t$-test confirms that ALTRUIST significantly outperforms GLUE, OVERWIND, and $L1$-SOUL with $p$-values of $6.21 \times 10^{-49}$, $3.04 \times 10^{-36}$, and $2.39 \times 10^{-19}$, respectively.

\begin{figure*}
	\centering
	\subfigure[B-mode]{{\includegraphics[width=.2\textwidth]{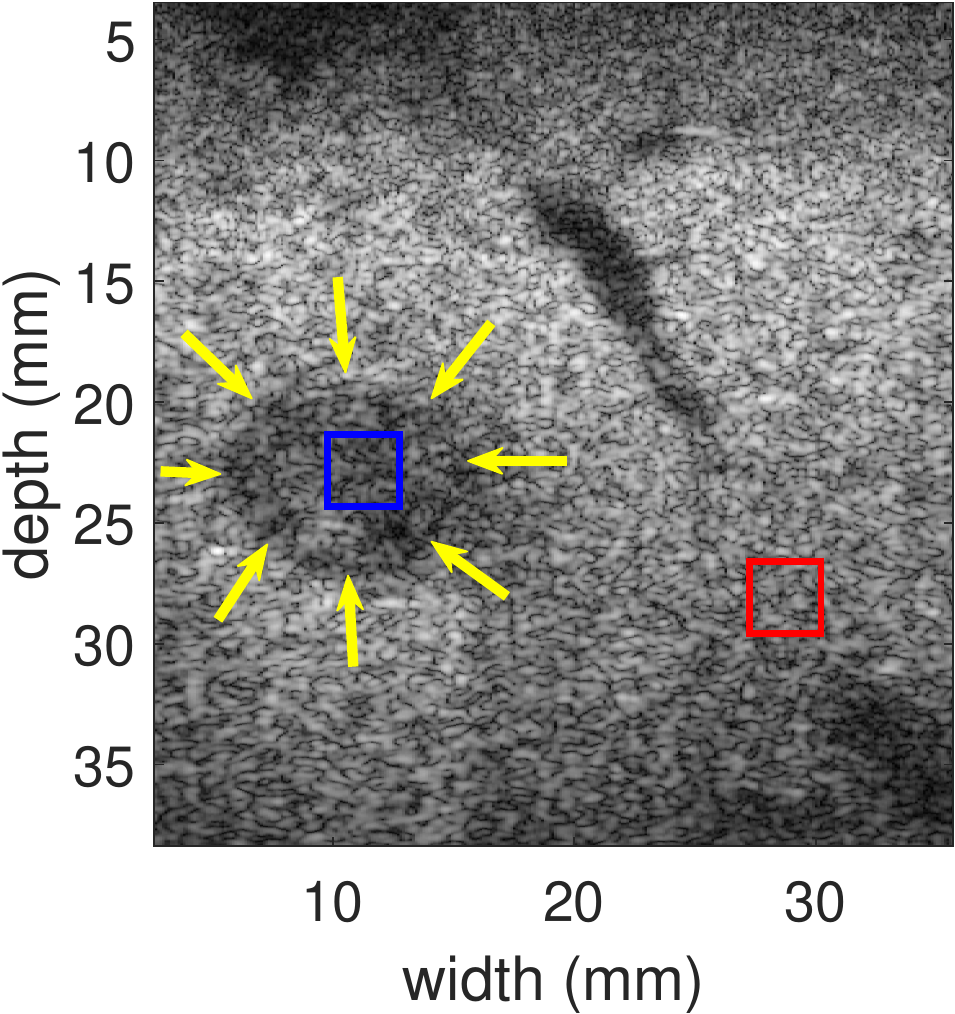}}}%
	\subfigure[GLUE]{{\includegraphics[width=.2\textwidth]{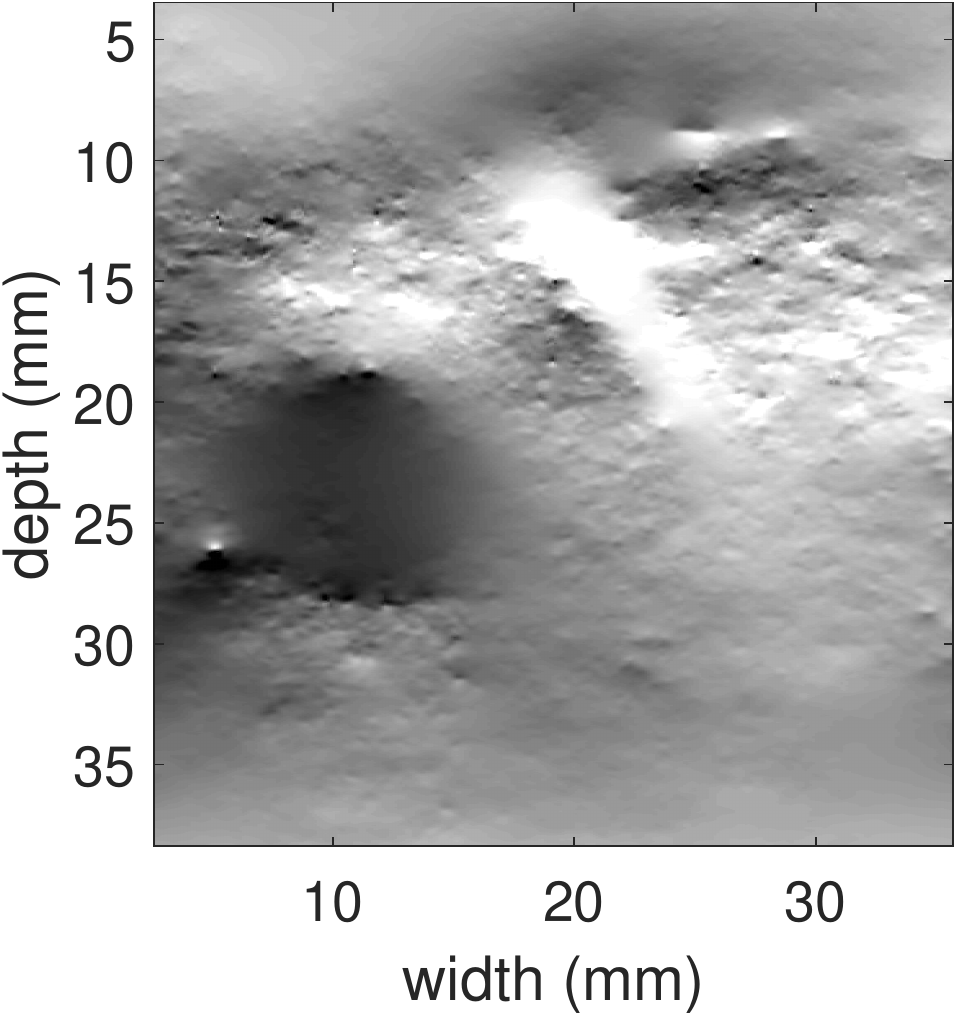}}}%
	\subfigure[OVERWIND]{{\includegraphics[width=.2\textwidth]{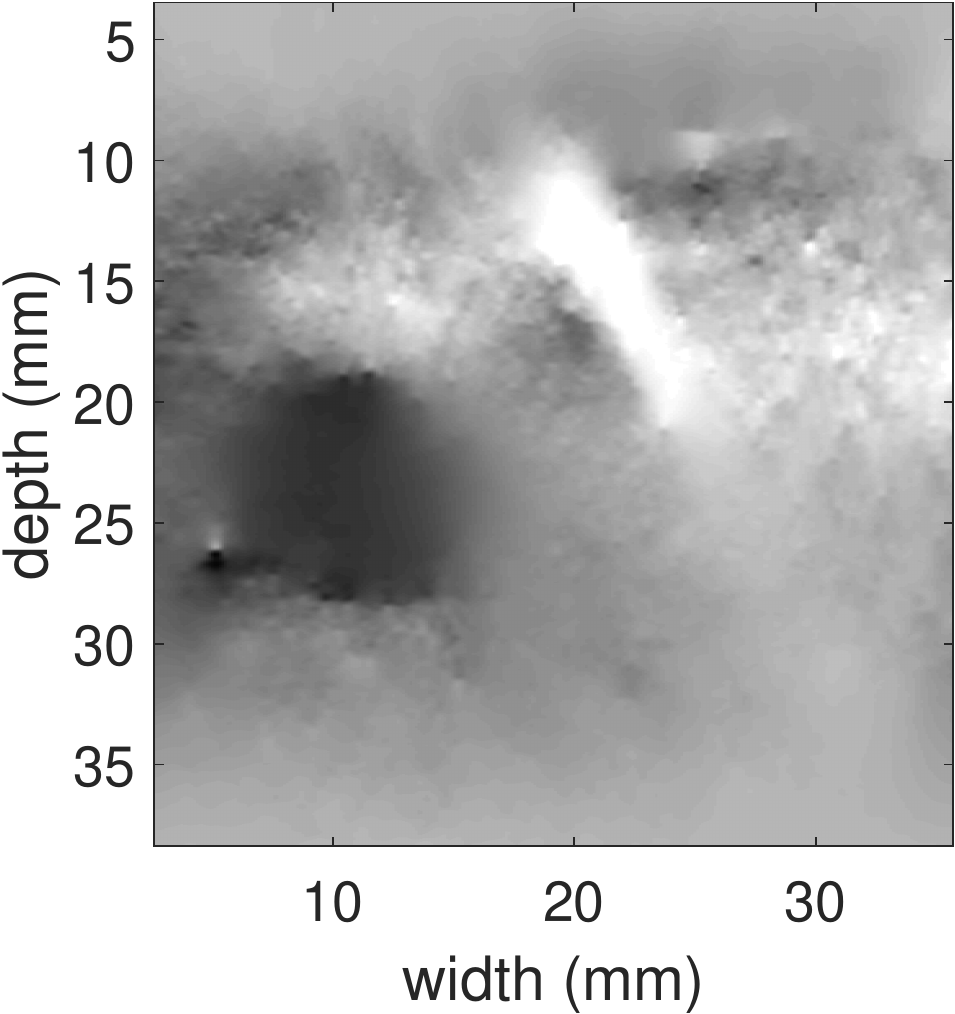} }}%
	\subfigure[$L1$-SOUL]{{\includegraphics[width=.2\textwidth]{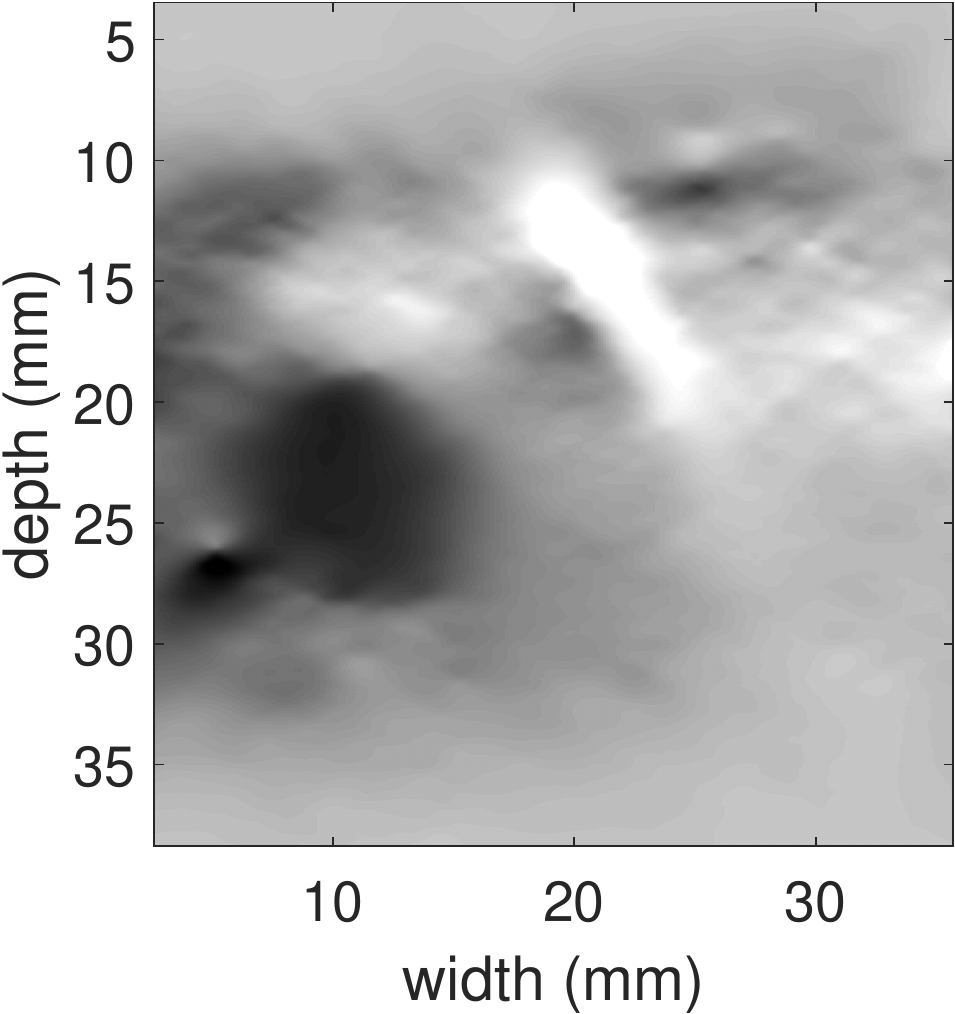} }}%
	\subfigure[ALTRUIST]{{\includegraphics[width=.2\textwidth]{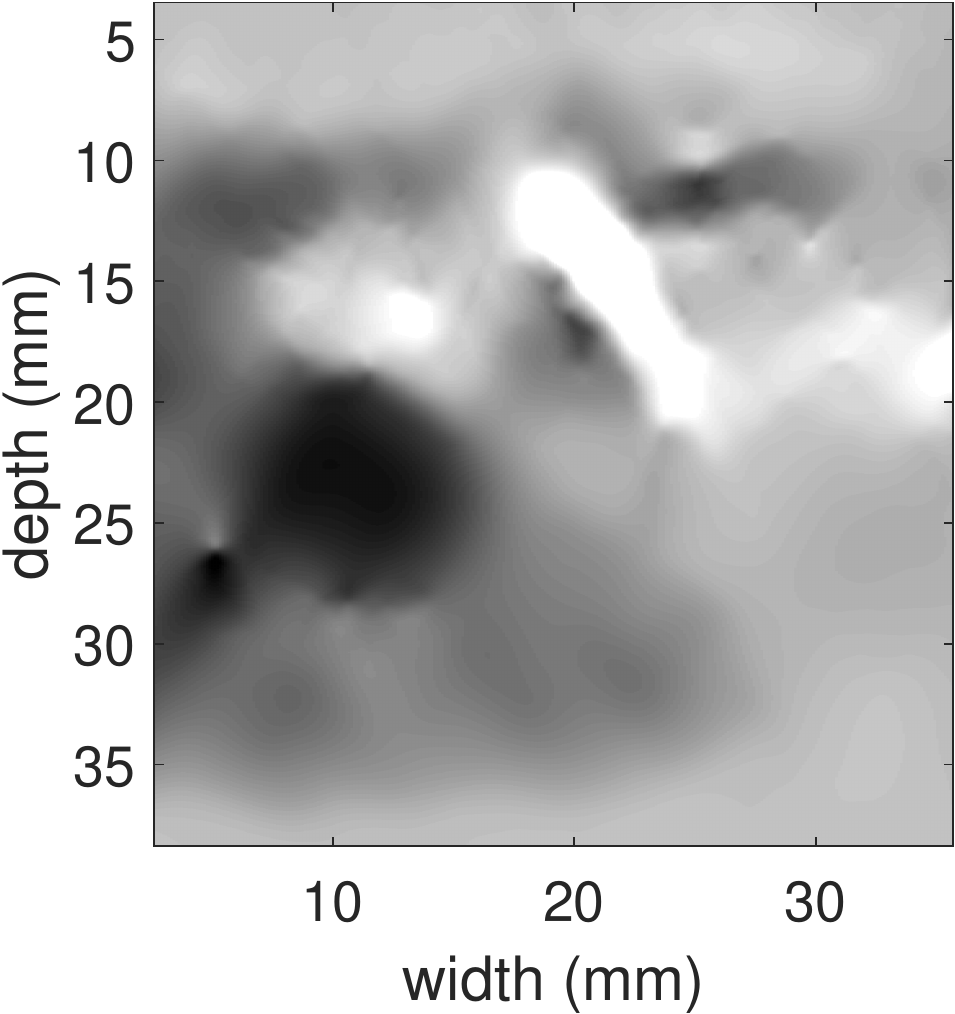} }}
	\subfigure[B-mode]{{\includegraphics[width=.2\textwidth]{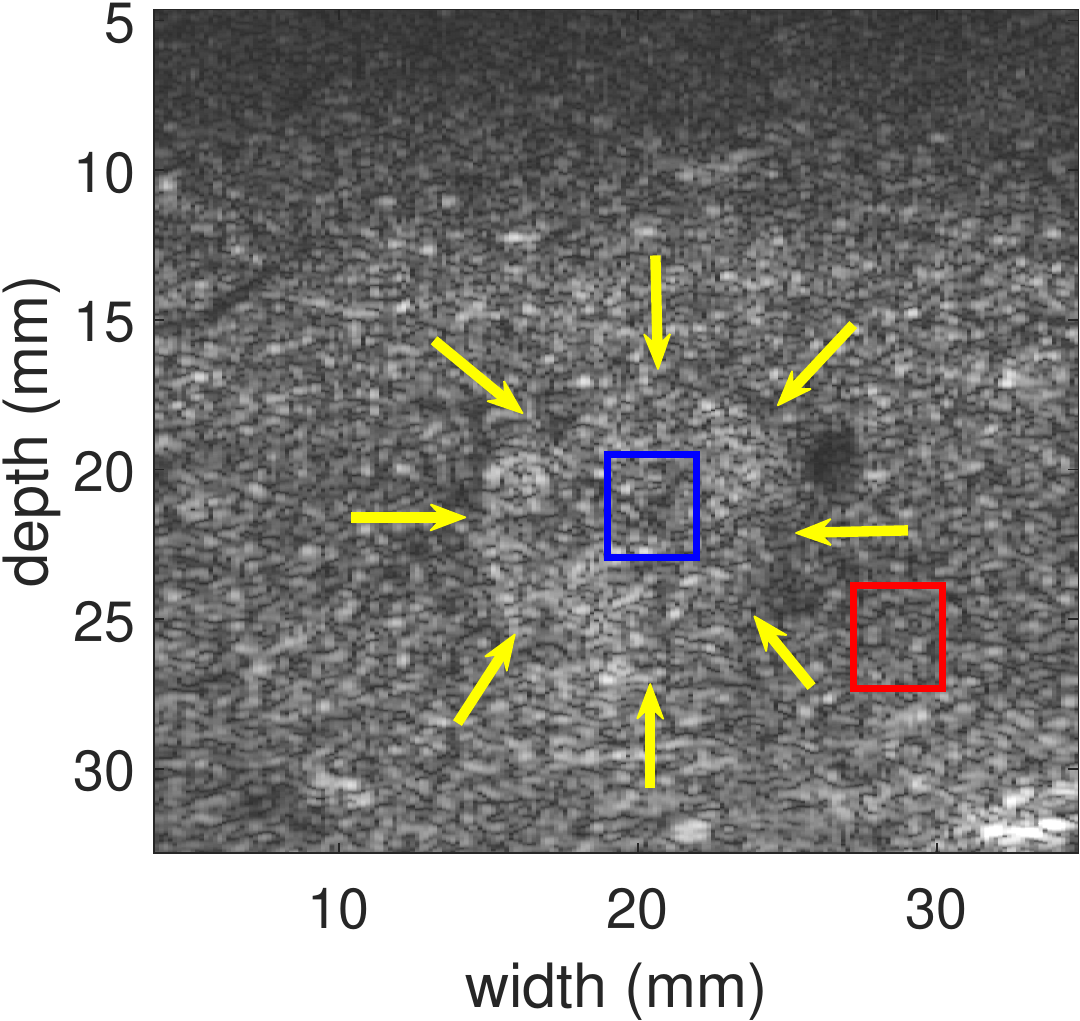}}}%
	\subfigure[GLUE]{{\includegraphics[width=.2\textwidth]{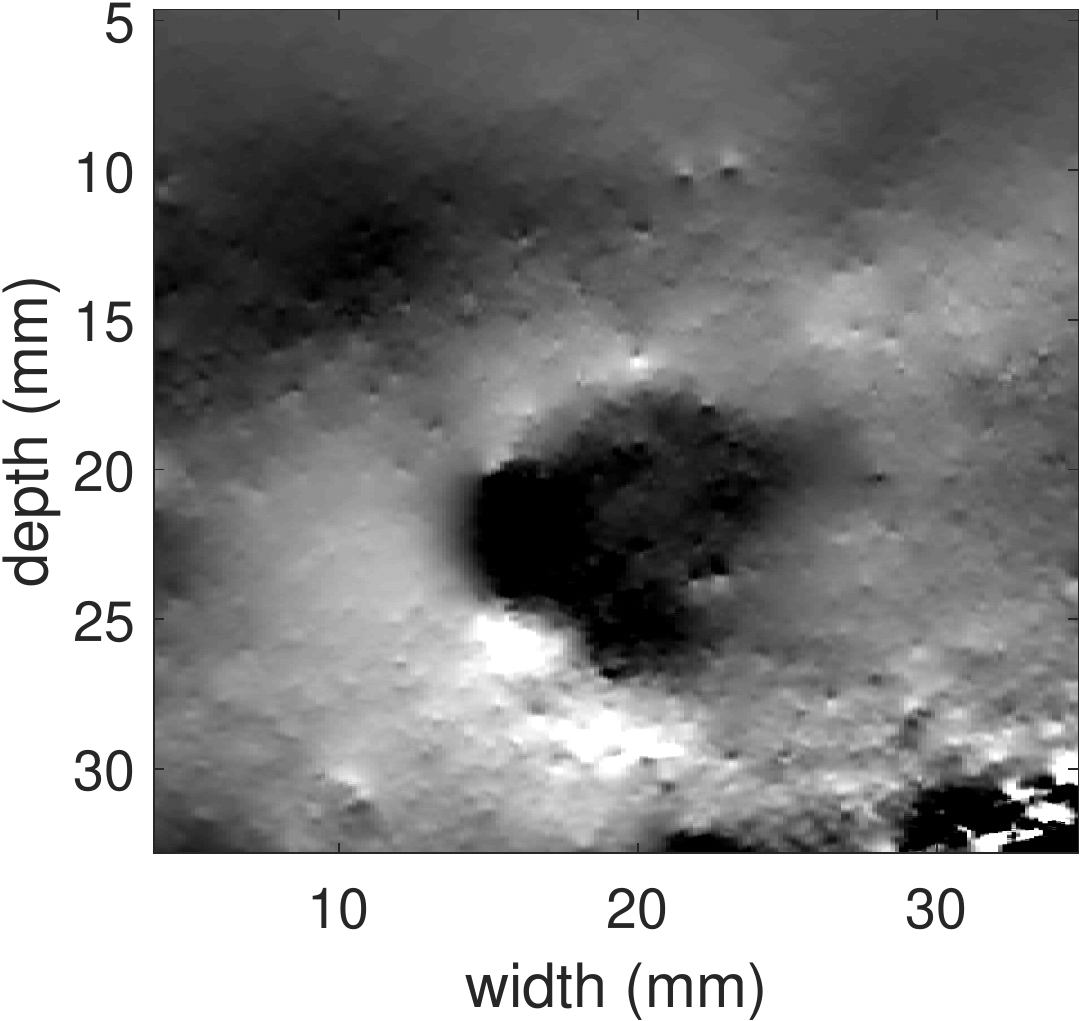}}}%
	\subfigure[OVERWIND]{{\includegraphics[width=.2\textwidth]{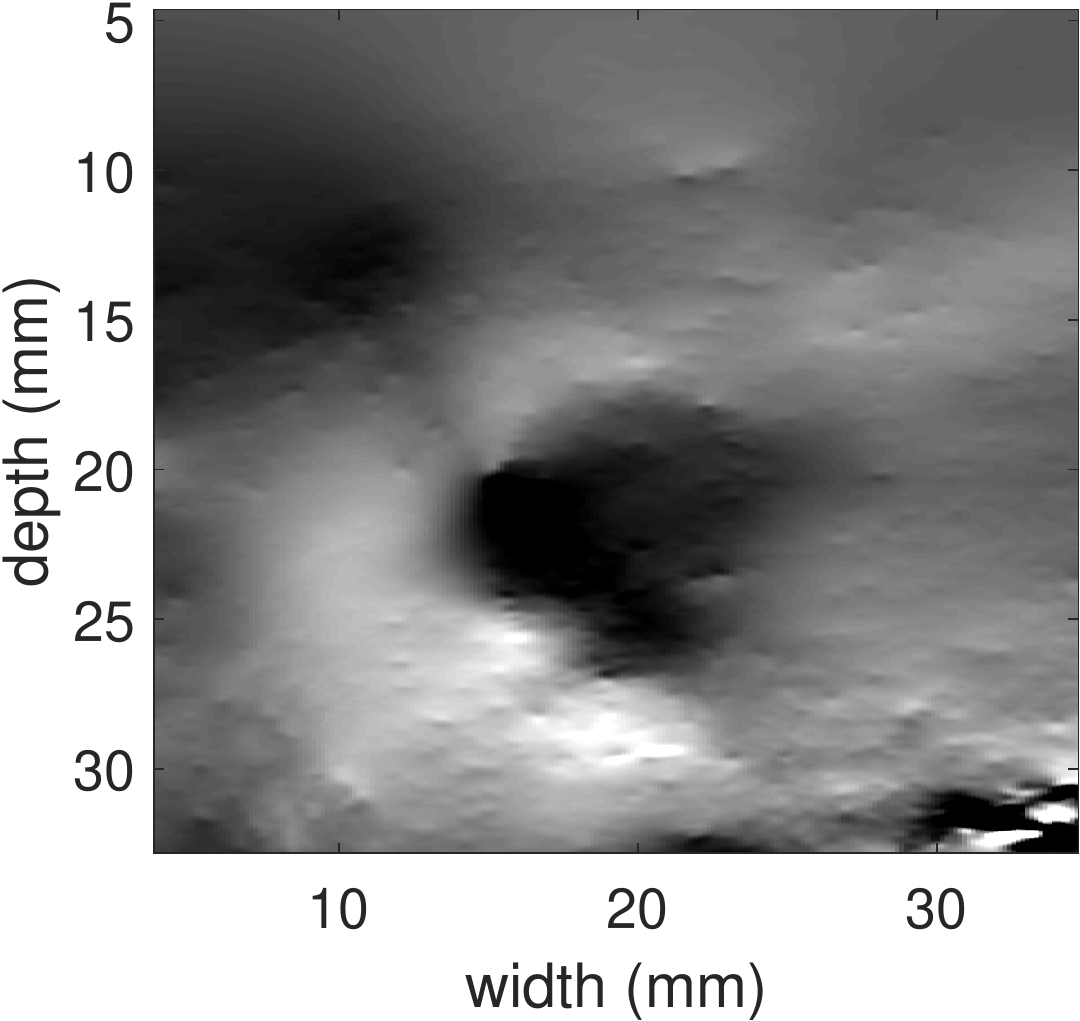} }}%
	\subfigure[$L1$-SOUL]{{\includegraphics[width=.2\textwidth]{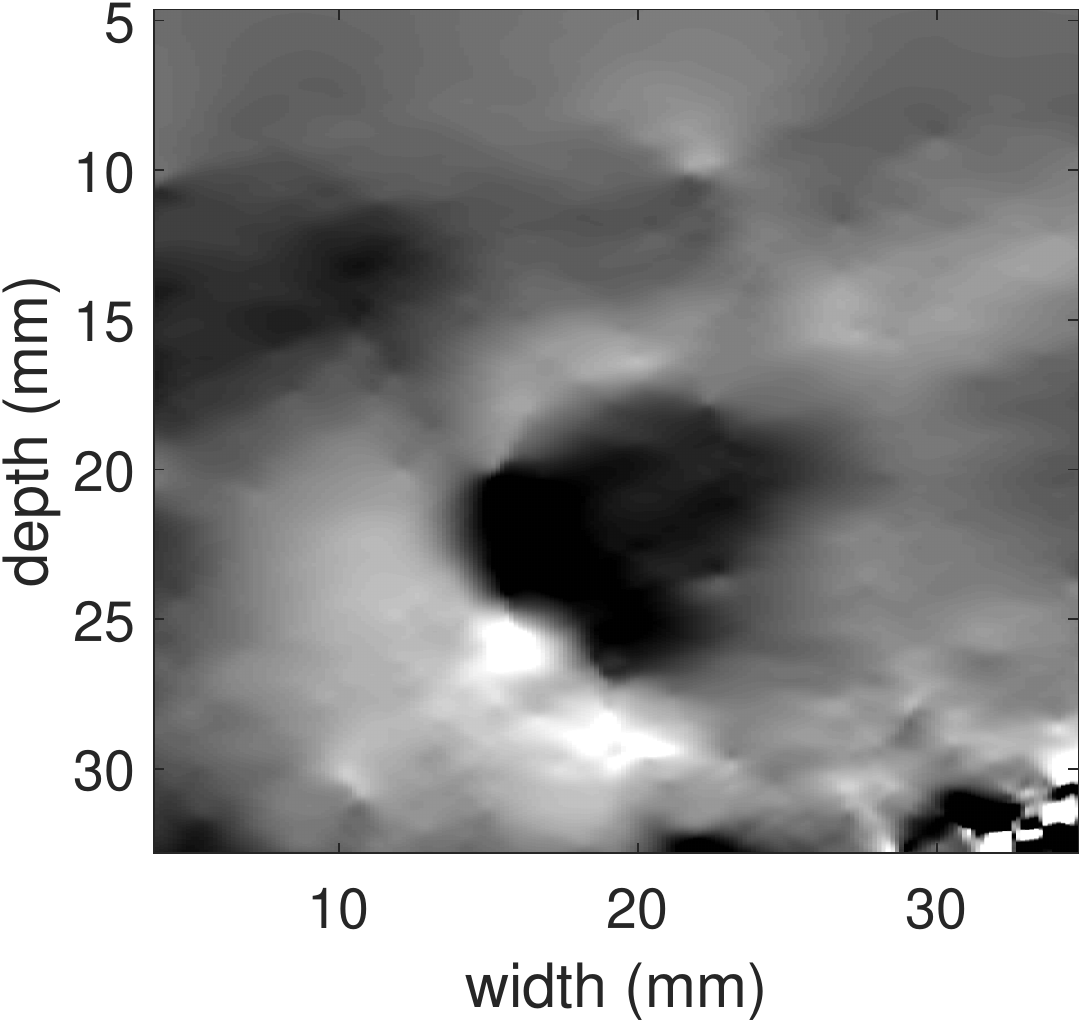} }}%
	\subfigure[ALTRUIST]{{\includegraphics[width=.2\textwidth]{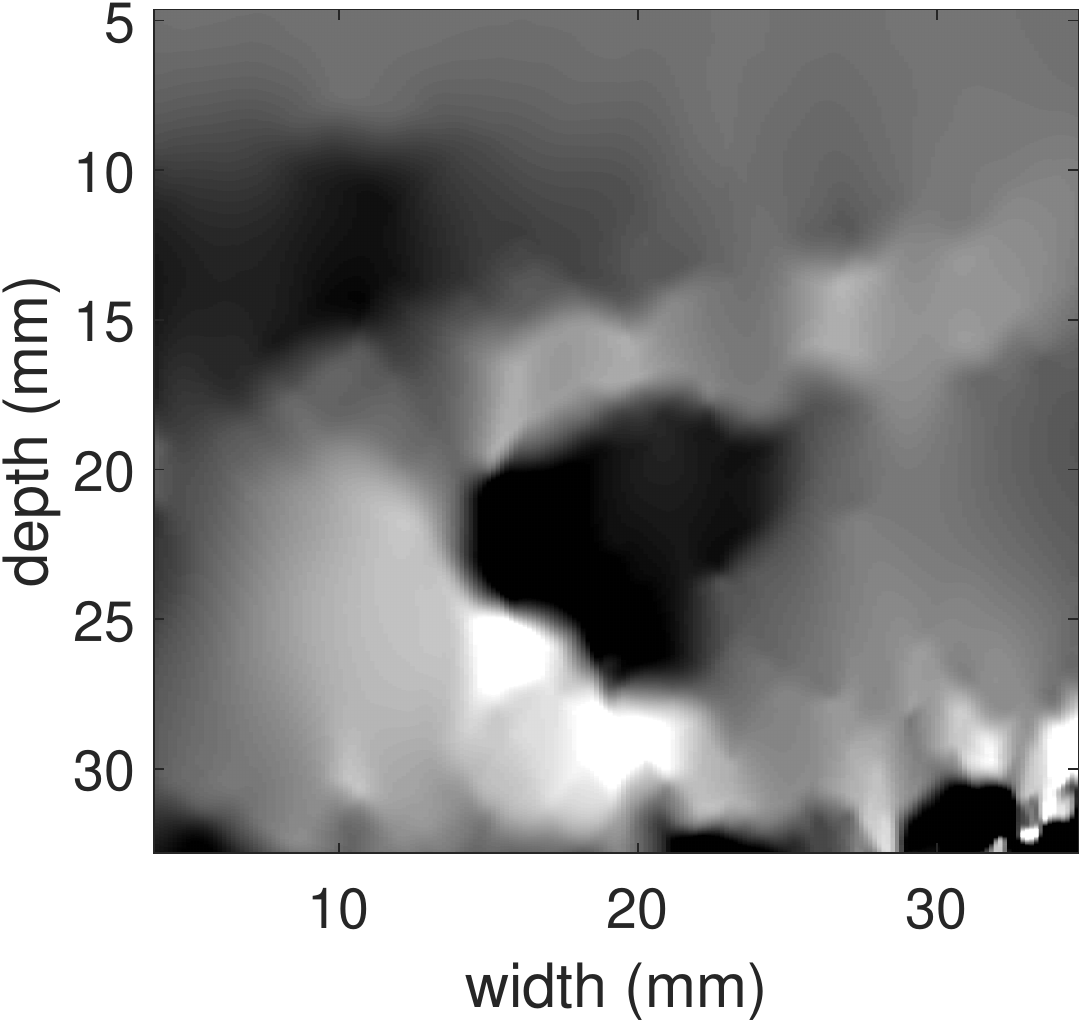} }}
	\subfigure[B-mode]{{\includegraphics[width=.2\textwidth]{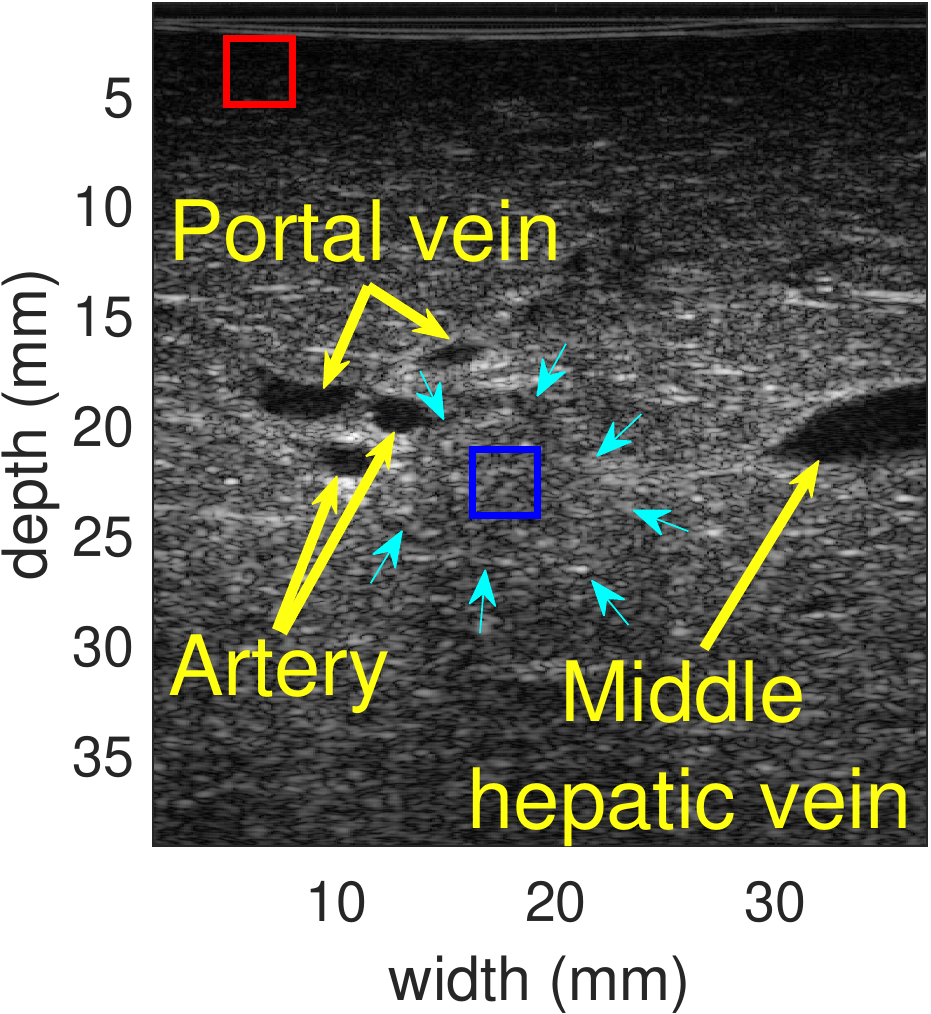}}}%
	\subfigure[GLUE]{{\includegraphics[width=.2\textwidth]{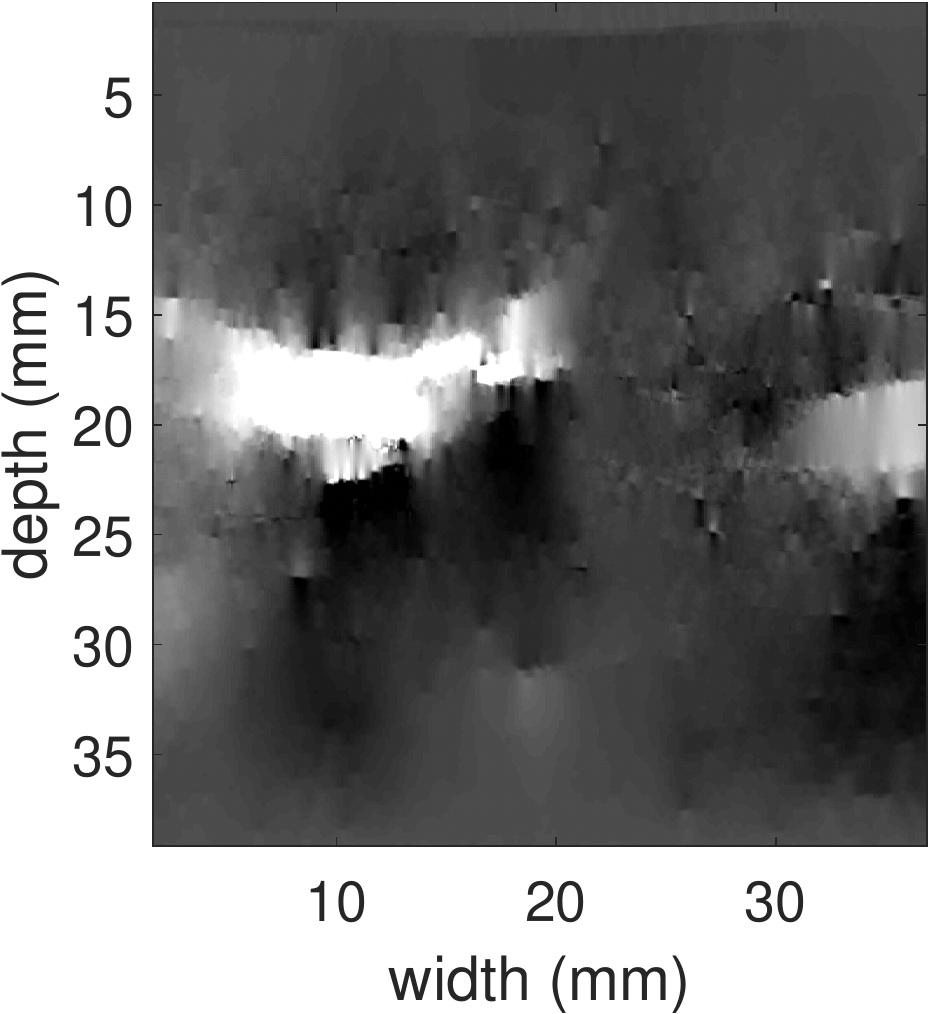}}}%
	\subfigure[OVERWIND]{{\includegraphics[width=.2\textwidth]{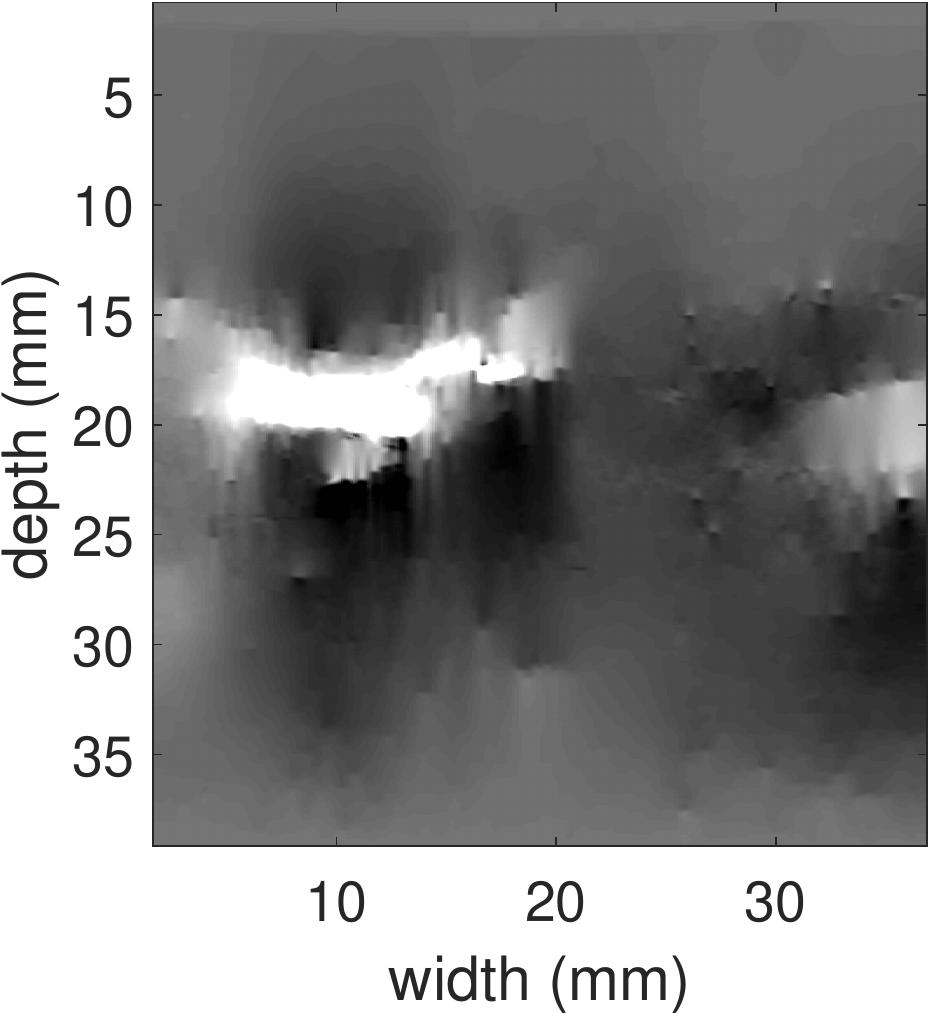} }}%
	\subfigure[$L1$-SOUL]{{\includegraphics[width=.2\textwidth]{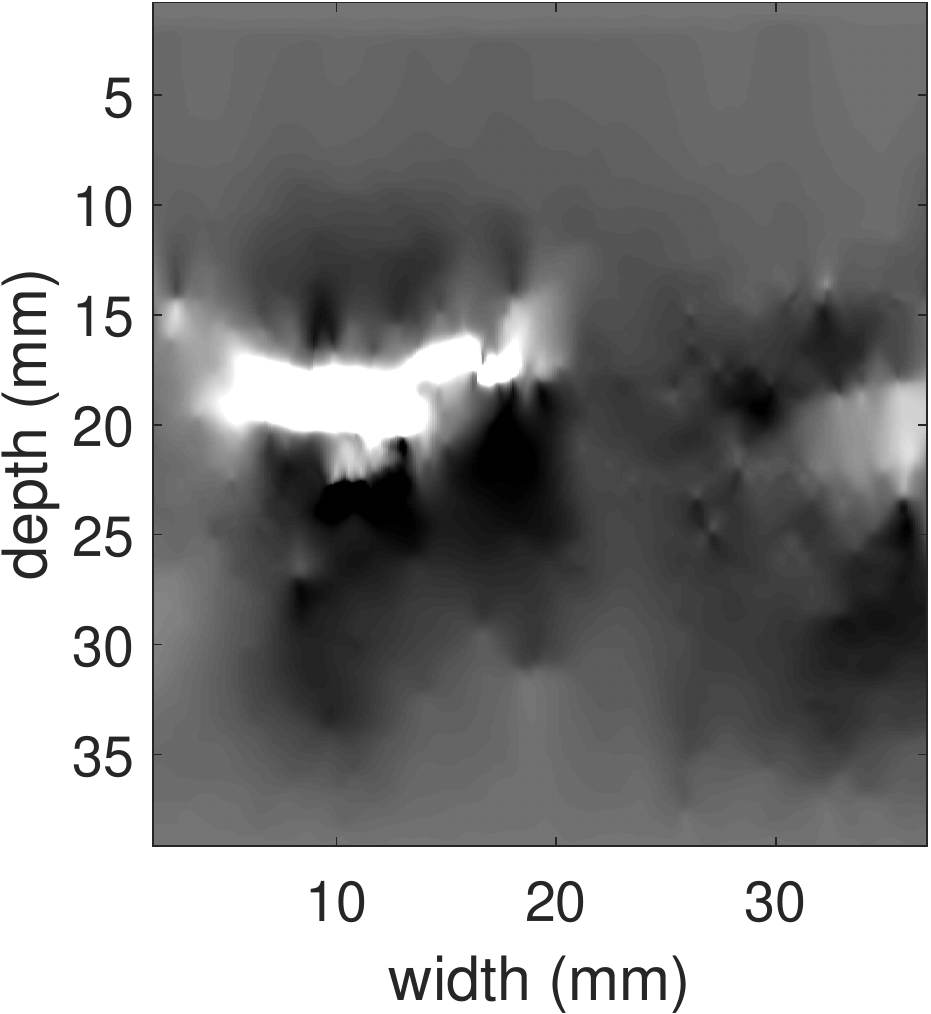} }}%
	\subfigure[ALTRUIST]{{\includegraphics[width=.2\textwidth]{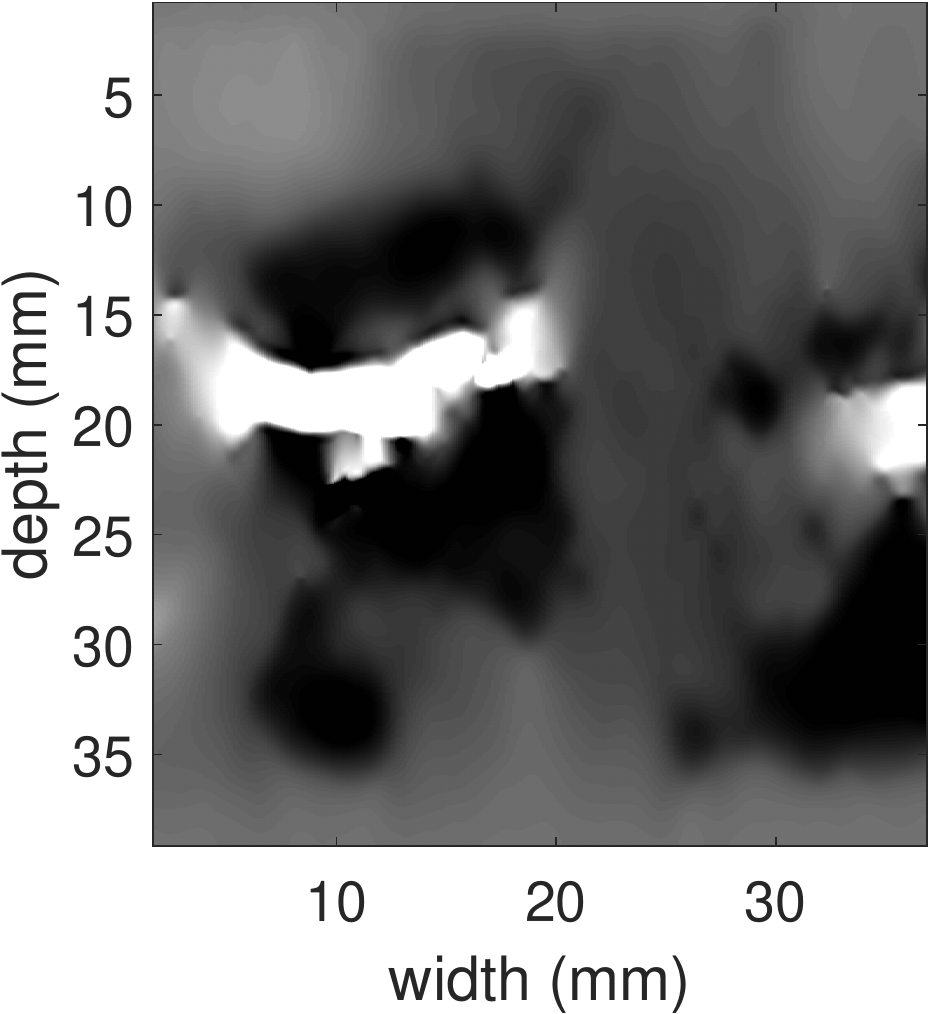} }}
	\subfigure[Axial strain, patient 1]{{\includegraphics[width=.25\textwidth]{Results/liver_p1/gray/color_p1}}}%
	\quad
	\subfigure[Axial strain, patient 2]{{\includegraphics[width=.25\textwidth]{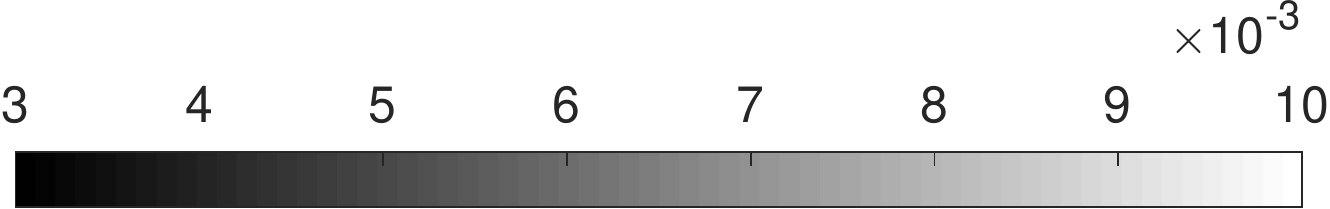}}}
	\quad
	\subfigure[Axial strain, patient 3]{{\includegraphics[width=.25\textwidth]{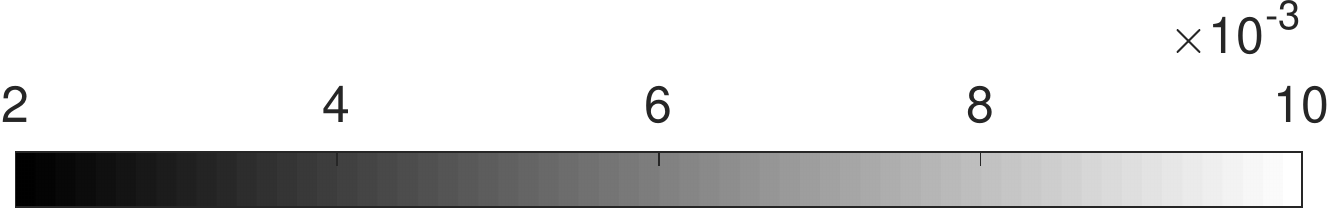}}}
	\caption{Axial strain results from the \textit{in vivo} liver datasets. Rows 1-3 correspond to patients 1, 2, and 3, respectively, whereas columns 1 to 5 present the B-mode and the strain images obtained from GLUE, OVERWIND, $L1$-SOUL, and ALTRUIST, respectively. The blue foreground and red background strain windows for calculating SNR, CNR, and SR are shown on (a), (f), and (k).}
	\label{liver}
\end{figure*}

\begin{table*}[tb]  
	\centering
	\caption{SNR, CNR, and SR for the \textit{in vivo} liver cancer datasets. CNR and SR are calculated utilizing the target and background windows shown in Figures~\ref{liver}(a), \ref{liver}(f), and \ref{liver}(k), whereas SNR values are calculated on the background windows.}
	\label{table_vivo}
	\begin{tabular}{c c c c c c c c c c c c c c c c}
		\hline
		\multicolumn{1}{c}{} &
		\multicolumn{3}{c}{Patient 1} &
		\multicolumn{1}{c}{} &
		\multicolumn{3}{c}{Patient 2} &
		\multicolumn{1}{c}{} &
		\multicolumn{3}{c}{Patient 3}\\
		\cline{2-4} 
		\cline{6-8}
		\cline{10-12} 
		$ $  $ $&    SNR & CNR & SR $ $  $ $&$ $  $ $ &$ $  $ $ SNR & CNR & SR $ $  $ $&$ $  $ $ &$ $  $ $ SNR & CNR & SR\\
		\hline
		GLUE &  24.83 &  17.65 & 0.43 && 13.83 & 7.37 & \textbf{0.54} && 103.81 & 10.76 & 0.54\\
		OVERWIND & 62.85 &  28.60 & 0.42 && 24.78 & 12.63 & 0.55 && 62.28 & 16.55 & 0.49\\
		$L1$-SOUL & 83.95 &  37.09 & 0.32 && 23.24 & 12.96 & \textbf{0.54} && 87.88 & 14.60 & 0.40\\
		ALTRUIST & \textbf{130.78}  & \textbf{48.79} & \textbf{0.24}  && \textbf{32.52} & \textbf{16.40} & \textbf{0.54} && \textbf{111.49} & \textbf{25.05} & \textbf{0.25}\\
		\hline
	\end{tabular}
\end{table*}

\begin{figure*}
	\centering
	\subfigure[Hard-inclusion simulated phantom]{{\includegraphics[width=.2\textwidth]{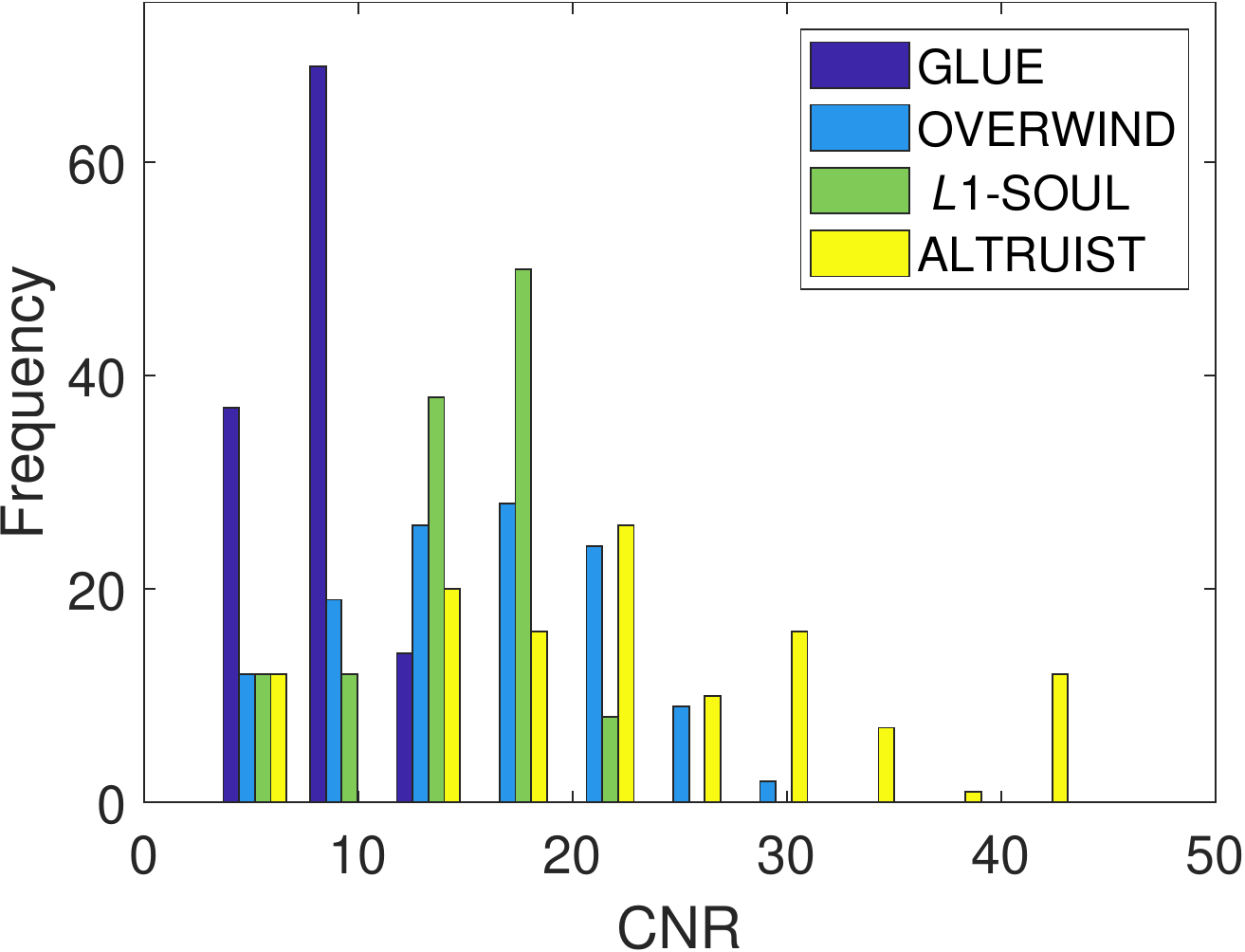}}}%
	\subfigure[Breast phantom]{{\includegraphics[width=.2\textwidth]{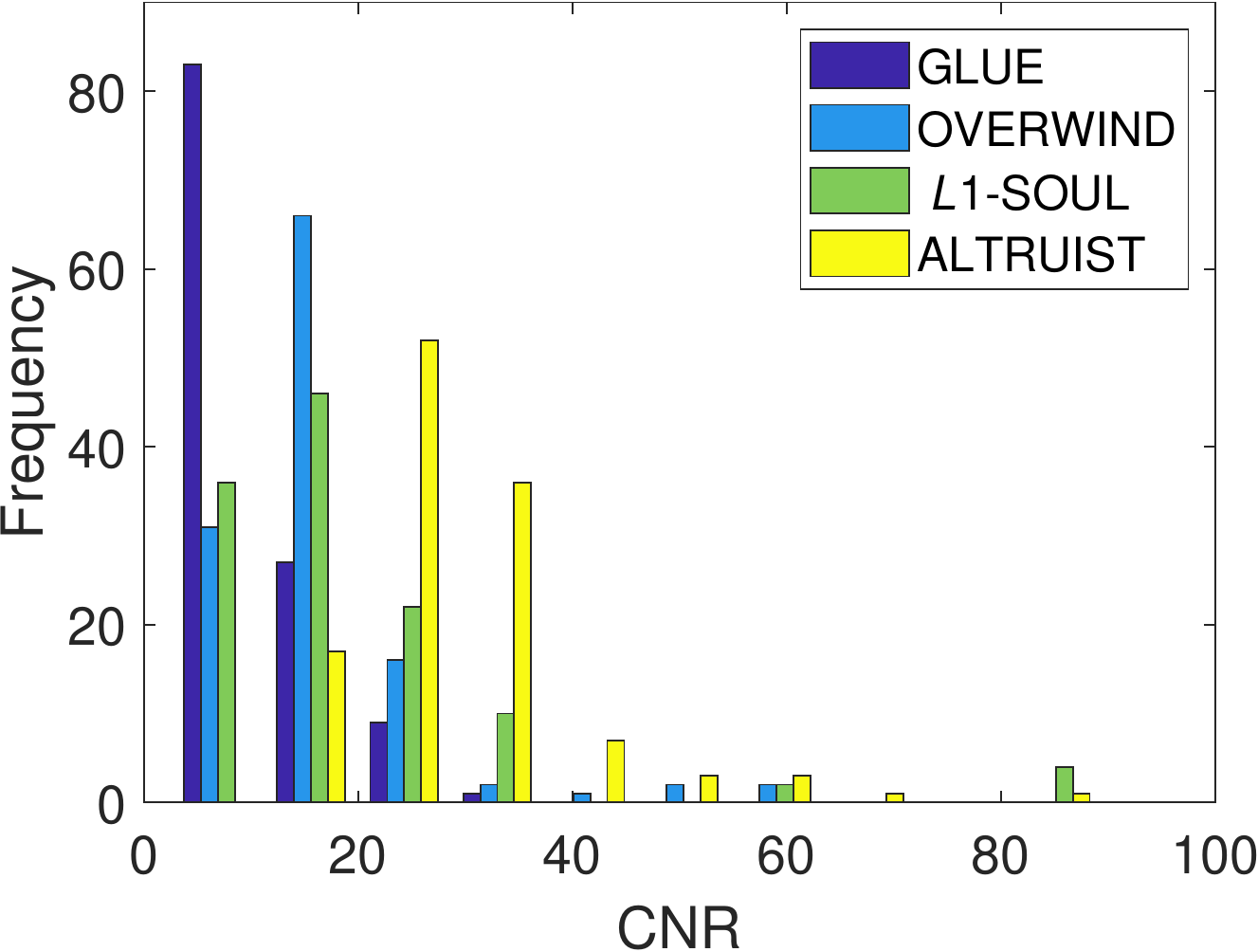}}}%
	\subfigure[Liver patient 1]{{\includegraphics[width=.2\textwidth]{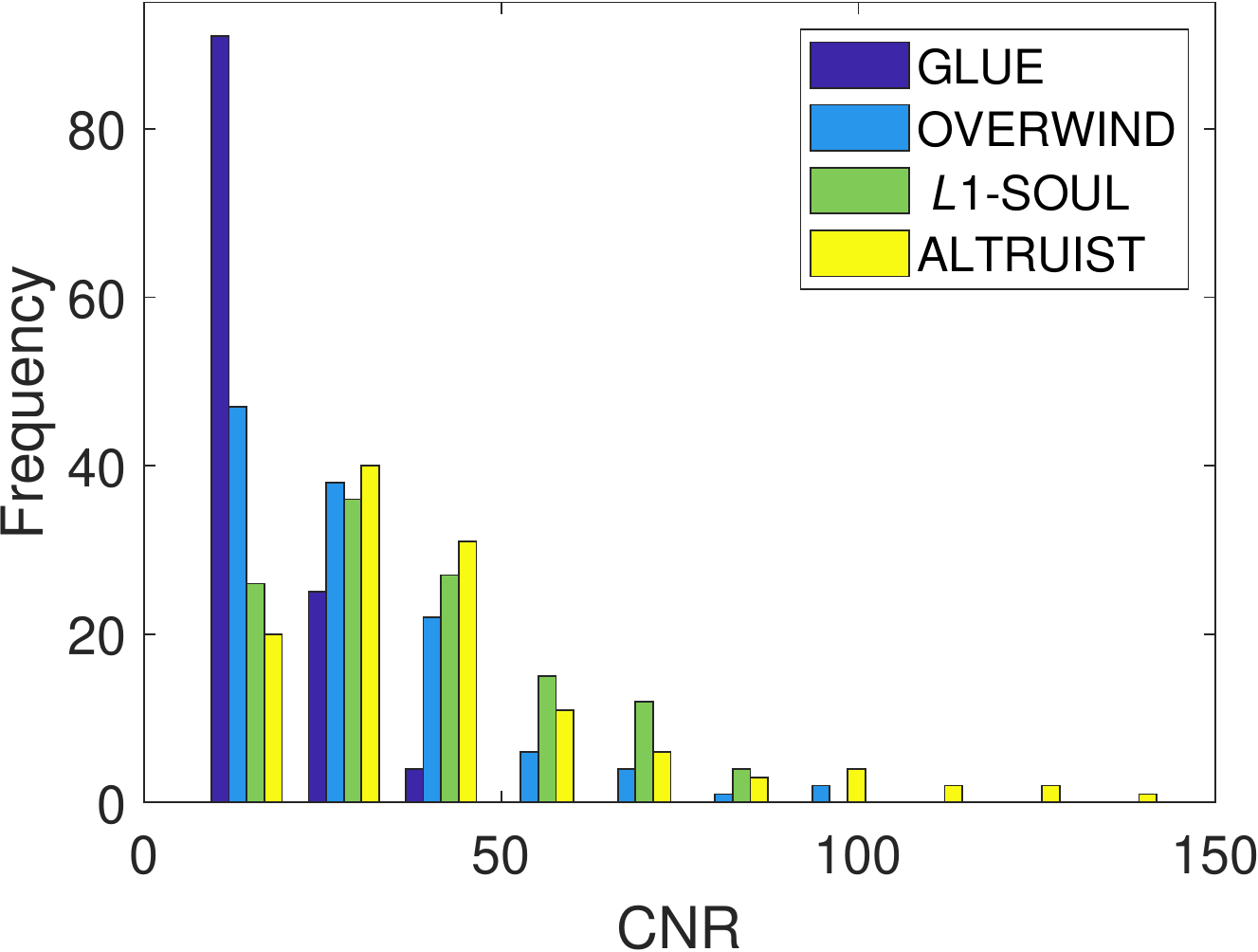}}}%
	\subfigure[Liver patient 2]{{\includegraphics[width=.2\textwidth]{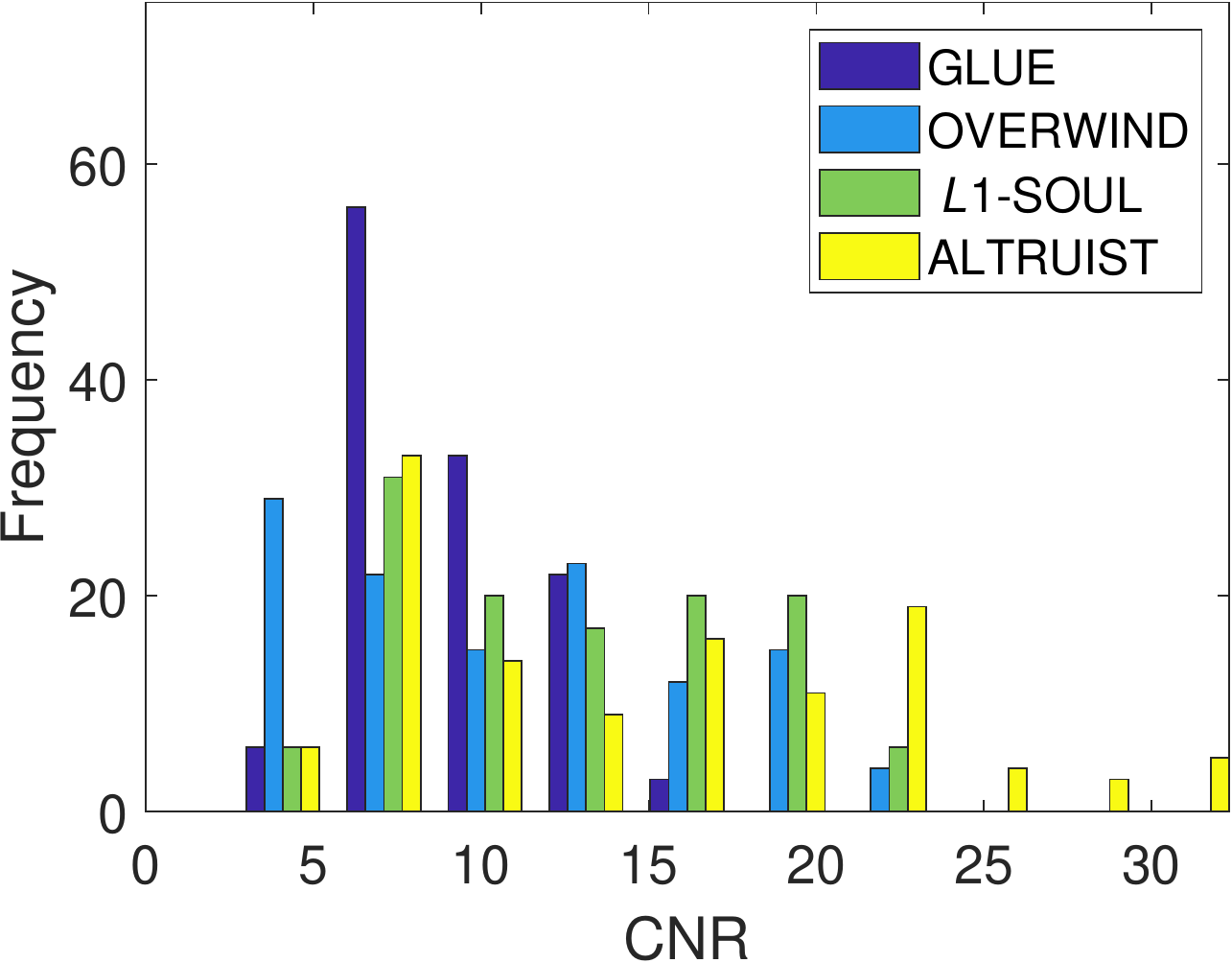}}}%
	\subfigure[Liver patient 3]{{\includegraphics[width=.2\textwidth]{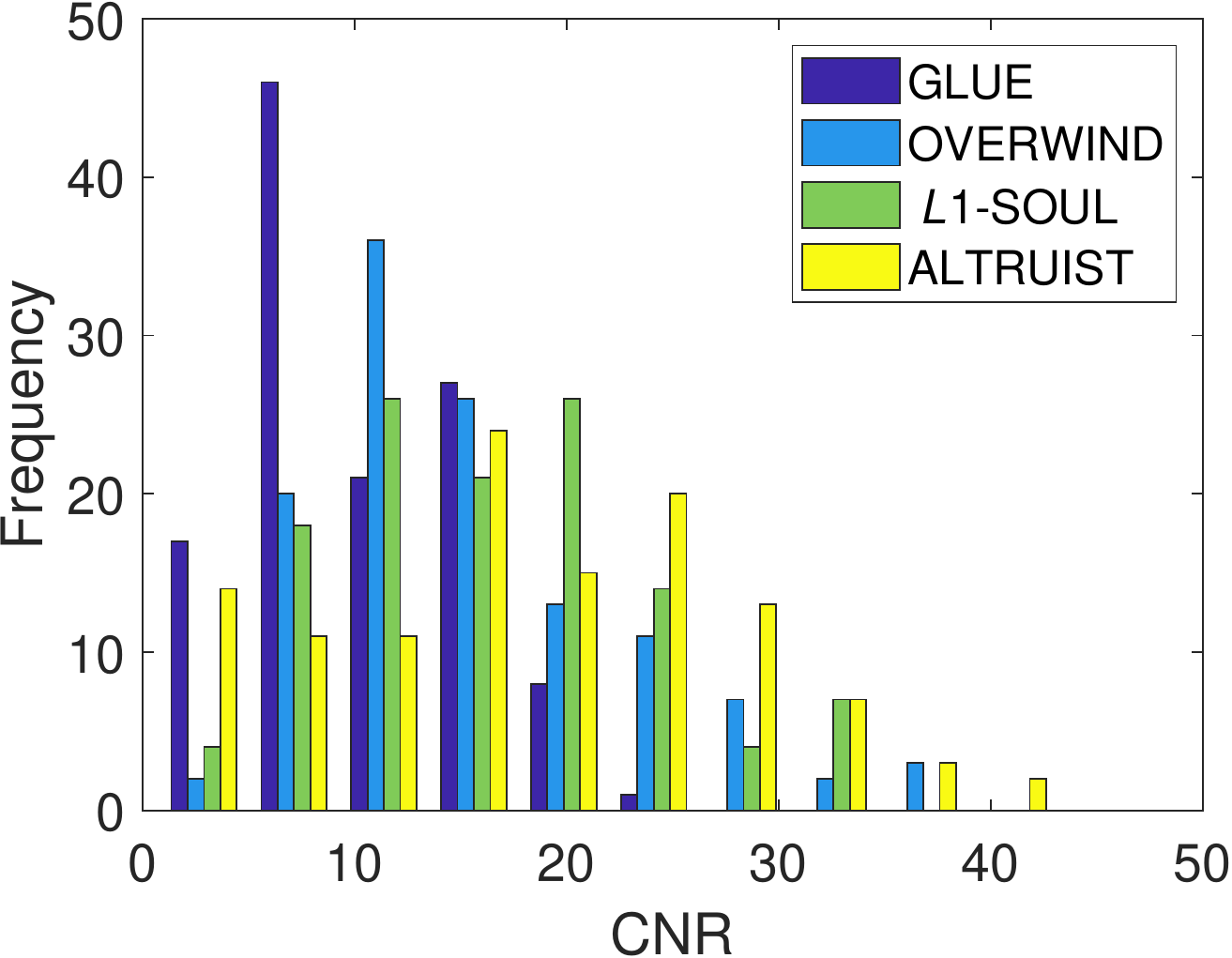}}}%
	\caption{CNR histograms obtained using 120 target-background window combinations. Columns 1 to 5 correspond to the simulated phantom with a hard inclusion, real breast phantom, and \textit{in vivo} liver cancer datasets from patients 1, 2, and 3, respectively.}
	\label{cnr_histograms}
\end{figure*}

\begin{figure}[h]
	\centering
	\subfigure[Residual map]{\includegraphics[width=.2\textwidth]{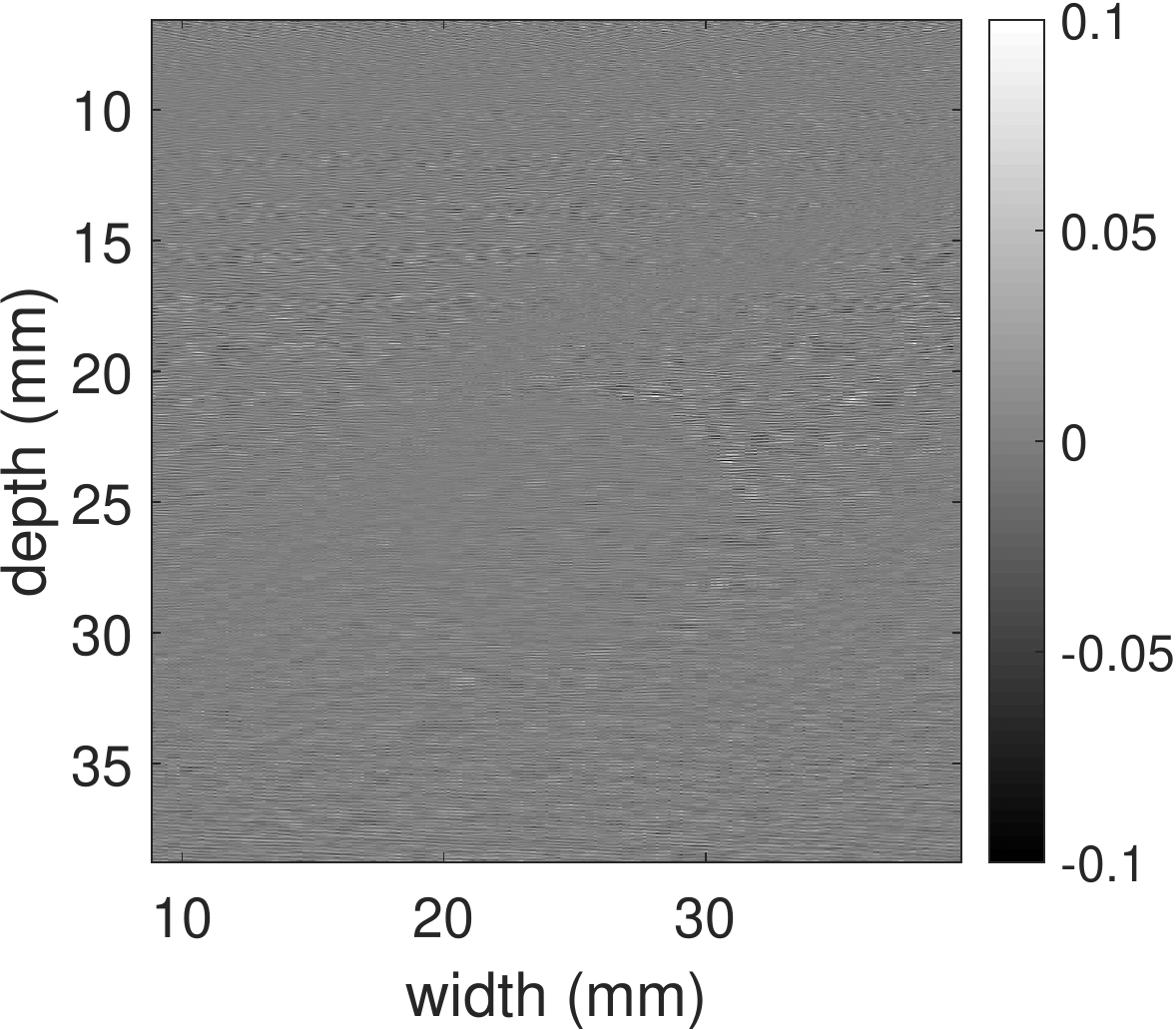}}%
	\subfigure[Residual histogram]{\includegraphics[width=.2\textwidth]{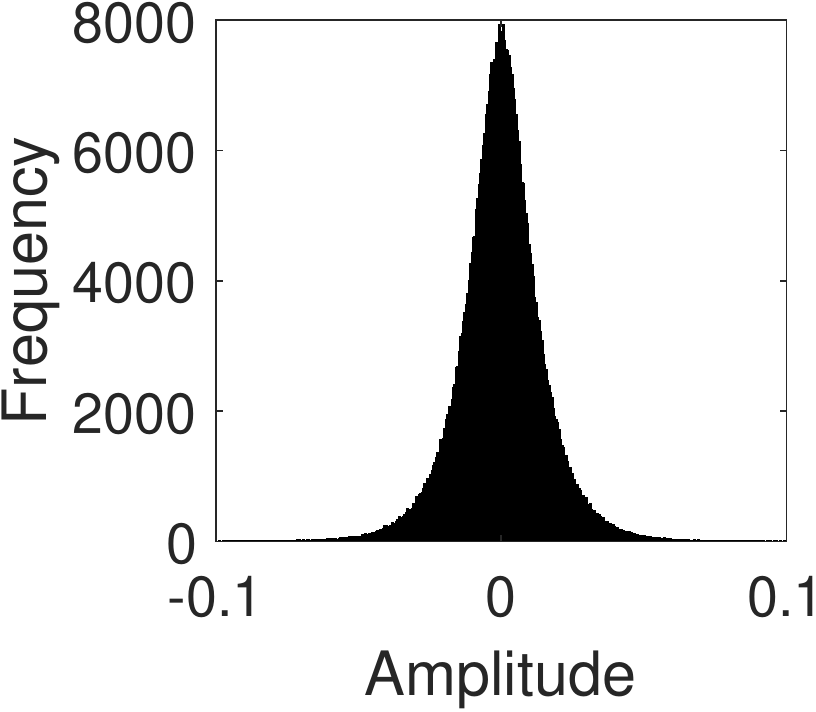}}%
	\caption{Amplitude residual between the pre-deformed and the warped post-deformed RF frames for the breast phantom dataset. Columns 1 and 2 correspond to the residual map and the histogram, respectively.}
	\label{res_hist}
\end{figure}


\begin{figure}[h]
	\centering
	\subfigure[Strain derivative map]{\includegraphics[width=.2\textwidth]{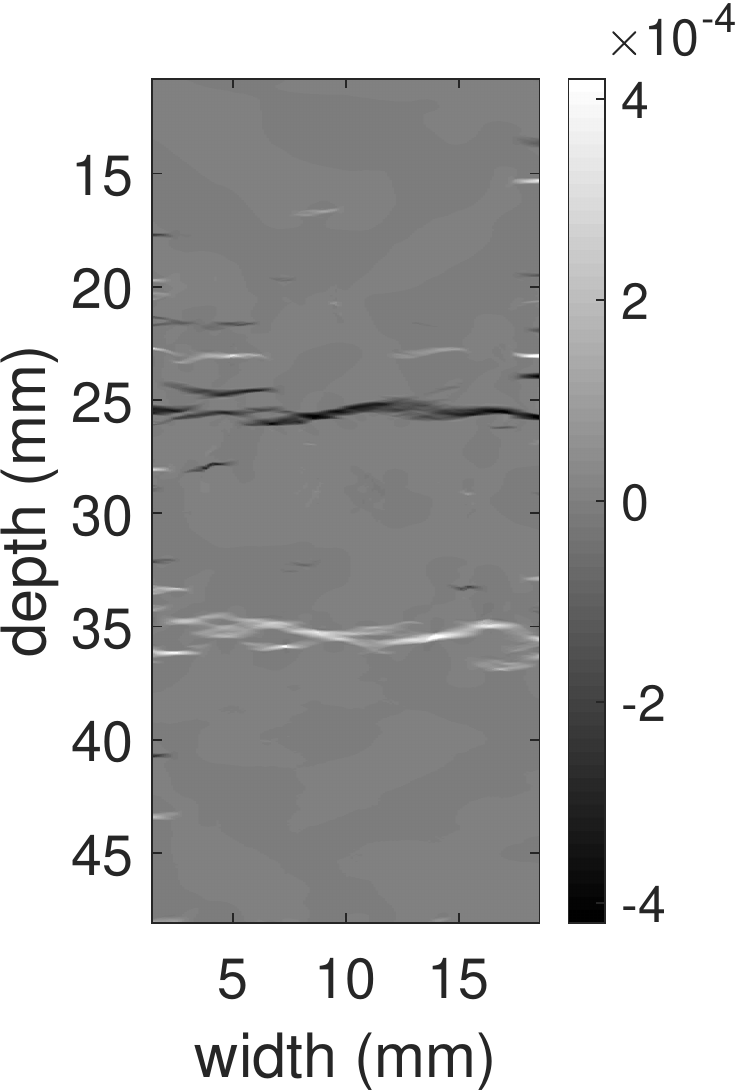}}%
	\subfigure[Strain derivative ESF]{\includegraphics[width=.2\textwidth]{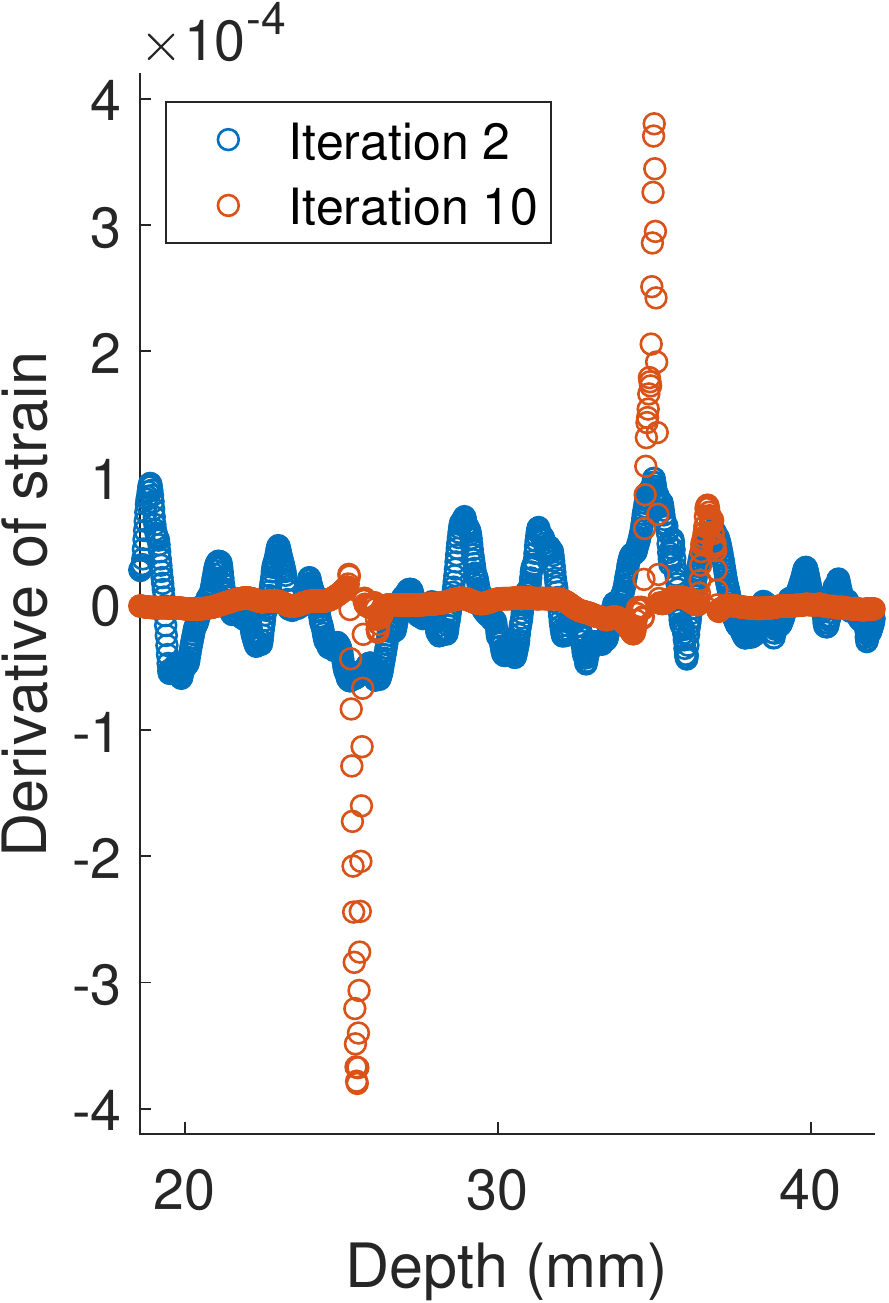}}%
	\caption{Axial derivative of the strain image for the low-contrast simulated layer phantom dataset. Columns 1 and 2 correspond to the strain derivative map and its ESF, respectively.}
	\label{strain_deriv}
\end{figure}

\subsection{\textit{In vivo} Liver Cancer Datasets}
Figure~\ref{liver} shows the B-mode and the axial strain estimates (see Figure 4 of the Supplementary Material for jet color map) obtained by GLUE, OVERWIND, $L1$-SOUL, and ALTRUIST. The patient one's B-mode image shows an echogenic contrast between the tumor and the healthy tissue. On the other hand, the target-background echogenic difference is negligible in case of the other datasets. For all patients, the strain images produced by all four techniques clearly show the pathologic region. In addition to extensive noise in the background, GLUE strain images lack edge clarity. Although OVERWIND yields better noise suppression and edge-preserving ability than GLUE, it is still noisy and suffers from insufficient target-background contrast. Moreover, both GLUE and OVERWIND underestimate the strain in shallow tissue regions in case of patient 2. The GLUE strain image for patient 3 appears to be darker than the ones estimated by the other three techniques. $L1$-SOUL partially resolves the issues of GLUE and OVERWIND providing a smoother background and darker tumor region. For all patients, ALTRUIST substantially outperforms the pervious techniques in terms of target-background contrast, smoothness in the homogeneous region and sharpness at the edges, which is substantiated by the quantitative metric values reported in Table~\ref{table_vivo}.

The high frequencies of ALTRUIST in high CNR values of the histograms (see Figures.~\ref{cnr_histograms}(c), \ref{cnr_histograms}(d), and \ref{cnr_histograms}(e)) demonstrate the superiority of ALTRUIST throughout the strain images. The paired $t$-tests corroborate this inspection by showing that ALTRUIST is statistically better than GLUE, OVERWIND, and $L1$-SOUL with $p$-values of $9.85 \times 10^{-22}$, $1.21 \times 10^{-8}$, and $1.07 \times 10^{-2}$, respectively, for patient 1; $9.08 \times 10^{-16}$, $9.57 \times 10^{-15}$, and $3.96 \times 10^{-6}$, respectively, for patient 2; $6.60 \times 10^{-22}$, $8.22 \times 10^{-4}$, and $5.5 \times 10^{-3}$, respectively, for patient 3.

\section{Discussion}
The penalty function formulated for this work consists of $L1$-norm continuity term and $L2$-norm data term. Ultrasound RF data generally contains additive Gaussian noise. Since $L2$-norm provides the maximum-likelihood estimator in such a scenario, the data term is devised in terms of the $L2$-norm of amplitude differences. Figure~\ref{res_hist} justifies our selection of the data term by demonstrating that the histogram of the amplitude residual between the pre-deformed and the warped post-deformed frames approximates a Gaussian distribution.

One of the strengths of ALTRUIST is that it incorporates the shrinkage function to make a discrete decision on the level of continuity at each RF sample. This task is vital to ensure continuity in the uniform region and sharp discontinuity at the boundaries. ALTRUIST determines the desired derivative values for the next iteration by shrinking the current iteration's displacement derivatives. Figure~\ref{strain_deriv} illustrates this fact for the low-contrast layer phantom dataset by showing how ALTRUIST's iterations gradually converge to the correct level of smoothness.

Both $L1$-SOUL and ALTRUIST formulate the second-order regularizer using only $\partial_{y}^{2}(\cdot)$ and $\partial_{x}^{2}(\cdot)$ terms to keep the cost function simple. However, penalizing $\partial_{xy}^{2}(\cdot)$ terms along with the existing second-order derivative terms is a valid regularization action and may lead to a better TDE, which will be investigated in a future work.   

ALTRUIST substantially outperforms the other three techniques in terms of estimation accuracy, structural similarity with the ground truth, visual contrast, and sharpness. In addition, the simulation experiment conducted in Figure 6 of the Supplementary Material indicates that ALTRUIST manifests superior performance in distinguishing two closely located targets. However, like $L1$-SOUL, it slightly distorts the elastic structure above the hard inclusion in Figure~\ref{hard_simu}. Moreover, ALTRUIST exhibits a spurious edge on the inclusion's left in Figure~\ref{perform_phan}(e), which does not correspond to any physical structure. This might happen due to ALTRUIST's over-sensitivity to a spatial transition of elasticity. A data-driven combination of mechanical-based and continuity constraints is a potential solution to these problems. Another option is to acquire RF frames with known loading conditions and incorporate FEM in an alternating optimization process. However, these modifications are beyond the scope of this work and call for further research.

This work adopts a manual parameter-selection scheme to optimize the performance of ALTRUIST. As described in the Supplementary Material, the parameter values for different datasets vary from each other. Noise distribution, imaging parameters, and the organ's quantitative properties such as attenuation coefficient and scatterer size and distribution are the major contributors to the difference in parameter values. While tuning the parameters, one should set the first- and second-order continuity weights to substantially lower values than $\zeta$. In most cases, optimal strain imaging performance is achieved by assigning lower values to the lateral parameters than the axial ones. As demonstrated in our previous work~\cite{soul,soulmate}, visual appearance of the strain images is robust to a moderate change in the parameter values. In the future, a semi-automatic parameter-optimization strategy will be devised where the continuity parameters in different directions will be combined into a single parameter based on the system's point-spread function (PSF), lateral sampling rate, and tissue characteristics. The optimal value of this parameter for each dataset will be obtained using L-curve~\cite{curve_2001}. Once the continuity parameters are obtained automatically, the optimal TDE performance can be achieved tuning only the shrinkage parameter.

\begin{figure}
		\centering
		\subfigure[SNR]{{\includegraphics[width=0.2\textwidth]{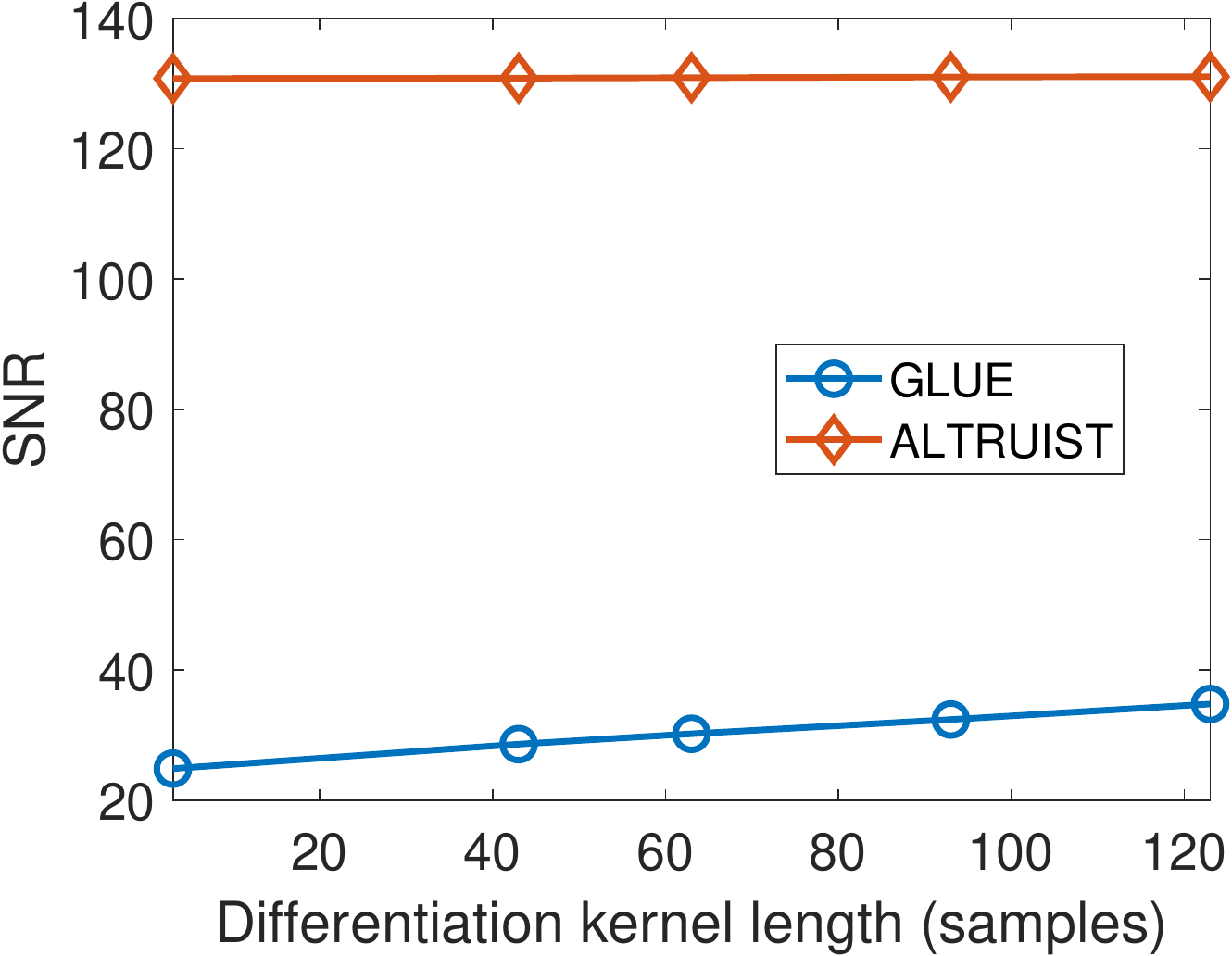} }}%
		\subfigure[CNR]{{\includegraphics[width=0.2\textwidth]{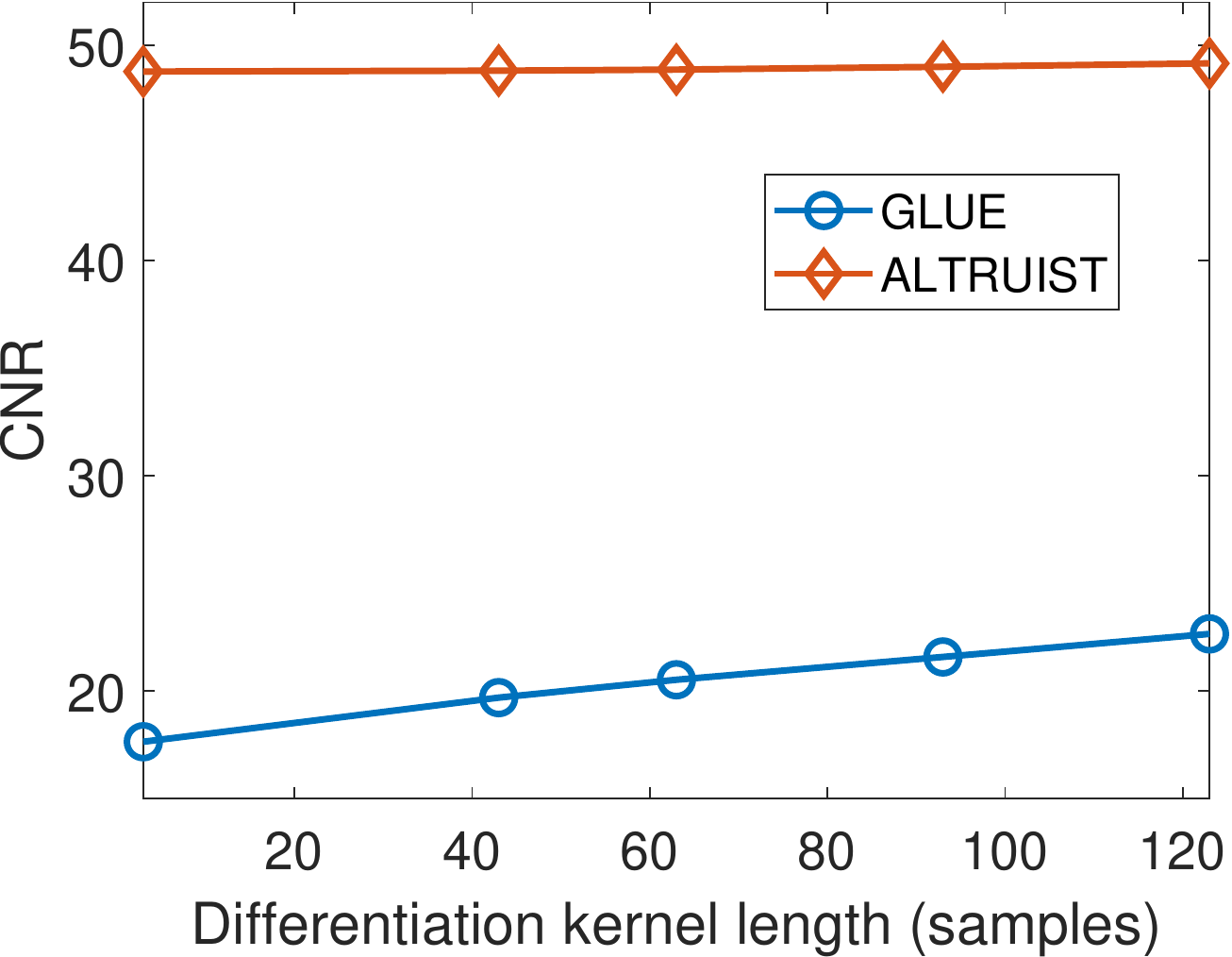} }}%
		\caption{SNR and CNR for the first liver dataset. Columns 1 and 2, respectively, show the SNR and CNR values for different differentiation kernel lengths. The strain windows used for SNR and CNR calculation are shown in Figure~\ref{liver}(a).}
		\label{snr_cnr}
\end{figure}

\begin{figure}[h]
		\centering
		\subfigure[GLUE, large kernel]{{\includegraphics[width=0.18\textwidth]{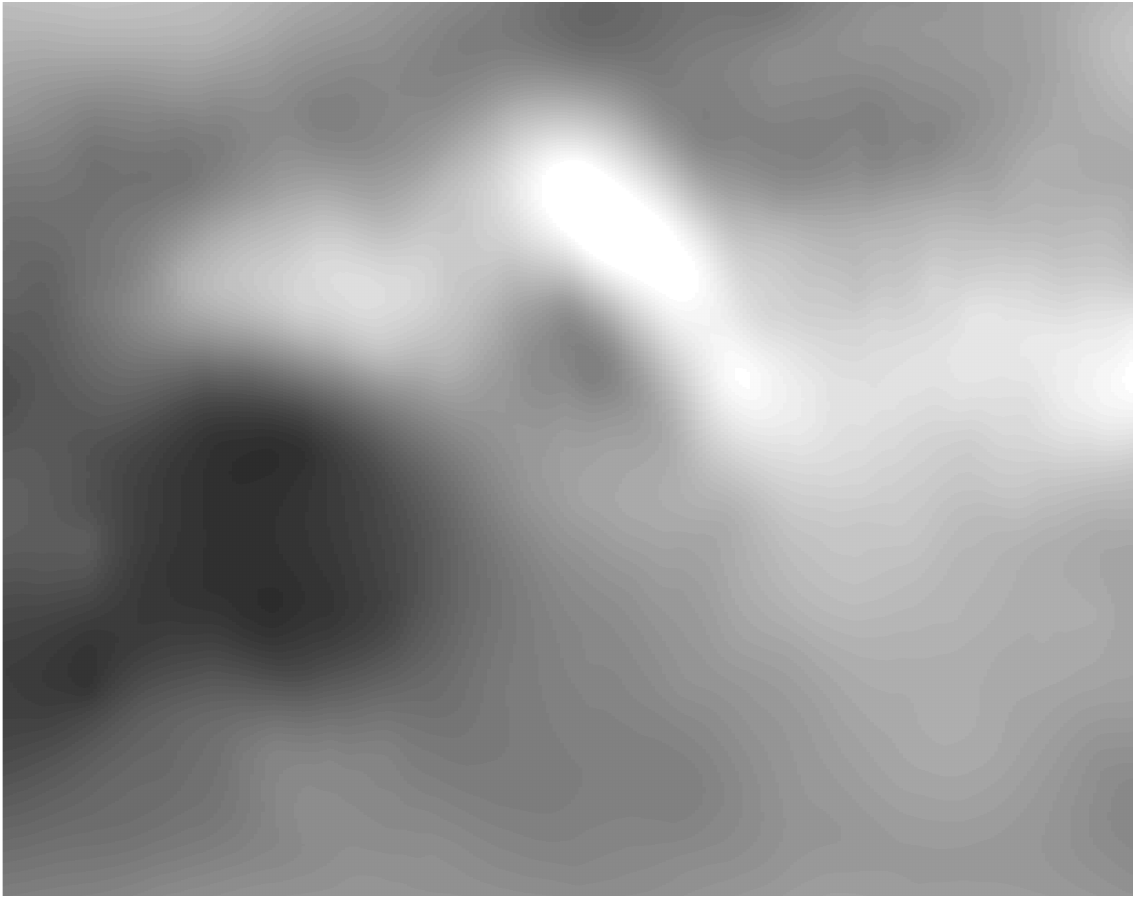} }}%
		\subfigure[ALTRUIST]{{\includegraphics[width=0.18\textwidth]{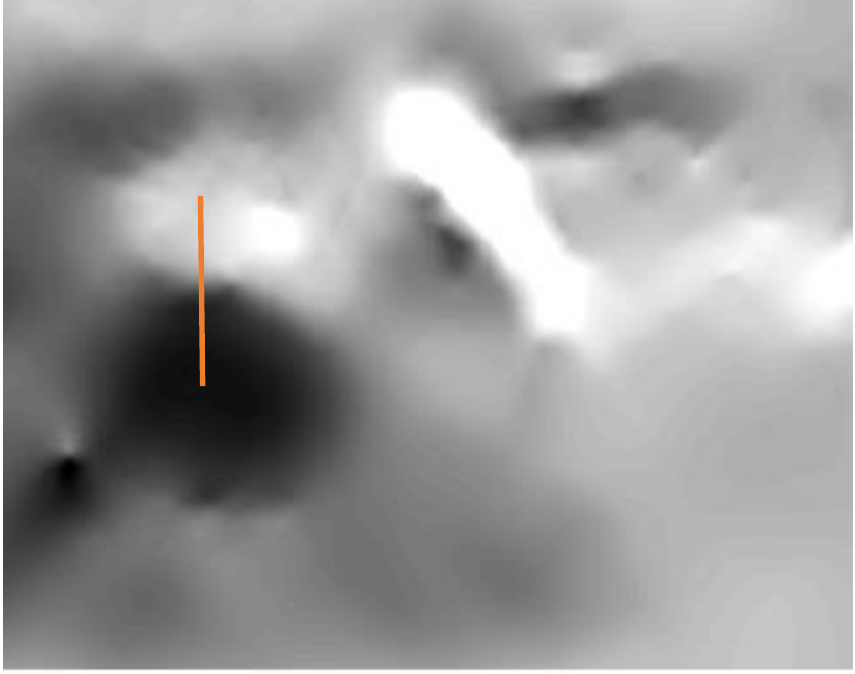} }}
		\subfigure[ESF]{{\includegraphics[width=0.33\textwidth]{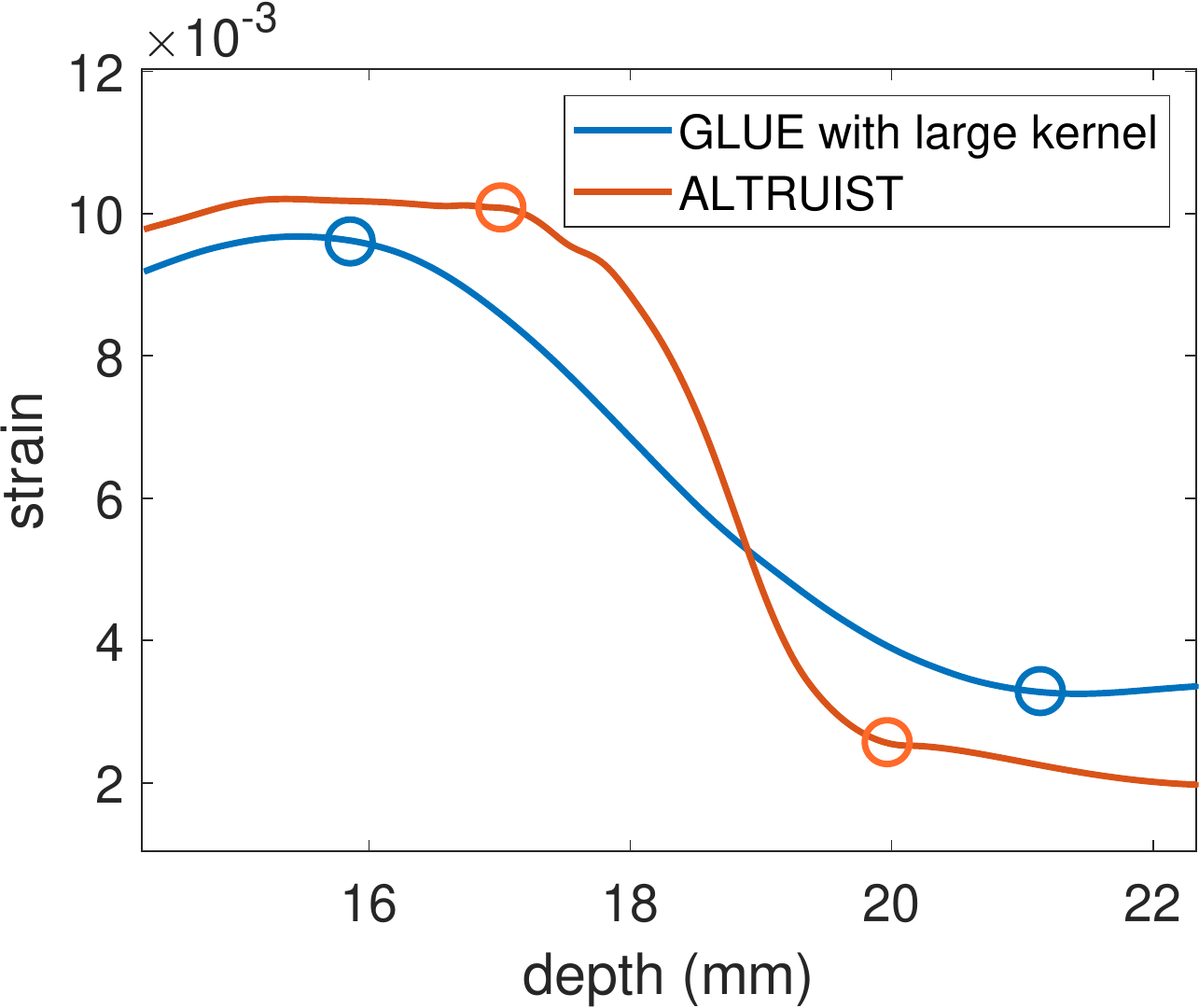} }}
		\caption{ESF for the first liver patient. (a) GLUE axial strain image, differentiation kernel length = 293 samples (b) ALTRUIST axial strain image, differentiation kernel length = 3 samples (c) ESF over the line shown in (b). The blue and orange circles indicate the beginning and ending of edge transitions.}
		\label{esf}
\end{figure}

The validation experiments conducted in this paper manifest that GLUE obtains the noisiest strain image among the techniques under consideration, which is a technical weakness of GLUE. The GLUE strain image can be artificially denoised by increasing the differentiation kernel length while calculating strain from displacement. This non-data-driven denoising step hurts one of the sole purposes of the comparative study that is to assess different TDE technique's inherent abilities to suppress noise while maintaining proper sharpness and contrast. We have carried out two experiments to deeply investigate the interplay between the differentiation kernel size and GLUE's performance. In the first experiment, we sweep the kernel length from 3 to 123 samples and calculate the corresponding SNR and CNR values. Figure~\ref{snr_cnr} demonstrates that a large differentiation kernel cannot compensate for GLUE's poor noise suppression ability, as ALTRUIST maintains substantially higher SNR and CNR values for all kernel lengths. In the second experiment, we use a differentiation kernel of 293 samples for GLUE so that its strain estimate achieves similar background smoothness as the optimal ALTRUIST strain image that uses a 3-sample kernel. Figure~\ref{esf} indicates that the GLUE strain image looses the interesting textures and appears to be washed-out due to an aggressive denoising measure. Figure~\ref{esf}(c) validates this observation by showing a notably wider ESF corresponding to GLUE. 

In this paper, we have formulated the mathematics and implemented ADMM from scratch to exploit its full capacity to optimize our $L1$-norm continuity-based cost function that is especially designed for ultrasonic strain imaging. Being an iterative technique, ALTRUIST's runtime varies depending on the number of iterations required to estimate the displacement field. While an automatic stopping criterion involving RF amplitude residual and sharpness of strain can be introduced, as the maiden attempt of exploiting ADMM for ultrasonic strain imaging, this work ascertains the optimal number of iterations empirically. In our experience, ALTRUIST usually requires 5 iterations (maximum 10) to provide the optimal TDE. While implemented on MATLAB R2018b using a standard computer, GLUE, OVERWIND, $L1$-SOUL, and ALTRUIST, respectively, require 0.49, 2.79, 5.33, and 6.02 seconds to obtain a $1000 \times 100$-sized displacement field. Therefore, ALTRUIST's execution time is similar to that of $L1$-SOUL, which also optimizes $L1$-norm-based first- and second-order continuity constraints.

The lack of carrier signal~\cite{intro26}, wider PSF~\cite{qiong_2017}, and low sampling rate~\cite{jianwen_2009} render displacement estimation in the lateral direction challenging. As a consequence, almost all displacement estimation algorithms exhibit poor lateral tracking ability. The techniques discussed in this work are also not suitable for high-quality lateral estimation since they are not designed to compensate for the lateral information which are not captured by the imaging system. Taking a mechanically-inspired relation among the axial, lateral, and shear strains such as the compatibility condition along with the RF data similarity constraint into account should potentially solve the lateral tracking issue by restoring the lost information using tissue deformation physics. However, this extension is beyond the scope of this work and therefore, further investigation is required to validate its potential.

Despite being well-regarded to be a promising modality to detect tissue abnormality, the wide clinical adoption of ultrasound elastography still remains a challenge due to a number of factors. First, the quality of elastograms generated by many existing techniques are not up-to-the-mark. Second, many elastography algorithms are not good candidates for real-time implementation. Third, the inter-frame motion often contains large out-of-plane component which is unsuitable for elastography. Consistent high performance in all experiments carried out in this study indicates that ALTRUIST resolves the issue associated with elastogram's quality. In addition, ALTRUIST is a fast strain imaging technique, where the runtime can be further accelerated by optimizing its implementation on a GPU. Finally, an appropriate motion pattern between the RF frames can be ensured by concatenating ALTRUIST with a CNN-based automatic frame selection~\cite{zayed2019automatic} technique. Featuring these consequential advancements, ALTRUIST and its extensions can be stepping stones to vast clinical adoption of ultrasound elastography.

This paper demonstrates ALTRUIST's performance only in axial strain imaging. However, ALTRUIST is a generalized TDE technique, which renders substantially higher displacement tracking accuracy than the existing algorithms. Therefore, ALTRUIST and its extensions can potentially be employed in many clinically important applications such as Young's modulus and Poisson's ratio reconstruction~\cite{islam2020non}, vascular permeability~\cite{islam2019estimation}, solid stress and fluid pressure imaging~\cite{islam2021non,islam2019estimation2}, \textit{etc.}, where displacement estimation is an inevitable task.               

Although ALTRUIST successfully obtains high-quality axial strain images, it is based on ADMM, which demands explicit optimizations of both data and prior functions. Therefore, ADMM might not be suitable for handling certain penalty functions which incorporate hard-to-optimize priors. Plug-and-Play (PnP)~\cite{venkatakrishnan2013plug} is an extension of ADMM which can potentially deal with such a situation. PnP converts objectives into actions and obtains the optimal solution from consensus equilibrium instead of solving the inverse problem explicitly. Our future work involves incorporating PnP to optimize a novel cost function consisting of $L2$ data norm and deep learning-based denoising prior.

\section{Conclusion}	
This paper proposes ALTRUIST, a novel algorithm for TDE in ultrasonic strain imaging. ALTRUIST incorporates ADMM for analytically optimizing a robust penalty function containing $L2$ data fidelity norm, $L1$ first- and second-order displacement derivative norms. ADMM ameliorates the issues regarding simultaneous optimization of $L2$ data and $L1$ continuity terms by employing different techniques for optimizing different components of the cost function. In addition, the optimization of the $L1$-norm using the shrinkage operator introduces adaptive smoothing to the displacement estimates, yielding data-driven decisions on the level of continuity at each sample. As such, ALTRUIST exhibits substantially superior performance over previous strain imaging techniques.                       

\section*{Acknowledgment}
This work is supported in part by Natural Sciences and Engineering Research Council of Canada (NSERC) Discovery Grant. Md Ashikuzzaman holds PBEEE and B2X Doctoral Research Fellowships awarded by the Fonds de Recherche du Québec - Nature et Technologies (FRQNT). Dr. Louis G. Johnson Foundation provided partial funding for the purchase of the Alpinion ultrasound machine. Authors thank Drs. E. Boctor, M. Choti, and G. Hager for allowing them to use the liver datasets and Dr. M. Mirzaei for his help tuning the parameters of OVERWIND and for providing the simulated FEM phantom.

\FloatBarrier
\balance
\bibliographystyle{IEEEtran}
\bibliography{ref}

\end{document}